\documentclass[aps, prd, twocolumn, superscriptaddress, nofootinbib, showpacs]{revtex4}

\usepackage{epsf}
\usepackage{bm}
\usepackage{amssymb}
\usepackage{amsmath}
\usepackage{amsfonts}
\usepackage{dcolumn}
\usepackage{color}
\usepackage{natbib}

\usepackage[]{psfig}
\usepackage[]{epsfig}
\usepackage{graphicx}

\newcommand{\z}{\phantom{0}}

\newcommand{\crc}{\mbox{\large $ \circ $}}
\newcommand{\blt}{\mbox{\large $ \bullet $}}
\newcommand{\tms}{\mbox{\footnotesize $ \times $}}

\newcommand{\coconut}{\textsc{CoCoNuT}}

\newcommand{\mmsun}{M_\odot}

\newcommand{\zz}{\phantom{00}}
\newcommand{\nummodel}{114}
\newcommand{\raiseentry}[1]{\smash{\raise 0.7 em \hbox{#1}}}

\newenvironment{equationarray*}
{\arraycolsep 0.14 em
\begin{eqnarray*}}
{\end{eqnarray*}}

\definecolor{darkgreen}{rgb}{0.008,0.417,0.067}

\def\mmsun{M_\odot}

\def\apj{Astrophys. J.}
\def\apjl{Astrophys. J. Lett.}
\def\apjs{Astrophys. J. Supp. Ser. }

\def\aap{Astron. Astrophys.}
\def\araa{Ann.\ Rev. Astron. Astroph.}
\def\physrep{Phys. Rep.}
\def\mnras{Mon. Not. Roy. Astron. Soc.}

\def\prl{Phys. Rev. Lett.}
\def\prd{Phys. Rev. D.}

\newcommand{\cf}{\textit{cf.}~}
\newcommand{\ie}{\textit{i.e.,}~}
\newcommand{\eg}{\textit{e.g.,}~}

\begin{document}

\title{Axisymmetric General Relativistic Simulations of the Accretion-Induced
  Collapse of White Dwarfs}

\author{E. B. Abdikamalov}
\affiliation{International School for Advanced Studies (SISSA) and
  INFN, Trieste, Italy}

\author{C.~D.~Ott}
\affiliation{TAPIR, California Institute of
  Technology, Pasadena, CA USA}
\affiliation{Niels Bohr International Academy, Niels Bohr Institute,
  Copenhagen, Denmark}
\affiliation{Center for Computation and Technology, Louisiana State University,
  LA, USA}

\author{L. Rezzolla}
\affiliation{Max-Planck-Institut f\"ur Gravitationsphysik,
  Albert-Einstein-Institut,
  Potsdam, Germany}
\affiliation{
  Department of Physics and Astronomy,
  Louisiana State University,
  Baton Rouge, LA, USA
}

\author{L. Dessart}
\affiliation{Laboratoire d'Astrophysique de Marseille,
   Marseille,
   France}

\author{H. Dimmelmeier}
\affiliation{Department of Physics, Aristotle University of
  Thessaloniki, Greece}

\author{A. Marek}
\affiliation{Max Planck Institute for Astrophysics, Garching, Germany}

\author{H.-T. Janka}
\affiliation{Max Planck Institute for Astrophysics, Garching, Germany}

\date{\today}


\begin{abstract}
  The accretion-induced collapse (AIC) of a white dwarf (WD) may lead
  to the formation of a protoneutron star and a collapse-driven
  supernova explosion. This process represents a path alternative to
  thermonuclear disruption of accreting white dwarfs in Type Ia
  supernovae. In the AIC scenario, the supernova explosion energy is
  expected to be small and the resulting transient short-lived, making
  it hard to detect by electromagnetic means alone. Neutrino and
  gravitational-wave (GW) observations may provide crucial information
  necessary to reveal a potential AIC. Motivated by the need for
  systematic predictions of the GW signature of AIC, we present
  results from an extensive set of general-relativistic AIC
  simulations using a microphysical finite-temperature equation of
  state and an approximate treatment of deleptonization during
  collapse. Investigating a set of 114 progenitor models in
  axisymmetric rotational equilibrium, with a wide range of rotational
  configurations, temperatures and central densities, and resulting
  white dwarf masses, we extend previous Newtonian studies and find
  that the GW signal has a generic shape akin to what is
  known as a ``Type~III'' signal in the literature.  Despite this
  reduction to a single type of waveform, we show that the emitted GWs
  carry information that can be used to constrain the progenitor and
  the postbounce rotation. We discuss the detectability of the emitted
  GWs, showing that the signal-to-noise ratio for current or
  next-generation interferometer detectors could be high enough to
  detect such events in our Galaxy. Furthermore, we contrast the GW
  signals of AIC and rotating massive star iron core collapse and find
  that they can be distinguished, but only if the distance to the
  source is known and a detailed reconstruction of the GW time series
  from detector data is possible.  Some of our AIC models form massive
  quasi-Keplerian accretion disks after bounce. The disk mass is very
  sensitive to progenitor mass and angular momentum distribution. In
  rapidly differentially rotating models whose precollapse masses are
  significantly larger than the Chandrasekhar mass, the
  resulting disk mass can be as large as $ \sim 0.8 M_\odot $.  Slowly
  and/or uniformly rotating models that are limited to masses near the
  Chandrasekhar mass produce much smaller disks or no disk at
  all. Finally, we find that the postbounce cores of rapidly spinning
  white dwarfs can reach sufficiently rapid rotation to develop a
  gravito-rotational bar-mode instability. Moreover, many of our
  models exhibit sufficiently rapid and differential rotation to
  become subject to recently discovered low-$ E_\mathrm{rot} / |W|
  $-type dynamical instabilities.
\end{abstract}


\maketitle

\section{Introduction}
\label{sec:introduction}

Single stars with main-sequence masses $M \lesssim 100\, \mmsun$ end
their nuclear-burning lives as electron-degenerate objects or with
central electron-degenerate cores. More specifically, the end state is
a carbon-oxygen or oxygen-neon white dwarf (WD) in the case of
low-mass stars (\ie with $M \lesssim 6-8\, \mmsun$), or a degenerate
oxygen-neon or iron core embedded in an extended non-degenerate
stellar envelope in the case of more massive stars (\ie $6-8\, \mmsun
\lesssim M \lesssim 100\,\mmsun$) (see, \eg
\cite{herwig:05,woosley_02_a,poelarends:08} and references
therein). Electron-degenerate spherically-symmetric objects become
unstable to radial contraction once their mass exceeds the
Chandrasekhar mass which, assuming zero temperature and no rotation,
is given by $M_{\mathrm{Ch}} = 1.4575\, (Y_e/0.5)^2\,\mmsun\,$, where
$Y_e$ is the number of electrons per baryon, or ``electron
fraction''~\cite{chandrasekhar:38,shapteu:83}. The \emph{effective}
Chandrasekhar mass $M_{\mathrm{Ch,eff}}$ of a WD or a stellar core
increases somewhat with WD/core entropy (\eg \cite{woosley_02_a}) and
can grow considerably by rotation, in which case it is limited only by
the onset of nonaxisymmetric instability (\eg
\cite{ostriker_68_b,yoon_04,shapteu:83}).

The iron core of a massive star is pushed over its Chandrasekhar limit
by the ashes of silicon shell burning and undergoes 
collapse to a protoneutron star (PNS), accelerated by
photodisintegration of heavy nuclei and electron capture
\cite{bethe:90}.  In a high-density sub-$M_\mathrm{ch}$ oxygen-neon
core of a less massive star, electron capture may decrease 
$M_\mathrm{Ch,eff}$, also leading to collapse
\cite{nomoto_84_a,gutierrez_05_a}.  In both cases, if an explosion
results, the observational display is associated with a Type II/Ibc
supernova (SN).

On the other hand, a carbon-oxygen WD can be pushed over its stability
limit through merger with or accretion from another WD
(double-degenerate scenario) or by accretion from a non-degenerate
companion star (single-degenerate scenario). Here, the WD generally
experiences carbon ignition and thermonuclear runaway, leading to a
Type Ia SN and leaving no compact remnant~\cite{livio_99_a}.  However,
at least theoretically, it is possible that massive oxygen-neon
WDs\footnote{Previously, such WDs were expected to have a significant
  central $^{24}{\mathrm{Mg}}$ mass fraction, hence were referred to
  as oxygen-neon-magnesium WDs. Recent work based on up-do-date input
  physics and modern stellar evolution codes suggests that the mass
  fraction of $^{24}{\mathrm{Mg}}$ is much smaller than previously
  thought (\eg~\cite{siess:06,iben:97,ritossa:96}).} formed by
accretion or merger, and, depending on initial mass, temperature, and
accretion rate, also carbon-oxygen WDs, may grow to reach their
$M_{\mathrm{Ch,eff}}$ or reach central densities sufficiently high
($\gtrsim 10^{9.7} - 10^{10}\,{\mathrm{g\,cm^{-3}}}$) for rapid
electron capture to take place, triggering collapse to a PNS rather
than thermonuclear
explosion~\cite{canal:76,saio:85,nomoto:86,mochkovitch_89_a,nomoto_91,saio:98,
  uenishi_03_a,saio_04,yoon_04,yoon_05,yoon_07_a,kalogera_00_a,
  gutierrez_05_a}. This may result in a peculiar, in most cases
probably sub-energetic, low-nickel-yield and short-lived
transient~\cite{nomoto:86,woosley_92_a,fryer_99_a,dessart_06_a,dessart_07_a,metzger:09b}.
This alternative to the Type Ia SN scenario is called
\textit{``accretion-induced collapse''} (AIC) and will be the focus of
this paper.

The details of the progenitor WD structure and formation and the
fraction of all WDs that evolve to AICs are presently uncertain.
Binary population synthesis models
\cite{yungelson_98_a,belczynski_05_a, kalogera_00_a} and constraints
on $r$-process nucleosynthetic yields from previous AIC simulations
\cite{fryer_99_a,qian:07} predict AIC to occur in the Milky Way at a
frequency of $\sim 10^{-5}$ to $\sim10^{-8}\,{\mathrm{yr}}^{-1}$ which
is $\sim 20-50$ times less frequent than the expected rate of standard
Type Ia SNe~(\eg~\cite{vdb:91,madau:98,scannapieco:05,mannucci:05}).
In part as a consequence of their rarity, but probably also due to
their short duration and potentially weak electromagnetic display,
AIC events have not been directly observed (but see \cite{perets:09},
who discovered a peculiar type-Ib SN in NGC 1032 that can be interpreted
as resulting from an AIC).

The chances of seeing a rare galactic AIC are dramatically boosted by
the possibility of guiding electromagnetic observations by the
detection of neutrinos and gravitational waves (GWs) emitted during
the AIC process and a subsequent SN explosion.  GWs, similar to
neutrinos, are extremely difficult to observe, but can carry ``live''
dynamical information from deep inside electromagnetically-opaque
regions. The inherent multi-D nature of GWs (they are lowest-order
quadrupole waves) makes them ideal messengers for probing multi-D
dynamics such as rotation, turbulence, or NS
pulsations~\cite{thorne:87,andersson:03,ott:09rev}.  The detection
prospects for a GW burst from an AIC are significantly enhanced if
theoretical knowledge of the expected GW signature of such an event is
provided by computational modelling. In reverse, once a detection is
made, detailed model predictions will make it possible to extract
physical information on the AIC dynamics and the properties of the
progenitor WD and, hence, will allow ``parameter-estimation'' of the
source.

Early spherically-symmetric (1D) simulations of
AIC~\cite{mayle_88_a,baron_87_b,woosley_92_a} and more recent
axisymmetric (2D) ones~\cite{fryer_99_a,dessart_06_a, dessart_07_a}
have demonstrated that the dynamics of AIC is quite similar to
standard massive star core collapse: During collapse, the WD separates
into a subsonically and homologously collapsing ($v \propto r$) inner
core and a supersonically collapsing outer core. Collapse is halted by
the stiffening of the equation of state (EOS) at densities near
nuclear matter density and the inner core rebounds into the still
infalling outer core. An unshocked low-entropy PNS of inner-core
material is formed. At its edge, a bounce shock is launched and
initially propagates rapidly outward in mass and radius, but 
loses energy to the dissociation of heavy nuclei as well as to
neutrinos that stream out from the optically thin postshock
region. The shock stalls and, in the AIC case (but also in the case of
the oxygen-neon core collapse in super-AGB stars
\cite{kitaura_06_a,burrows:07c}), is successfully revived by the
deposition of energy by neutrinos in the postshock region (\ie the
``delayed-neutrino mechanism''~\cite{bethewilson:85,bethe:90}) or by a
combination of neutrino energy deposition and magnetorotational
effects in very rapidly rotating WDs~\cite{dessart_07_a}.  But even
without shock revival, explosion would occur when the WD surface layer
is eventually accreted through the shock.  Following the onset of
explosion, a strong long-lasting neutrino-driven wind blows off the
PNS surface, adding to the total explosion energy and establishing
favorable conditions for $r$-process
nucleosynthesis~\cite{woosley_92_a,fryer_99_a,
  dessart_06_a,dessart_07_a,arcones:07}. If the progenitor WD was
rotating rapidly (and had a rotationally-enhanced
$M_\mathrm{Ch,eff}$), a quasi-Keplerian accretion disk of outer-core
material may be left after the explosion~\cite{dessart_06_a}.
Metzger~et~al.~\cite{metzger:09,metzger:09b} recently proposed that
this may lead to nickel-rich outflows that could significantly enhance
the AIC observational display.

Rotating iron core collapse and bounce is the most extensively studied
and best understood GW emission process in the massive star collapse
context (see, \eg~\cite{dimmelmeier_08_a} and the historical overview
in~\cite{ott:09rev}). However, most massive stars (perhaps up to $\sim
99\%$ in the local universe) are likely to be rather slow rotators
that develop little asphericity during collapse and in the early
postbounce phase~\cite{heger:05,woosley:06b,ott:06spin} and produce
PNSs that cool and contract to neutron stars with periods above $\sim
10\,\mathrm{ms}$ and parameter $\beta = E_\mathrm{rot}/|W|\lesssim
0.1\%$ \cite{ott:06spin}, where $E_\mathrm{rot}$ is the rotational
kinetic energy and $|W|$ is the gravitational binding energy.  This
does not only reduce the overall relevance of this emission process,
but also diminishes the chances for postbounce gravito-rotational
nonaxisymmetric deformation of the PNS which could boost the overall
GW emission~\cite{ott:09rev}. Axisymmetric rapidly rotating stars
become unstable to nonaxisymmetric deformations if a nonaxisymmetric
configuration with a lower total energy exists at a given $ \beta $
(see~\cite{stergioulas:03} for a review).  The classical high-$\beta$
instability develops in Newtonian stars on a dynamical timescale at
$\beta \gtrsim \beta_\mathrm{dyn} \simeq 27\%$ (the
general-relativistic value is $\beta \gtrsim 25\%$
\cite{baiotti_07_a,Manca07}). A ``secular'' instability, driven by
fluid viscosity or GW backreaction, can develop already at $ \beta
\gtrsim \beta_{\mathrm{sec}} \simeq 14\% $ \cite{stergioulas:03}.
Slower, but strongly differentially rotating stars may also be subject
to a nonaxisymmetric dynamical instability at $ \beta $ as small as $
\sim 1\% $. This instability at low $\beta$ was observed in a number
of recent 3D simulations (\eg \cite{centrella_01_a, shibata:04a,
  ott_05_a, ou_06_a, ott:06phd, cerda_07_b, ott_07_a,
  ott_07_b,scheidegger:08}), and may be related to corotation
instabilities in disks, but its nature and the precise conditions for
its onset are presently not understood \cite{watts:05,saijo:06}.

Stellar evolution theory and pulsar birth spin estimates suggest that
most massive stars are rotating rather slowly (\eg \cite{heger:05,
  ott:06spin}, but also \cite{cantiello:07,woosley:06b} for
exceptions). Hence, rotating collapse and bounce and nonaxisymmetric
rotational instabilities are unlikely to be the dominant GW emission
mechanisms in most massive star collapse events \cite{ott:09rev}. The
situation may be radically different in AIC: Independent of the
details of their formation scenario, AIC progenitors are expected to
accrete significant amounts of mass and angular momentum in their
pre-AIC evolutions \cite{yoon_04,saio_04,yoon_05,uenishi_03_a,
  piersanti_03_a}. They may reach values of $\beta$ of up to $\sim
10\%$ \emph{prior} to collapse, according to the recent work of Yoon
\& Langer~\cite{yoon_04,yoon_05}, who studied the precollapse stellar
structure and rotational configuration of WDs with sequences of 2D
rotational equilibria.  Depending on the distribution of angular
momentum in the WD, rotational effects may significantly affect the
collapse and bounce dynamics and lead to a large time-varying
quadrupole moment of the inner core, resulting in a strong burst of
GWs emitted at core bounce. In addition, the postbounce PNS may be
subject to the high-$\beta$ rotational instability (see
\cite{liu_01_a,liu_02_a} for an investigation via equilibrium
sequences of PNSs formed in AIC) or to the recently discovered
low-$\beta$ instability.

Most previous (radiation-)hydrodynamic studies of AIC have either been
limited to 1D~\cite{mayle_88_a,baron_87_b,woosley_92_a} or were 2D,
but did not use consistent 2D progenitor models in rotational
equilibrium~\cite{fryer_99_a}.  Fryer, Hughes and Holz~\cite{fryer:02}
presented the first estimates for the GW signal emitted by AIC based
on one model of~\cite{fryer_99_a}. Drawing from the Yoon \& Langer AIC
progenitors~\cite{yoon_04,yoon_05},
Dessart~et~al.~\cite{dessart_06_a,dessart_07_a} have recently
performed 2D Newtonian AIC simulations with the multi-group
flux-limited diffusion (MGFLD) neutrino radiation-(M)HD code
VULCAN/2D \cite{livne:93,livne:07,burrows:07a}. They chose two
representative WD configurations for slow and rapid rotation with
central densities of $ 5 \times 10^{10} \ {\mathrm{g\,cm}}^{-3} $ and
total masses of $ 1.46 M_\odot $ and $ 1.92 M_\odot $. Both models
were set up with the differential rotation law of
\cite{yoon_04,yoon_05}. The $1.46 M_\odot$ model had zero rotation in
the inner core and rapid outer core rotation while the $1.92 M_\odot$
was rapidly rotating throughout (ratio $\Omega_\mathrm{max,initial} /
\Omega_\mathrm{center,initial} \sim
1.5$). Dessart~et~al.~\cite{dessart_06_a,dessart_07_a} found that
rapid electron capture in the central regions of both models led to
collapse to a PNS within only a few tens of milliseconds and reported
successful neutrino-driven~\cite{dessart_06_a} and magnetorotational
explosions~\cite{dessart_07_a} with final values of $\beta$ (\ie a few
hundred milliseconds after core bounce) of $\sim 6\%$ and $\sim
26\%$, for the $1.46 \mmsun$ and $1.92 \mmsun$ models,
respectively\footnote{These numbers are for the non-MHD simulations
  of~\cite{dessart_06_a}. In the MHD models of ~\cite{dessart_07_a},
  an $\Omega$-dynamo builds up toroidal magnetic field, reducing the
  overall rotational energy and $\beta$.}. The analysis in
\cite{dessart_06_a,ott:06phd,ott:09rev} of the GW signal of the
Dessart~et~al.\ models showed that the morphology of the AIC rotating
collapse and bounce gravitational waveform is reminiscent of the
so-called Type~III signal first discussed by Zwerger \&
M\"uller~\cite{zwerger_97_a} and associated with small inner core
masses and a large pressure reduction at the onset of collapse in the
latter's polytropic models.

In this paper, we follow a different approach than
Dessart~et~al.~\cite{dessart_06_a,dessart_07_a}. We omit their
detailed and computationally-expensive treatment of neutrino radiation
transport in favor of a simple, yet effective deleptonization scheme
for the collapse phase \cite{liebendoerfer_05_a}. This simplification,
while limiting the accuracy of our models at postbounce times $\gtrsim
5-10\,\mathrm{ms}$, (i) enables us to study a very large set of
precollapse WD configurations and their resulting AIC dynamics and GW
signals and, importantly, (ii) allows us to perform these AIC
simulations in \emph{general relativity}, which is a crucial
ingredient for the accurate modeling of dynamics in regions of strong
gravity inside and near the PNS. Furthermore, as demonstrated by
\cite{dimmelmeier_02_b,dimmelmeier_07_a,dimmelmeier_08_a}, general
relativity is required for qualitatively and quantitatively correct
predictions of the GW signal of rotating core collapse.

We focus on the collapse and immediate postbounce phase of AIC and
perform an extensive set of \nummodel\ 2D general-relativistic
hydrodynamics simulations.  We analyze systematically the AIC dynamics
and the properties of the resulting GW signal. We explore the
dependence of nonrotating and rotating AIC on the precollapse WD
rotational setup, central density, core temperature, and core
deleptonization, and study the resulting PNS's susceptibility to
rotational nonaxisymmetric deformation.  Furthermore, motivated by the
recent work of Metzger~et~al.\ \cite{metzger:09,metzger:09b}, who
discussed the possible enhancement of the AIC observational signature
by outflows from PNS accretion disks, we study the dependence of disk
mass and morphology on WD progenitor characteristics and rotational
setup.

We employ the general-relativistic hydrodynamics code \coconut
~\cite{dimmelmeier_02_b,dimmelmeier_05} and neglect MHD effects since
they were shown to be small in the considered phases unless the
precollapse magnetic field strength is extremely large ($B \gtrsim
10^{12}\, {\mathrm G}$, \eg
\cite{obergaulinger_06_a,dessart_07_a,burrows_07_b}).  We employ a
finite-temperature microphysical nuclear EOS in combination with the
aforementioned deleptonization treatment of \cite{liebendoerfer_05_a}.
The precollapse 2D rotational-equilibrium WDs are generated according
to the prescription of Yoon \& Langer~\cite{yoon_04,yoon_05}.

The plan of the paper is as follows. In Sec.~\ref{sec:methods}, we
introduce the numerical methods employed and discuss the generation of
our 2D rotational-equilibrium precollapse WD models as well as the
parameter space of WD structure and rotational configuration
investigated.  In Sec.~\ref{sec:colldyn}, we discuss the
overall AIC dynamics and the properties of the quasi-Keplerian
accretion disks seen in many models. Sec.~\ref{sec:GW} is devoted to a
detailed analysis of the GW signal from rotating AIC. There, we also
assess the detectability by current and future GW observatories and
carry out a comparison of the GW signals of AIC and massive star iron
core collapse.  In Sec.~\ref{sec:rotinst}, we study the postbounce
rotational configurations of the PNSs in our models and assess the
possibility for nonaxisymmetric rotational instabilities. In
Sec.~\ref{sec:summary}, we present a critical summary and outlook.


\section{Methods and Initial Models}
\label{sec:methods}

\subsection{The General-Relativistic Hydrodynamics Code}
\label{sec:evolution_code}

We perform our simulations in $ 2 + 1 $ dimensions using the
\coconut\ code~\cite{dimmelmeier_02_a, dimmelmeier_05} which adopts
the conformally-flat approximation of general
relativity~\cite{isenberg:08}. This has been shown to be a very good
approximation in the context of stellar collapse to PNSs
\cite{ott_07_a,ott_07_b,cerda:05}. \coconut\ solves the metric
equations as formulated in \cite{CorderoCarrion:2008nf} using spectral
methods as described in \cite{dimmelmeier_05}. The relativistic
hydrodynamics equations are solved via a finite-volume approach,
piecewise parabolic reconstruction, and the HLLE approximate Riemann
solver \cite{einfeldt:88}. \coconut\ uses Eulerian spherical
coordinates $ \{r, \theta\} $ and for our purposes assumes
axisymmetry. For the computational grid, we choose 250
logarithmically-spaced, centrally-condensed radial zones with a
central resolution of $ 250 $ m and 45 equidistant angular zones
covering $ 90^\circ $. We have performed test calculations with
different grid resolutions to ascertain that the grid setup specified
above is appropriate for our simulations. The space between the
surface of the star and outer boundary of the finite difference grid
is filled with an artificial atmosphere. We assume a constant density
and stationary atmosphere in all zones where density drops below a
prescribed threshold of $ 7 \times 10^5 \, {\mathrm{g \, cm}^{-3}} $,
a value marginally larger than the lowest density value in the EOS
table employed in our calculations (cf. Sec.~\ref{sec:eos}). The
atmosphere is reset after each timestep in order to ensure that it
adapts to the time-dependent shape of the stellar surface.  For
further details of the formulations of the hydrodynamics and metric
equations as well as their numerical implementation in \coconut, the
reader is referred
to~\cite{dimmelmeier_08_a,dimmelmeier_05,CorderoCarrion:2008nf}.

The version of \coconut\ employed in this study does not include a
nuclear reaction network. Hence, we, like
Dessart~et~al.~\cite{dessart_06_a,dessart_07_a}, ignore nuclear
burning which may be relevant in the outer core of AIC progenitors
where material is not in nuclear statistical equilibrium (NSE), but
still sufficiently hot for oxygen/neon/magnesium burning to occur.
This approximation is justified by results from previous work of
\cite{woosley_92_a,kitaura_06_a} who included nuclear burning and did
not observe a strong dynamical effect.


\subsubsection{Equation of State} 
\label{sec:eos}

We make use of the finite-temperature nuclear EOS of Shen et al.
(``Shen-et-al EOS'' in the following,~\cite{shen_98_a,shen_98_b})
which is based on a relativistic mean-field model and is extended with
the Thomas-Fermi approximation to describe the homogeneous phase of
matter as well as the inhomogeneous matter composition. The parameter
for the incompressibility of nuclear matter is $ 281 $ MeV and the
symmetry energy has a value of $ 36.9 $ MeV. The Shen-et-al EOS is
used in tabulated fashion and in our version (equivalent to that used
in~\cite{marek_05_a,dimmelmeier_08_a}) includes contributions from
baryons, electrons, positrons and photons.

The Shen-et-al EOS table used in our simulation has 180, 120, and 50
equidistant points in $\log_{10} \rho$, $\log_{10} T$, and $Y_e$,
respectively.  The table ranges are $6.4 \times 10^{5}\, {\mathrm{g \,
    cm^{-3}}} \le \rho \le 1.1 \times 10^{15}\,{\mathrm{g \,
    cm^{-3}}}$, $0.1\,{\mathrm{MeV}} \le T \le 100.0\,{\mathrm{MeV}}$,
and $0.015 \le Y_e \le 0.56$.  Our variant of the Shen-et-al EOS
assumes that NSE holds throughout the entire $\{\rho,T,Y_e\}$
domain. In reality, NSE generally holds only at $T \gtrsim 0.5 $ MeV.
At lower temperatures, a nuclear reaction network and, the advection
of multiple chemical species and accounting for their individual
ideal-gas contributions to the EOS is necessary for a correct
thermodynamic description of the baryonic component of the
fluid. However, since the electron component of the EOS is vastly
dominant in the central regions of AIC progenitors (and also in the
central regions of iron cores), the incorrect assumption of NSE at low
temperatures can lead to only a small error in the overall
(thermo)dynamics of the collapse and early postbounce phase.


\subsubsection{Deleptonization during Collapse and Neutrino Pressure}
\label{sec:deleptonization}

To account for the dynamically highly important change of the electron
fraction $Y_e$ by electron capture during collapse, we employ the
approximate prescription proposed by
Liebend\"orfer~\cite{liebendoerfer_05_a}.  Liebend\"orfer's scheme is
based on the observation that the local $Y_e$ of each fluid element
during the contraction phase can be rather accurately parametrized
from full radiation-hydrodynamics simulations as a function of density
alone. Liebend\"orfer demonstrated the effectiveness of this
parametrization in the case of spherical symmetry, but also argued
that it should still be reliable to employ a parametrization
$\overline{Y_e\!}(\rho)$ obtained from a 1D radiation-hydrodynamics
calculation in a 2D or 3D simulation, since electron capture depends
more on local matter properties and less on the global dynamics of the
collapsing core. On the basis of this argument, a 
$\overline{Y_e\!}(\rho)$ parametrization was applied in the rotating iron core
collapse calculations of
\cite{ott_07_a,ott_07_b,dimmelmeier_07_a,dimmelmeier_08_a,scheidegger:08}.
Here, we use the same implementation as discussed in
\cite{dimmelmeier_08_a} and track the changes in $Y_e$ up to the point
of core bounce. After bounce, we simply advect $Y_e$. Furthermore, as
in \cite{dimmelmeier_08_a}, we approximate the pressure contribution
due to neutrinos in the optically-thick regime ($\rho \gtrsim 2 \times
10^{12}\, {\mathrm{g\,cm^{-3}}}$) by an ideal Fermi gas, following the
prescription of \cite{liebendoerfer_05_a}.  This pressure contribution
and the energy of the trapped neutrino radiation field are included in
the matter stress-energy tensor and coupled with the hydrodynamics
equations via the energy and momentum source terms specified
in~\cite{ott_07_b}.

\begin{figure}
  \centerline{\includegraphics[width = 86 mm, angle = 0]{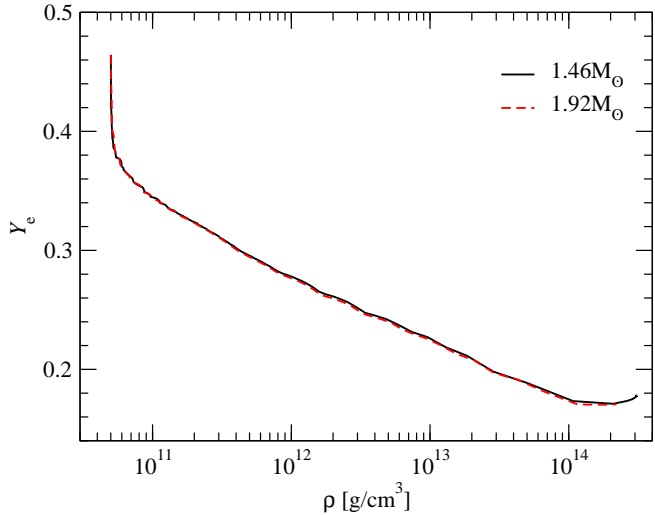}}
  \caption{Average electron fraction $ Y_e $ in the innermost
    $2$ km in the collapsing WD as a function of density obtained from
    2D MGFLD simulations with the VULCAN/2D code for models $ 1.46
    M_\odot $ and $ 1.92 M_\odot $ of
    Dessart~et~al.~\cite{dessart_06_a}.  Both models were set up with
    the same initial dependence of temperature on density and a
    temperature $T_{\mathrm 0} = 1.0 \times 10^{10}$ K (see
    Sec.~\ref{sec:initial_model} for details).} 
  \label{fig:ye_vs_rho_VULCAN}
\end{figure}

The deleptonization scheme described here is applicable only until
core bounce and can neither track the postbounce neutrino burst (see,
\eg \cite{thompson_03_a}) nor neutrino cooling/heating and the postbounce
deleptonization of the PNS. The dynamics in the very early postbounce
evolution (up to $\sim 10\,\mathrm{ms}$) are unlikely to be dramatically
affected by this limitation, but it should be kept in mind when
interpreting results from later postbounce times.

For our AIC simulations we obtain $\overline{Y_e\!}(\rho)$ data from
the 2D Newtonian radiation-hydrodynamics simulations carried out by
Dessart~et~al.~\cite{dessart_06_a} with the VULCAN/2D
code~\cite{burrows_07_b,livne:93,livne:07} in its MGFLD variant.  We
use these data because of their ready availability, but point out that
the microphysics \cite{brt:06} used in VULCAN/2D does not yet include
the updated electron capture rates of \cite{langanke_00_a}. Moreover,
VULCAN/2D presently does not treat velocity-dependent terms in the
transport equation and neglects neutrino--electron scattering, both of
which my have some impact on the evolution of $Y_e$ in the collapse
phase \cite{thompson_03_a,buras_06_a}. In
Fig.~\ref{fig:ye_vs_rho_VULCAN} we plot representative
$\overline{Y_e\!}(\rho)$ trajectories obtained from VULCAN/2D AIC
simulations. At nuclear density, these data predict $Y_e \sim 0.18$
which is low compared to $Y_e \gtrsim 0.22-0.26$ seen in simulations
of iron core collapse \cite{buras_06_a,thompson_03_a,
  liebendoerfer_05_a,dimmelmeier_08_a} and oxygen-neon core collapse
\cite{kitaura_06_a}. This difference is not fully understood, but
({\it i}) could be physical and due to the WD initial data used here
and in \cite{dessart_06_a} or ({\it ii}) may be related to the
radiation transport approximations and microphysics treatment in
VULCAN/2D. To measure the importance of these uncertainties in
$\overline{Y_e\!}(\rho)$, we perform calculations with systematic
variations of $\overline{Y_e\!}(\rho)$ due either to changes in the
precollapse WD temperature or to an ad-hoc scaling (see
Sec.~\ref{sec:tempye}).

Since AIC progenitors may be extremely rapidly rotating, it is not
clear that  the $\overline{Y_e\!}(\rho)$
parametrization is indeed independent of the specific model and rotational
setup.  The $\overline{Y_e\!}(\rho)$ trajectories shown in
Fig.~\ref{fig:ye_vs_rho_VULCAN} result from the collapse simulations
of the slowly rotating $1.46\, M_\odot $ model and of the rapidly
rotating $1.92\, M_\odot$ model of Dessart~et~al.  The very close
agreement of the two curves suggests that rotational effects have only
a small influence on the prebounce deleptonization and confirm the
supposition of~\cite{liebendoerfer_05_a} at the level of the MGFLD and
microphysics treatment in VULCAN/2D.  All $\overline{Y_e\!}(\rho)$
data used in this study are available from~\cite{stellarcollapseAIC}.


\subsection{Precollapse White Dwarf Models}
\label{sec:initial_model}


For constructing 2D WD models in rotational equilibrium with a given
rotation law, we follow~\cite{yoon_05} and employ the self-consistent
field (SCF) method~\cite{ostriker_68_a, ostriker_68_b, hachisu_86,
  komatsu_89_a} in Newtonian gravity. For the purpose of the SFC
method, we assume that the WD is cold and has a constant $Y_e$ of $0.5$.
After finding the 2D equilibrium configuration, we impose a temperature
and $Y_e$ distribution motivated by previous work \cite{dessart_06_a,
woosley_92_a}. Ideally, the WD initial model should be evolved in a
multi-dimensional stellar evolution code with a finite-temperature
EoS and accounting for weak processes such as neutrino cooling and
electron capture. Due to the unavailability of such self-consistent
AIC progenitors, we resort to the treatment that we discuss in detail
in the remainder of this section.

\subsubsection{Implementation of the Self-Consistent Field Method}

Our implementation of the Newtonian SFC method has been tested by
reproducing the WD models presented in~\cite{hachisu_86, yoon_05}, and
finding excellent agreement. The compactness parameter $GM/Rc^2$ of
the highest-density WD models considered here reaches $\sim 5 \times
10^{-3}$, hence general-relativistic effects at the precollapse stage
are small and the error introduced by Newtonian WD models is therefore
negligible. Hereafter we will assume that the \textit{Newtonian mass}
of the equilibrium model represents the \textit{baryon mass} accounted
for when solving the general-relativistic equations.

The equation governing the stellar equilibrium is given by 
\begin{equation}
  \label{eqq:wd_equilibrium}
  \int \rho^{-1} \ d P + \Phi - \int \Omega^2 \ \varpi \ d \varpi = C
  \ ,
\end{equation}
where $ \Phi $ is the gravitational potential, $ \Omega $ is the
angular velocity, $ \varpi $ is the radial cylindrical coordinate and
$ C $ is a constant that will be determined from boundary conditions
using the SCF iterations as discussed below. 

White dwarfs are stabilized against gravity by electron degeneracy
pressure. For constructing precollapse WDs, we assume complete
degeneracy for which the WD EOS (\eg \cite{ostriker_68_b}) is
given by
\begin{equation}
\label{eqq:eos_el_gas}
P = A[x(2x^2-3)(x^2+1)^{1/2}+3\sinh^{-1}x]\,;\, ~ x = \left (\rho / B
\right)^{1/3}, 
\end{equation}
where $A = 6.01 \times 10^{22} ~ {\mathrm{dyn ~ cm^{-2}}} $ and $B =
9.82 \times 10^5\, Y^{-1}_e ~ {\mathrm{g ~ cm^{-3}}} $. We set $Y_e =
0.5$, assuming at this stage
that no electron capture has taken place. The integral $ \int
\rho^{-1} d P $ in Eq.~(\ref{eqq:wd_equilibrium}) is the enthalpy $H$
which, given our choice of WD EOS, can be expressed analytically as
\begin{equation}
  \label{eqq:enthalpy_wd}
  H = \frac{8A}{B} \left[ 1 + \left( \frac{\rho}{B} \right)^{2 / 3}
    \right]\,\,.
\end{equation}
With this, Eq.~(\ref{eqq:wd_equilibrium}) trivially becomes
\begin{equation}
  \label{eqq:wd_equilibrium_2}
  H = C - \Phi + \int \Omega^2 \ \varpi \ d \varpi \ .
\end{equation}

Following the SCF method, we proceed to first produce a trial density
distribution $ \rho(r, \theta) $ and impose a rotation law (discussed
in the following Sec.~\ref{sec:rotlaw}). 

We then calculate $ C $ by using the value for the maximum density and
the angular velocity at the center of the star $ \Omega( \varpi = 0) =
\Omega_{\mathrm{c,i}} $.
Based on the trial density distribution,
we calculate $ H $ via Eq.~(\ref{eqq:wd_equilibrium_2}) and then
update the density distribution based on $H$ using the analytic
expression (\ref{eqq:enthalpy_wd}). This updated density distribution
in turn results in a new value for $H$. We iterate this procedure
until all the maximum absolute values of three relative differences of
$ H $, $ \Omega $ and $ \rho $ become less than $10^{-3}$.


\subsubsection{Progenitor Rotational Configuration} 
\label{sec:rotlaw}

Our axisymmetric progenitor WD models are assumed to be either in
\emph{uniform} rotation or to follow the \emph{differential} rotation
law proposed by Yoon \& Langer \cite{yoon_05}. The latter argued that
the rotation law of a WD that accretes matter at high rates ($ >
10^{-7}\, M_\odot \, {\mathrm{yr}}^{-1}$) is strongly affected by 
angular momentum transport via the dynamical shear instability (DSI)
in the inner region, and due to the secular shear instability (as well
as Eddington-Sweet circulations \cite{tassoul_00}) in the outer
layers. According to their results, the shear rate in the core remains
near the threshold value for the onset of the DSI. This results in a
characteristic rotation law which has an absolute maximum in the
angular velocity just above the shear-unstable core. We define $
\varpi_{\mathrm p} $ as the position of this maximum. This position is
linked to layers with a density as low as several percent of the WD
central density so that
\begin{equation}
\rho_\mathrm{i}(\varpi = \varpi_{\mathrm p}, \ z = 0) = f_{\mathrm p}
\rho_{\mathrm{c,i}}, 
\end{equation}
and where, following~\cite{dessart_06_a, yoon_05}, we choose $
f_{\mathrm p} =\{ 0.05, \ 0.1 \}$ in our models. (Note that the
differential rotation law adopted for the models
of~\cite{dessart_06_a} had $f_{\mathrm p} = 0.05 $). In the inner
regions with $ \varpi < \varpi_{\mathrm p} $, we have
\begin{equation}
\label{eqq:omega_inner_core}
\Omega(\varpi) = \Omega_{\mathrm{c,i}} + \int_0^\varpi \frac{f_{\mathrm{sh}}
  \sigma_{\mathrm{DSI, crit}}}{\varpi'}{d \varpi'},
\end{equation}
where $ \Omega_{\mathrm{c,i}} $ is the angular velocity at the center and $
\sigma_{\mathrm{DSI, crit}} $ is the threshold value of the shear rate
in the inner core for the onset of the DSI.  $f_{\mathrm{sh}}$ is a
dimensionless parameter ($\le 1.0$) describing the deviation of the
shear rate from $ \sigma_{\mathrm{DSI, crit}} $. We compute $
\sigma_{\mathrm{DSI, crit}} $ assuming homogeneous chemical composition
and constant temperature, in which case $ \sigma_{\mathrm{DSI, crit}} $
can be estimated as (\cf Eq.(7) of~\cite{yoon_04}):
\begin{eqnarray}
  \label{eqq:sigma_dsi}
  \sigma_{\mathrm{DSI, crit}}^2 & \simeq & 0.2  \left( \frac{g}{10^9 ~
    {\mathrm{cm ~ s^{-2}}}} \right) \\ \nonumber & \times & \left(
  \frac{\delta}{0.01} \right) \left( \frac{H_\mathrm{p}}{8 \times 10^7 ~
    {\mathrm{cm}} } \right)^{-1} \left( \frac{ \nabla_{\mathrm{ad}} }{ 0.4
  } \right) \ ,
\end{eqnarray}
where $ g $ is the free-fall acceleration, $ H_\mathrm{p} $ is the pressure
scale height ($ =-dr / d \ln P $), $ \nabla_{\mathrm{ad}} $ is the
adiabatic temperature gradient ($ = -(\partial \ln T / \partial \ln
P)_s $ where $s$ is the specific entropy) and $ \delta = (\partial \ln
\rho / \partial \ln T)_P $. The quantities $ \delta $, $ H_\mathrm{p}$
and $ \nabla_{\mathrm{ad}} $ are computed using the routines of
Blinnikov~et~al.~\cite{blinnikov_96}.

At the equatorial surface, the WD is assumed to rotate at a certain
fraction $ f_{\mathrm K} $ of the local Keplerian angular velocity $
\Omega_{\mathrm K}$:
\begin{equation}
\label{eqq:omega_surface}
\Omega(R_\mathrm{e}) = f_{\mathrm K} \Omega_{\mathrm K}(R_\mathrm{e}),
\end{equation}
where $ R_\mathrm{e} $ is the equatorial radius of the WD
and where we have set $
f_{\mathrm K} = 0.95$. In the region between $ \varpi_{\mathrm p} $
and $ R_\mathrm{e} $, we again follow~\cite{yoon_05} and adopt the following
rotation law: 
\begin{equation}
\label{eqq:omega_outer_core}
\Omega(\varpi) / \Omega_{\mathrm K} = \Omega(\varpi_{\mathrm p}) /
\Omega_{\mathrm K} (\varpi_{\mathrm p}) + {\cal C} (\varpi -
\varpi_{\mathrm p})^a\,,
\end{equation}
where the constant $ {\cal C} $ is determined for a given value of $ a
$ as  
\begin{equation}
{\cal C} = \frac{f_{\mathrm K} - \Omega(\varpi_{\mathrm p}) /
  \Omega_{\mathrm K} (\varpi_{\mathrm p})} {(R_\mathrm{e} -
  \varpi_{\mathrm p})^a}\,.   
\end{equation}
The choice of the exponent $ a $ does not have a strong impact on the
WD structure because of the constraints imposed by $
\Omega(\varpi_{\mathrm p}) $ and $ \Omega(R_\mathrm{e})$ at each
boundary. In our study, we adopt $ a = 1.2 $. For further details, we
refer the reader to Sec.~2.2 of \cite{yoon_05}.

Saio \& Nomoto~\cite{saio_04} argued that
turbulent viscosity resulting from a combination of a baroclinic
instability (see, \eg \cite{pedlosky_87}; neglected 
by Yoon \& Langer \cite{yoon_04,yoon_05})
and the DSI is so efficient in transporting angular momentum
that the angular velocity becomes nearly uniform in the WD interior,
while only surface layers with mass $ \lesssim 0.01 M_\odot $ rotate
differentially~\cite{saio_04}. Piro~\cite{piro_08_a}, who also
considered angular momentum transport by magnetic stresses, confirmed
these results. Hence, in order to study the suggested
case of uniform precollapse WD rotation, we complement our
differentially rotating WD models with a set of uniformly rotating AIC
progenitors.

\subsubsection{Initial Temperature Profile} 
\label{sec:temp}

Because our initial models are constructed by imposing hydrostatic
equilibrium (Eq.~(\ref{eqq:wd_equilibrium})) with a barotropic EOS
(Eq.~(\ref{eqq:eos_el_gas})), the WD structure is independent of
temperature. However, the latter is needed as input for the
finite-temperature nuclear EOS used in our AIC simulations.
We follow Dessart~et~al.~\cite{dessart_06_a} and impose a
scaling of the temperature with density according to
\begin{equation}
  \label{eqq:temp_profile}
  T (\varpi, z) = T_{\mathrm 0} \left[ \rho_{\mathrm{c,i}} / \rho (\varpi,
    z) \right]^{0.35} ,
\end{equation}
where $(\varpi, z)$ are cylindrical coordinates and
$\rho_\mathrm{0}$ is the density at which the stellar temperature equals 
$T_\mathrm{0}$.

\subsubsection{Initial Electron Fraction Profile} 
\label{IEFP}

For the purpose of constructing AIC progenitor WDs in rotational
equilibrium we assume that no electron capture has yet taken place and
set $Y_e = 0.5$. A real AIC progenitor, however, will have seen some
electron captures on Ne/Mg/Na nuclei (\eg~\cite{gutierrez_05_a})
before the onset of dynamical collapse. In addition, electrons will be
captured easily by free protons that are abundant at the temperatures
of the WD models considered here. Hence, a $Y_e$ of $0.5$ is rather
inconsistent with real WD evolution.
Dessart~et~al.~\cite{dessart_06_a}, who started their simulations with
$Y_e = 0.5$ models, observed an early burst of electron capture.  This
led to a significant initial drop of $Y_e$ that leveled off after
$5-10$ ms beyond which the $Y_e$ profile evolved in qualitatively
similar fashion to what is known from iron core collapse (see
Fig.~\ref{fig:ye_vs_rho_VULCAN} which depicts this drop of $Y_e$ at
low densities). To account for this, we adopt as initial
$\overline{Y}_e(\rho)$ a parametrization obtained from the equatorial
plane of the models of Dessart~et~al.~\cite{dessart_06_a} at $\sim 7$
ms into their evolution when the initial electron capture burst has
subsided. We use these $\overline{Y}_e(\rho)$ data for the $Y_e$
evolution of the low-density ($\rho < \rho_\mathrm{c,i}$) part of the
WD during collapse.


\subsection{Parameter Space}
\label{sec:parameter_space}

The structure and thermodynamics of the AIC progenitor and the
resulting dynamics of the collapse depend on a variety of parameters
that are constrained only weakly by theory and observation 
(\eg \cite{nomoto_91,yoon_04,yoon_05}). Here, we study 
the dependence on the central density, rotational configuration and
the core temperature. In the following we lay out our parameter
choices and discuss the nomenclature of our initial models whose key
properties we summarize in Table~\ref{tab:initial_models}.

\subsubsection{Progenitor White Dwarf Central Density}
\label{sec:parameter_space:rhoc}

In order to investigate the impact of the precollapse central WD
density $\rho_\mathrm{c,i}$ on the collapse dynamics, we consider
sequences of WD models with central densities in the range from $ 4
\times 10^9 \, {\mathrm{g \, cm}^{-3}} $ to $ 5 \times 10^{10} \,
       {\mathrm{g \, cm^{-3}}} $. This range of densities is motivated
       by previous studies arguing that WDs in this range of
       $\rho_\mathrm{c,i}$ may experience AIC
       \cite{nomoto_91,yoon_04,dessart_06_a}.

We therefore choose a set of four central densities, \ie $4 \times
10^9, \ 1 \times 10^{10}, \ 2 \times 10^{10}, \ 5 \times
10^{10}\,~{\mathrm{g\,cm}}^{-3}$, and correspondingly begin our model
names with letters ${\rm A,~B,~C,~D}$. We perform AIC simulations of
nonrotating (spherically-symmetric) WDs with central density choices
A-D and restrict the rotating models to the limiting central density
choices A and D.

\begin{figure}
  \centerline{\includegraphics[width = 86 mm, angle = 0]{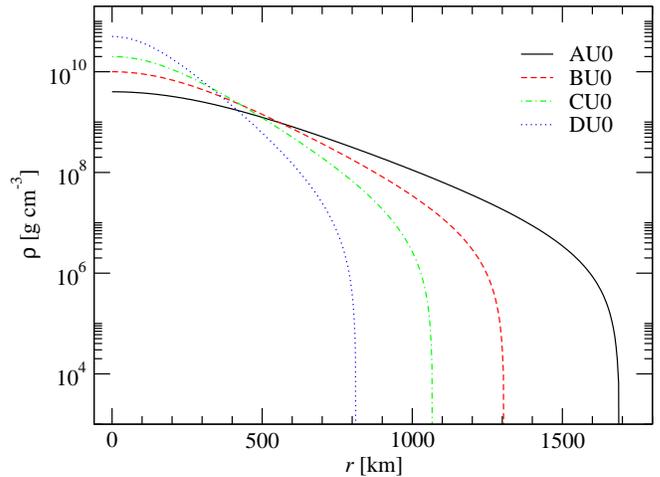}}
  \caption{The radial profile of the rest-mass density for nonrotating
    white dwarf models AU0, BU0, CU0 and DU0.}
  \label{fig:rho_e_non_rotating}
\end{figure}

In Fig.~\ref{fig:rho_e_non_rotating} we plot radial density profiles
of our nonrotating WD models to show the strong dependence of the WD
compactness on the choice of central density. This aspect will prove
important for the understanding of the collapse dynamics of 
rapidly rotating models.

\begin{figure}
  \centerline{\includegraphics[width = 86 mm, angle = 0]{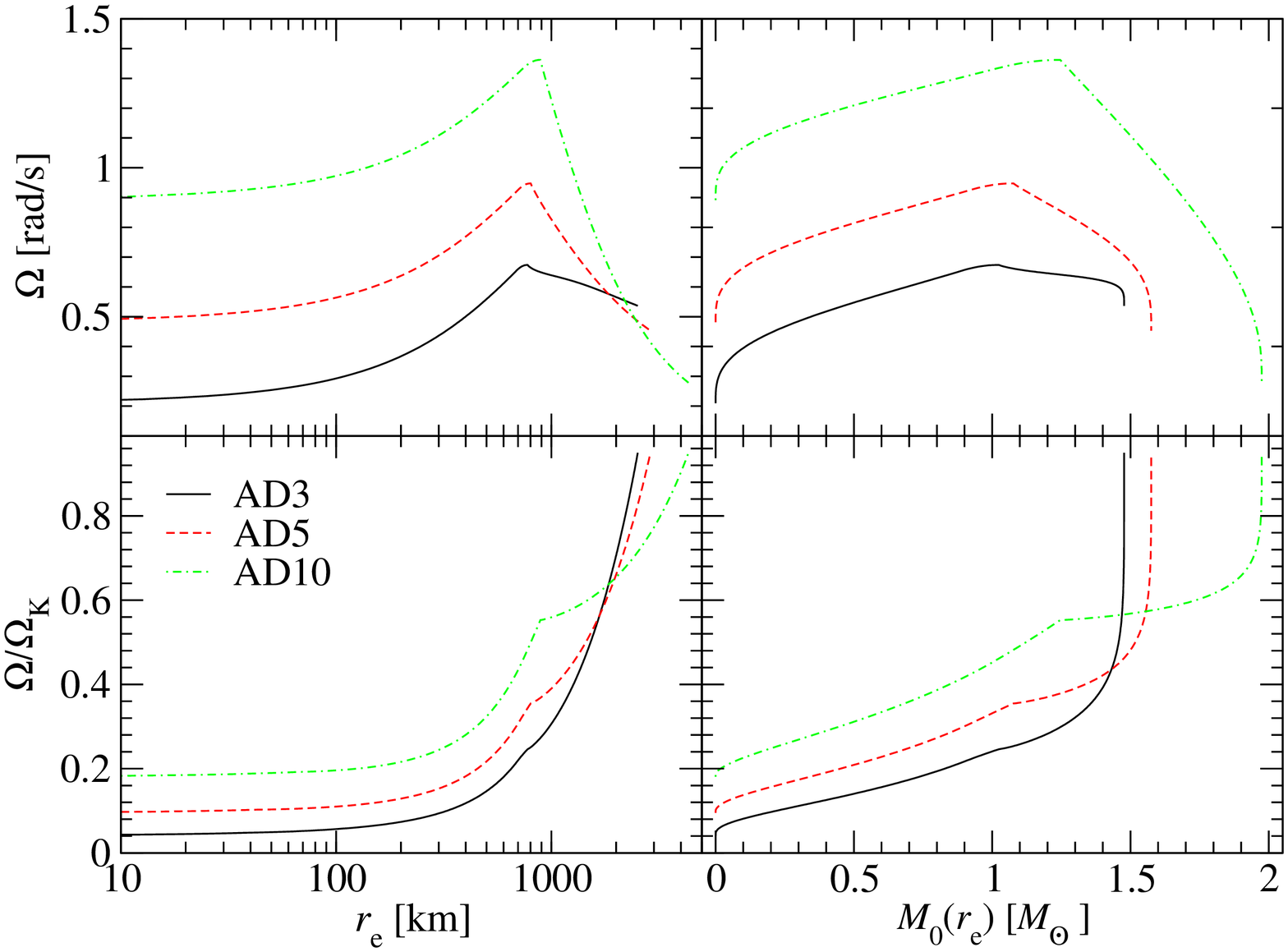}}
  \caption{Upper panels: angular velocity as a function of equatorial
    radius (left panel) and enclosed mass coordinate (right panel) for
    three representative precollapse WD models AD3, AD5 and
    AD10. Lower panels: Angular velocity normalized to the local
    Keplerian value as a function of equatorial radius and enclosed
    mass for the same models.}
  \label{fig:precollapse_omega}
\end{figure}

\begin{figure*}[t]
  \centerline{\epsfxsize = 6.5 cm
              \epsfbox{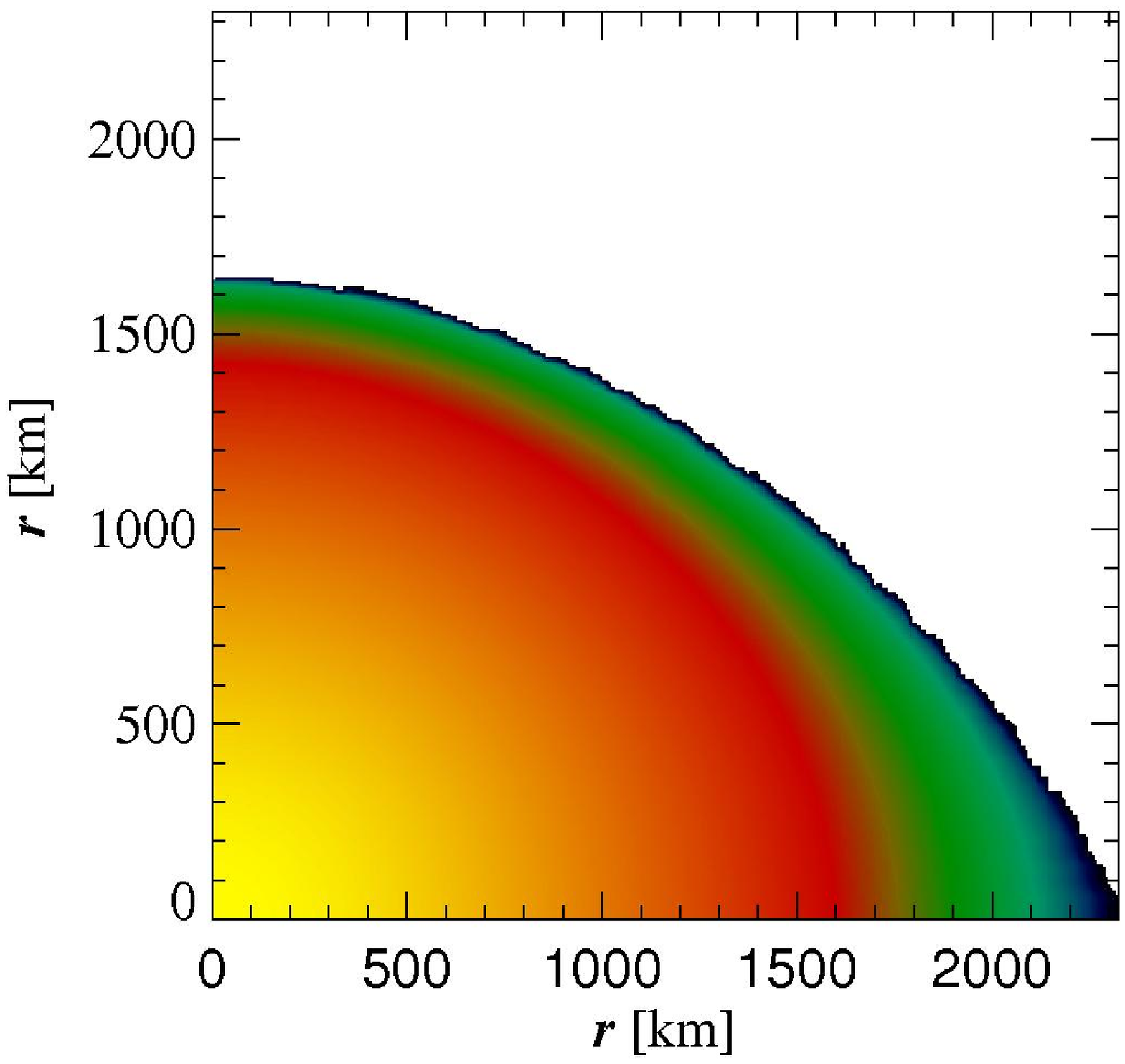}
              \hspace{-0.8cm}
              \epsfxsize = 6.5 cm
              \epsfbox{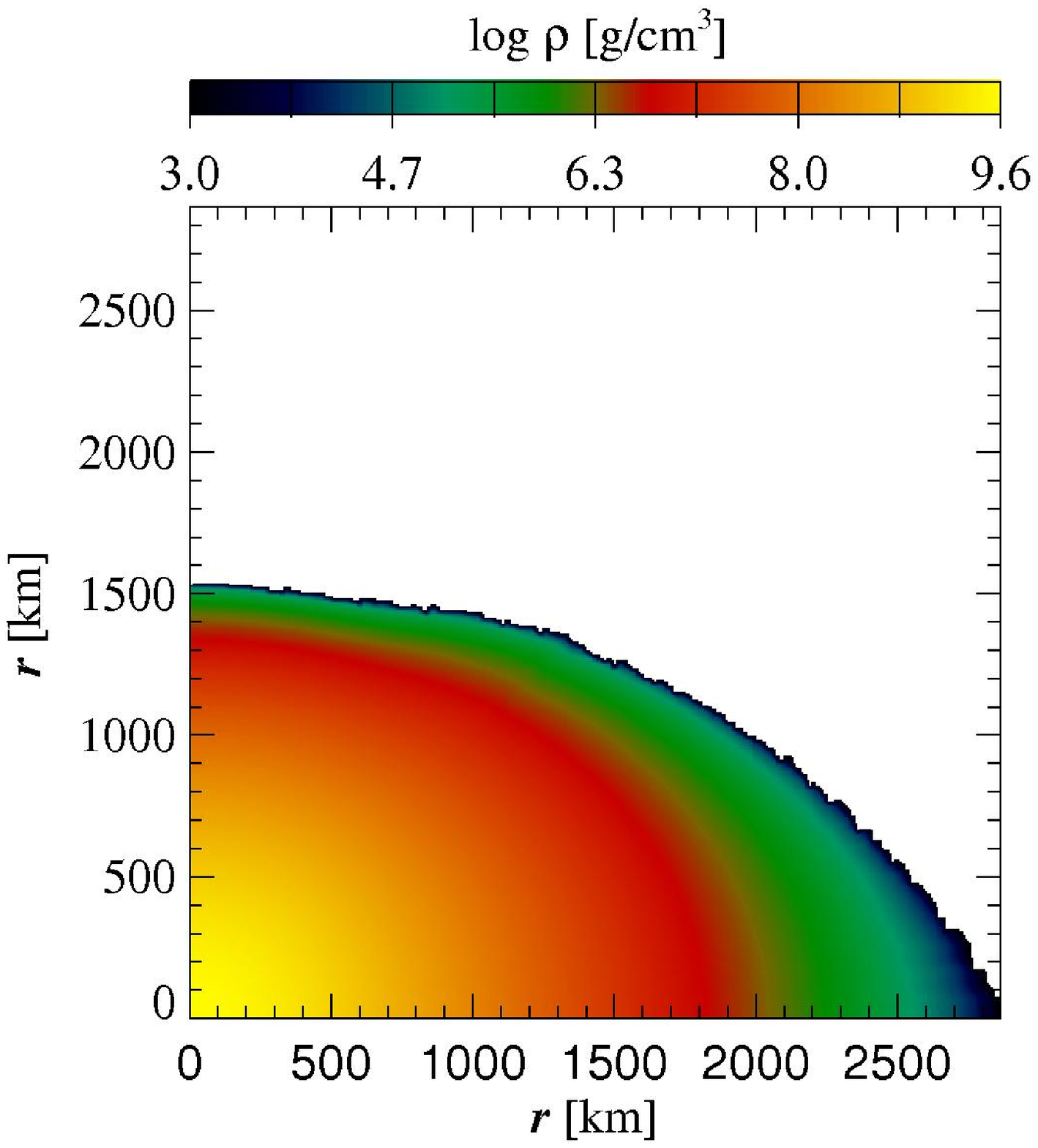}
              \hspace{-0.8cm}
              \epsfxsize = 6.5 cm
              \epsfbox{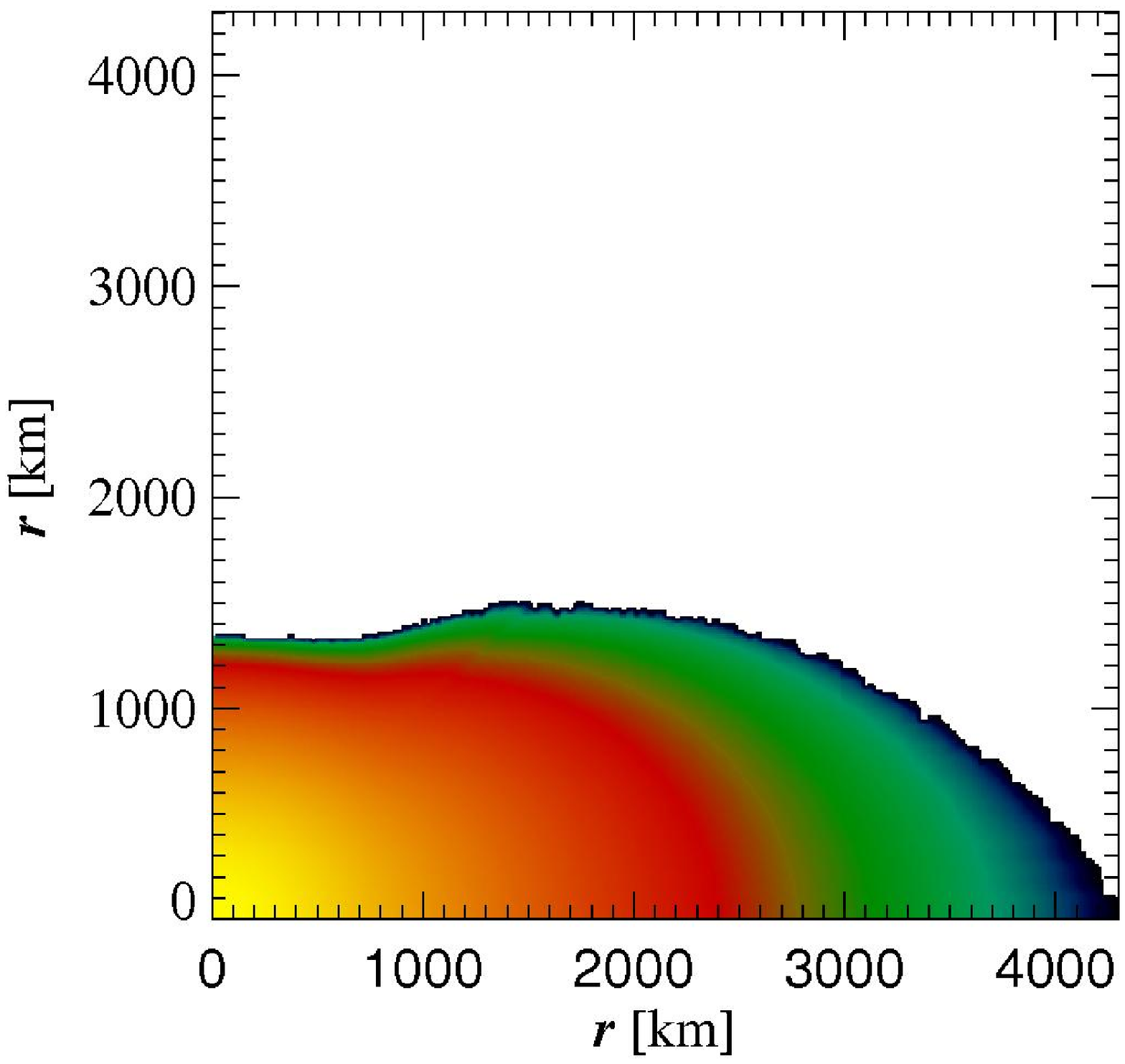}}
  \caption{Colormap of the rest mass density for the precollapse white
    dwarf models AD1 (right panel) AD5 (center panel), and AD10 (right
    panel). The apparent ruggedness of the WD surface layers is a
    results of the finite resolution of our computational grid and the
    mapping procedure in the visualization tool. The ruggedness has no
    influence on the collapse and postbounce dynamics of the inner core.}
  \label{fig:contour_rho} 
\end{figure*}

\begin{figure}
  \centerline{\includegraphics[width = 86 mm, angle = 0]{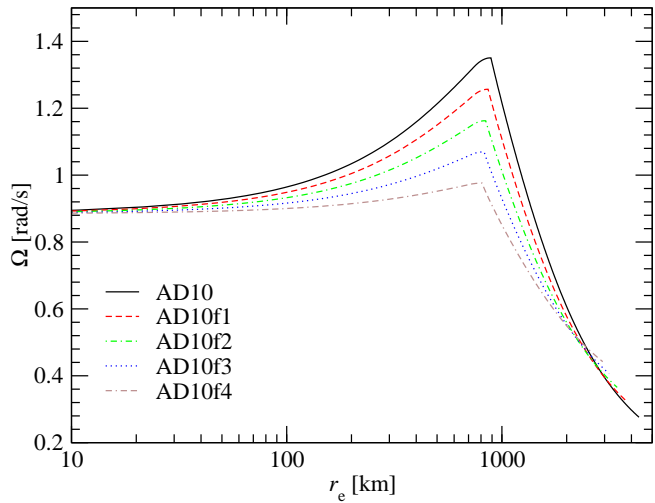}}
  \caption{Angular velocity as a function of equatorial radius for
    model AD10 and varying values of the dimensionless shear parameter
    $f_{\mathrm{sh}}$ that controls the rate at which the angular
    velocity increases with $ \varpi $ in the region $ \varpi <
    \varpi_{\mathrm p}$.}
  \label{fig:omega_vs_fsh}
\end{figure}

\begin{figure}
  \centerline{\includegraphics[width = 86 mm, angle = 0]{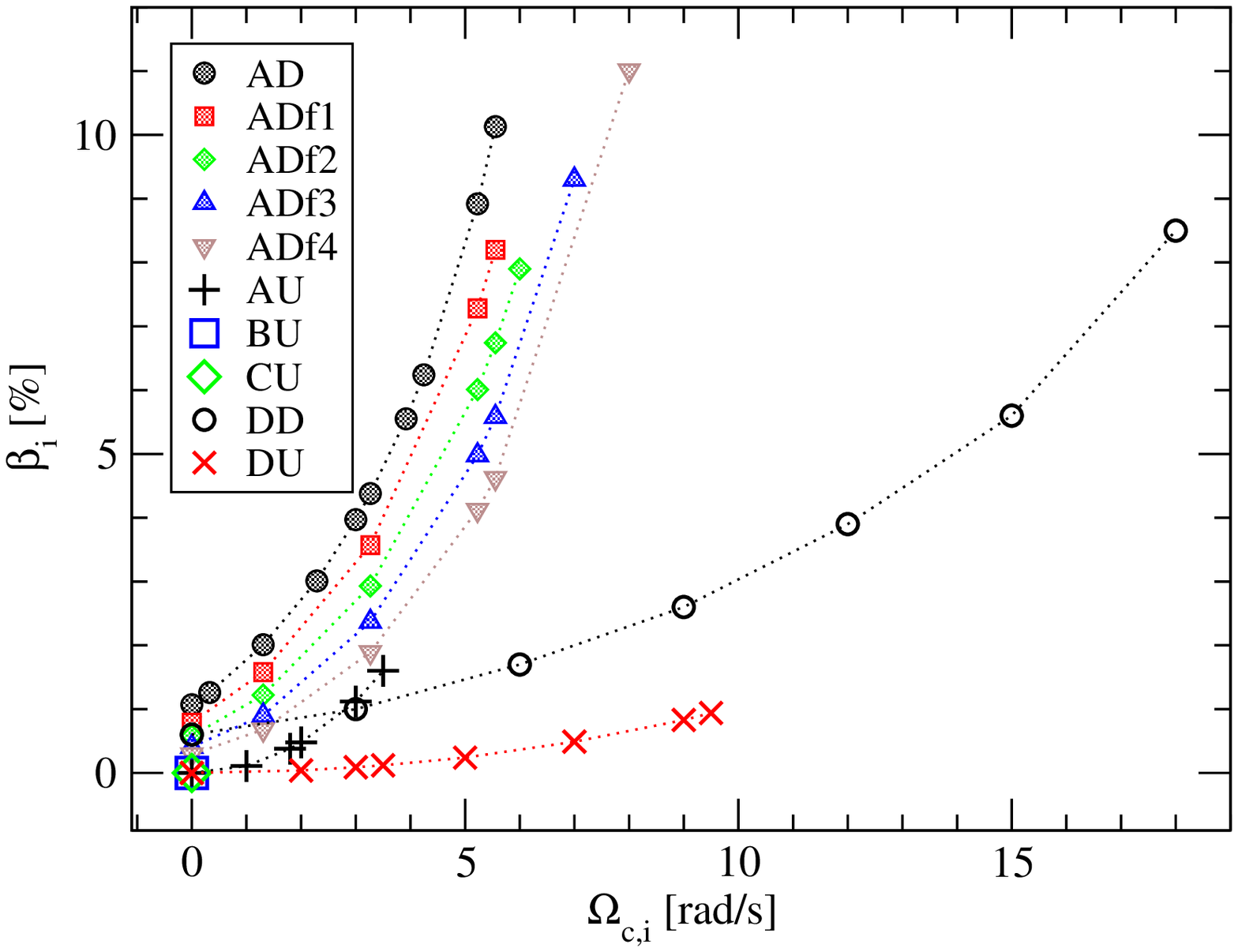}}
  \caption{Parameter $\beta_\mathrm{i}$ versus central angular
    velocity $ \Omega_{\mathrm{c,i}} $ for our AIC progenitor WD model
    set.}
  \label{fig:beta_vs_omega_precollapse}
\end{figure}

\subsubsection{Progenitor White Dwarf Rotational Configuration}
\label{sec:progenitor_rotation}

Since the rotational configuration of AIC progenitor WDs is
constrained only poorly, we consider uniformly rotating ($
\Omega_\mathrm{i} = \Omega_{\mathrm{c,i}} $ everywhere) as well as a
variety of differentially rotating WD configurations. To denote the
general rotation type, we use the letter U (D) for uniform
(differential) rotation as the second letter in each model name.

The low-density uniformly rotating model sequence AU\{1-5\} is set up
with initial angular velocities $ \Omega_{\mathrm{c,i}} $ from $1$ to
$3.5\,{\mathrm{rad\,s}}^{-1}$, where the latter value corresponds to
rotation very close to the mass-shedding limit.  The more compact
uniformly-rotating sequence DU\{1-7\} is set up with precollapse
$\Omega_\mathrm{c,i}$ from $2$ to $9.5\,{\mathrm{rad\,s}}^{-1}$,
where, again, the latter value corresponds to near-mass-shedding
rotation.

Model sequences AD\{1-10\}, DD\{1-7\} are differentially rotating
according to the rotation law discussed in Sec.~\ref{sec:rotlaw} and
specified by Eqs.~(\ref{eqq:omega_inner_core}) and
(\ref{eqq:omega_outer_core}), with the parameter choice
$f_{\mathrm{sh}} = 1$ and $f_{\mathrm p} = 0.1$ for the AD sequence
and $f_{\mathrm{sh}} = 1$ and $f_{\mathrm p} = 0.05$ for the DD
sequence. We recall that $f_{\mathrm p}$ is the fraction of the
central density where the angular velocity has a global maximum. While
$f_{\mathrm p} = 0.1$ is the standard choice of \cite{yoon_05}, we
adopt $f_{\mathrm p} = 0.05$ for the high-density sequence DD to
be in line with the parameter choices made for the models of
Dessart~et~al.~\cite{dessart_06_a}. Test calculations with AD models
show that the variation of $f_{\mathrm p}$ between $0.05$ and $0.10$
affects the rotational configuration of the outer WD layers only and
does not have any appreciable effect on the AIC dynamics. For the
AD\{1-10\} sequence, we chose $\Omega_{\mathrm{c,i}}$ in the range from
$0$ to $5.6 \ {\mathrm{rad\,s}}^{-1}$, resulting in maximum angular
velocities $\Omega_{\mathrm{max,i}}$ in the range of $ 2.88 $ to $ 8.49
\ {\mathrm{rad\,s}}^{-1}$. The higher-density DD\{1-7\} sequence
rotates with $\Omega_{\mathrm{c,i}} $ in the range from $0$ to
$18\,{\mathrm{rad\,s}}^{-1}$, corresponding to maximum $\Omega$ in the
range of $ 7.69 $ to $ 25.84 \,{\mathrm{rad \, s}}^{-1}$. The values of
$\Omega_{\mathrm{c,i}}$ and $\Omega_{\mathrm{max,i}}$ for the individual
AD and DD models are given in Table~\ref{tab:initial_models}. As
representative examples resulting from our assumed rotation law, we
plot in Fig.~\ref{fig:precollapse_omega} for models AD3, AD5, and AD10
the angular velocity and the ratio of the angular velocity to the
local Keplerian value as a function of cylindrical radius and of the
enclosed rest mass. In Fig.~\ref{fig:contour_rho}, we plot the
colormaps of the rest mass density on the $ r-\theta $ plane for
the representative precollapse WD models AD1, AD5 and 
AD10.

In order to study the effect of variations in the degree of
differential rotation, we vary the dimensionless shear parameter
$f_{\mathrm{sh}}$ for a subsequence of AD models and append suffices
$f\{1-4\}$ to their names corresponding to $f_{\mathrm{sh}} =
\{0.8,0.6,0.4,0.2\}$, respectively. Figure~\ref{fig:omega_vs_fsh}
shows the behavior of the initial angular velocity distribution with
decreasing $f_{\mathrm{sh}}$ in the rapidly differentially rotating
model AD10.

An important point to mention is the large range of precollapse WD
masses covered by our models.  Depending on the initial central
density and the rotational setup, our WDs masses range from
a sub-$M_\mathrm{Ch}$ value of $1.39\,M_\odot$ in the nonrotating
low-$\rho_\mathrm{c,i}$ model AU0 to a rotationally-supported
super-$M_\mathrm{Ch}$ mass of $2.05\,M_\odot$ in the rapidly
differentially rotating model AD13f4. The maximum mass
in our sequence of uniformly rotating WDs is $1.462\,M_\odot$
and is obtained in model DU7.

To conclude the discussion of our initial rotational configurations,
we present in Fig.~\ref{fig:beta_vs_omega_precollapse} for all models
the initial values ($\beta_\mathrm{i}$) of the parameter $\beta$ as a
function of their precollapse central angular velocity
$\Omega_{\mathrm{c,i}}$.  Differentially rotating WD models can reach
$\beta_\mathrm{i}$ of up to $\sim 10\%$ while staying below the
mass-shedding limit. This number is more than a factor of $2$ larger
than what seems possible in massive star iron core collapse (see,
\eg \cite{dimmelmeier_08_a}), making these rapidly rotating AIC
progenitor models potential candidates for a dynamical nonaxisymmetric
rotational instability during their postbounce AIC evolution (see
Sec.~\ref{sec:rotinst}).

\subsubsection{Progenitor White Dwarf Core Temperature and $\overline{Y_e}({\rho})$ parametrization}
\label{sec:tempye}

We use Eq.~(\ref{eqq:temp_profile}) to set up the initial temperature
distribution as a function of density.  Dessart~et~al.\ chose
$\rho_\mathrm{0} = \rho_\mathrm{c,i}$ ($= 5\times
10^{10}\,\mathrm{g\,cm}^{-3}$ in their models) and $ T_{\mathrm 0} =
10^{10} $ K for their $ 1.46 M_\odot$ model, and $ T_{\mathrm 0} = 1.3
\times 10^{10}$ K for their $ 1.92 M_\odot$ model. These values (i)
are similar to what was used in the earlier work of Woosley \&
Baron~\cite{woosley_92_a} and (ii) work well with the tabulated EOS
employed and the assumption of NSE, but may be higher than the
temperatures prevailing in accreting precollapse WDs in nature~(see,
\eg \cite{gutierrez_05_a, saio_04, yoon_04}).

While the fluid pressure is affected very little by different
temperature distributions, this is not the case for the free proton
fraction which increases strongly with $T$ in the range from $10^9$ to
$ 10^{10}$ K and at precollapse core densities. This increase of the
proton fraction can lead to enhanced electron capture during AIC and
in this way may have a significant influence on the AIC dynamics. In
order to test the sensitivity of our AIC simulations on the assumed $
T_{\mathrm 0} $, we not only study models with $ T_{\mathrm 0} =
10^{10}$ K (at $\rho_\mathrm{0} = 5\times10^{10}\,\mathrm{g\,cm}^{-3}$,
herafter the \textit{``high-$T$'' models}), but perform also
simulations for models set up with $ T_{\mathrm 0} = 5 \times 10^{9}$
K (at $\rho_\mathrm{0} = 5\times10^{10}\,\mathrm{g\,cm}^{-3}$,
hereafter the \textit{``low-$T$'' models}). To obtain the
$\overline{Y_e\!}(\rho)$ parametrization (see
Sec.~\ref{sec:deleptonization}) for the latter temperature, we re-ran
with VULCAN/2D the $1.46 M_\odot$ AIC model of Dessart~et~al. up to
core bounce with the same setup as discussed in~\cite{dessart_06_a}, but
using the lower value of $T_\mathrm{0}$.  We do not indicate the two
different initial temperatures in the model names, but list the
results obtained in the two cases side-by-side in
Table~\ref{tab:collapse_models}.

In addition to variations in deleptonization due to differences in the
precollapse WD thermodynamics, we must also consider the possibility of
unknown systematic biases that lead to small values of $Y_e$ in the
inner core at bounce (see Sec.~\ref{sec:deleptonization}). In order to
study the effect that larger values of $Y_e$ in the inner core have on
the AIC dynamics, we perform a set of test calculations with scaled
$\overline{Y_e}(\rho)$ trajectories. We implement this by making use
of the fact that $Y_e(\rho)$ is to good approximation a linear
function of $\log (\rho)$ (see Fig.~\ref{fig:ye_vs_rho_VULCAN}).  We
change the slope of this function between $\rho = 5\times 10^{10}\,
\mathrm{g\,cm}^{-3}$ and $\rho = 2.5\times 10^{14}\,
\mathrm{g\,cm}^{-3}$ by increasing $Y_e(\rho = 2.5\times 10^{14}\,
\mathrm{g\,cm}^{-3}$) by $10\%$ and $20\%$. We pick these particular
scalings, since the $20\%$ increase yields inner-core values of $Y_e$
at bounce that are very close to those obtained in 1D Boltzmann
neutrino transport simulations of oxygen-neon core collapse
\cite{kitaura_06_a,mueller:09phd}.  The $10\%$ scaling yields values
in between those of \cite{dessart_06_a} and
\cite{kitaura_06_a,mueller:09phd} and, hence, allows us to study
trends in AIC dynamics with variations in deleptonization in between
constraints provided by simulations. We will not list the results of
these tests in our summary tables, but discuss them wherever the
context requires their consideration (\ie
Sect.~\ref{sec:nonrotating_collapse_dynamics},~\ref{sec:rotating_collapse_dynamics},~\ref{sec:GW},~\ref{sec:peak_amplitude}).



\begin{table*}
  \small
  \centering
  \caption{Summary of the initial WD models: $
    \Omega_{\mathrm{c,i}} $ is the central angular velocity and $
    \Omega_{\mathrm{max,i}} = \Omega (\varpi_{\mathrm p}) $, $ M_0 $ is the
    total rest-mass and $ J $ is the total angular momentum. $
    |W_\mathrm{i}| $ and $ E_\mathrm{rot, i} $ are the gravitational energy
    and rotational kinetic energy of the WD,
    respectively. $R_\mathrm{e}$ and $R_\mathrm{p}$ are the equatorial
    and polar radii.}   
  \label{tab:initial_models}
  \begin{tabular}{@{}l@{~~~}c@{~~~}c@{~~~}c@{~~~}c@{~~~}c@{~~~}c@{~~~}c@{~~~}c@{~~~}c@{~~~}c@{}}
    \hline \\ [-1 em]
    Initial &
    $ \Omega_{\mathrm{c,i}} $ &
    $ \Omega_{\mathrm{max,i}} $ &
    $ \rho_{\mathrm{c,i}} $ &
    $ M_0 $ &
    $ J $ &
    $ |W_{\mathrm i}| $ &
    $ E_{\mathrm{rot, i}} $ &
    $ \beta_{\mathrm i} $ &
    $ R_\mathrm{e} $ &
    $ R_{\mathrm e} / R_{\mathrm p} $ \\
    model &
    [rad/s] &
    [rad/s] &
    [$ 10^{10} {\mathrm{\ g\ cm}}^{-3}$ ] &
    [$ M_\odot $] &
    [$ 10^{50} {\mathrm{ \ ergs}} $] & 
    [$ 10^{50} {\mathrm{ \ ergs}} $] &
    [$ 10^{50} {\mathrm{ \ ergs}} $] &
    [\%]&
    [km] &
    \\
    \hline \\ [-0.5 em]
    AU0 & 0.000 & 0.000 & 0.4 & 1.390 & 0.00 & \z37.32 & \zz0.00 & 0.00 & 1692 & 1.000\\
    AU1 & 1.000 & 1.000 & 0.4 & 1.394 & 0.09 & \z37.50 & \zz0.04 & 0.11 & 1710 & 0.988\\
    AU2 & 1.800 & 1.800 & 0.4 & 1.405 & 0.16 & \z37.90 & \zz0.15 & 0.38 & 1757 & 0.953\\
    AU3 & 2.000 & 2.000 & 0.4 & 1.409 & 0.18 & \z38.04 & \zz0.18 & 0.48 & 1775 & 0.943\\
    AU4 & 3.000 & 3.000 & 0.4 & 1.437 & 0.29 & \z39.04 & \zz0.44 & 1.12 & 1938 & 0.848\\
    AU5 & 3.500 & 3.500 & 0.4 & 1.458 & 0.36 & \z39.78 & \zz0.64 & 1.60 & 2172 & 0.748\\ [0.3 em]

    BU0 & 0.000 & 0.000 & 1.0 & 1.407 & 0.00 & \z51.44 & \z0.00 & 0.00 & 1307 & 1.000\\ [0.3 em]

    CU0 & 0.000 & 0.000 & 2.0 & 1.415 & 0.00 & \z65.28 & \z0.00 & 0.00 & 1069 & 1.000\\ [0.3 em]

    DU0 & 0.000 & 0.000 & 5.0 & 1.421 & 0.00 & \z89.11 & \z0.00 & 0.00 &\z813 & 1.000\\
    DU1 & 2.000 & 2.000 & 5.0 & 1.423 & 0.03 & \z89.25 & \z0.03 & 0.04 &\z817 & 0.995\\
    DU2 & 3.000 & 3.000 & 5.0 & 1.425 & 0.05 & \z89.42 & \z0.08 & 0.09 &\z822 & 0.988\\
    DU3 & 3.500 & 3.500 & 5.0 & 1.426 & 0.06 & \z89.53 & \z0.11 & 0.12 &\z825 & 0.983\\
    DU4 & 5.000 & 5.000 & 5.0 & 1.432 & 0.09 & \z90.00 & \z0.22 & 0.24 &\z840 & 0.963\\
    DU5 & 7.000 & 7.000 & 5.0 & 1.442 & 0.13 & \z90.87 & \z0.44 & 0.49 &\z871 & 0.920\\
    DU6 & 9.000 & 9.000 & 5.0 & 1.458 & 0.17 & \z92.12 & \z0.77 & 0.83 &\z931 & 0.853\\
    DU7 & 9.500 & 9.500 & 5.0 & 1.462 & 0.18 & \z92.50 & \z0.86 & 0.94 &\z956 & 0.828\\ [0.3 em]

    AD1 & 0.000 & 2.881 & 0.4 & 1.434 & 0.28 & \z38.74 &  \z0.41 & 1.07 & 2344 & 0.71\\
    AD2 & 0.327 & 3.204 & 0.4 & 1.443 & 0.31 & \z39.04 &  \z0.49 & 1.26 & 2382 & 0.69\\
    AD3 & 1.307 & 4.198 & 0.4 & 1.477 & 0.42 & \z40.25 &  \z0.81 & 2.01 & 2521 & 0.64\\
    AD4 & 2.287 & 5.174 & 0.4 & 1.526 & 0.56 & \z42.01 &  \z1.27 & 3.01 & 2707 & 0.58\\
    AD5 & 3.000 & 5.903 & 0.4 & 1.575 & 0.69 & \z43.77 &  \z1.74 & 3.97 & 2888 & 0.54\\
    AD6 & 3.267 & 6.173 & 0.4 & 1.595 & 0.75 & \z44.47 &  \z1.95 & 4.38 & 2964 & 0.53\\
    AD7 & 3.920 & 6.833 & 0.4 & 1.659 & 0.93 & \z46.77 &  \z2.59 & 5.55 & 3200 & 0.47\\ 
    AD8 & 4.247 & 7,161 & 0.4 & 1.706 & 1.05 & \z48.45 &  \z3.02 & 6.24 & 3366 & 0.44\\
    AD9 & 5.227 & 8.155 & 0.4 & 1.884 & 1.58 & \z54.69 &  \z4.88 & 8.92 & 4008 & 0.35\\
    AD10& 5.554 & 8.485 & 0.4 & 1.974 & 1.87 & \z57.80 &  \z5.85 &10.13 & 4338 & 0.313\\ [0.3 em]

    DD1 & 0.000 & 7.688 & 5.0 & 1.446 & 0.13 & \z90.85 & \z0.51 & 0.60 & 1097 & 0.73\\
    DD2 & 3.000 & 10.70 & 5.0 & 1.467 & 0.19 & \z92.52 & \z0.95 & 1.00 & 1156 & 0.69\\
    DD3 & 6.000 & 13.73 & 5.0 & 1.498 & 0.26 & \z95.05 & \z1.61 & 1.70 & 1238 & 0.63\\
    DD4 & 9.000 & 16.74 & 5.0 & 1.544 & 0.35 & \z98.70 & \z2.57 & 2.60 & 1353 & 0.56\\
    DD5 & 12.00 & 19.77 & 5.0 & 1.612 & 0.48 &  104.08 & \z4.01 & 3.90 & 1528 & 0.48\\
    DD6 & 15.00 & 22.81 & 5.0 & 1.716 & 0.68 &  111.95 & \z6.31 & 5.60 & 1819 & 0.39\\
    DD7 & 18.00 & 25.84 & 5.0 & 1.922 & 1.10 &  126.69 &  10.77 & 8.50 & 2430 & 0.28\\ [0.3 em]

    AD1f1 & 0.000 & 2.305 & 0.4 & 1.422 & 0.23 & \z38.33 & \z0.30 & 0.79 & 2283 & 0.730\\
    AD1f2 & 0.000 & 1.723 & 0.4 & 1.413 & 0.19 & \z38.03 & \z0.22 & 0.58 & 2233 & 0.753\\
    AD1f3 & 0.000 & 1.152 & 0.4 & 1.406 & 0.15 & \z37.79 & \z0.16 & 0.41 & 2188 & 0.770\\
    AD1f4 & 0.000 & 0.576 & 0.4 & 1.401 & 0.11 & \z37.64 & \z0.11 & 0.29 & 2151 & 0.785\\ [0.3 em]

    AD3f1 & 1.307 & 3.610 & 0.4 & 1.457 & 0.36 & \z39.56 & \z0.62 & 1.58 & 2434 & 0.673\\
    AD3f2 & 1.307 & 3.032 & 0.4 & 1.441 & 0.31 & \z39.01 & \z0.47 & 1.22 & 2361 & 0.700\\
    AD3f3 & 1.307 & 2.457 & 0.4 & 1.428 & 0.26 & \z38.56 & \z0.36 & 0.90 & 2298 & 0.723\\
    AD3f4 & 1.307 & 1.883 & 0.4 & 1.417 & 0.21 & \z38.20 & \z0.26 & 0.70 & 2243 & 0.745\\ [0.3 em]

    AD6f1 & 3.267 & 5.574 & 0.4 & 1.555 & 0.64 & \z43.13 & \z1.54 & 3.57 & 2798 & 0.555\\
    AD6f2 & 3.267 & 5.003 & 0.4 & 1.522 & 0.55 & \z41.96 & \z1.23 & 2.93 & 2666 & 0.590\\
    AD6f3 & 3.267 & 4.423 & 0.4 & 1.494 & 0.47 & \z41.00 & \z0.97 & 2.37 & 2554 & 0.625\\
    AD6f4 & 3.267 & 3.842 & 0.4 & 1.472 & 0.41 & \z40.20 & \z0.76 & 1.89 & 2462 & 0.655\\ [0.3 em]

    AD9f1 & 5.227 & 7.564 & 0.4 & 1.772 & 1.23 & \z50.92 & \z3.71 & 7.28 & 3574 & 0.400\\
    AD9f2 & 5.227 & 6.978 & 0.4 & 1.691 & 1.00 & \z48.13 & \z2.89 & 6.01 & 3264 & 0.448\\
    AD9f3 & 5.227 & 6.392 & 0.4 & 1.630 & 0.84 & \z46.00 & \z2.29 & 4.98 & 3029 & 0.493\\
    AD9f4 & 5.227 & 5.808 & 0.4 & 1.584 & 0.71 & \z44.38 & \z1.83 & 4.12 & 2851 & 0.533\\ [0.3 em]

    AD10f1 & 5.554 & 7.896 & 0.4 & 1.833 & 1.41 & \z53.06 & \z4.35 & 8.20 & 3793 & 0.370\\
    AD10f2 & 5.554 & 7.305 & 0.4 & 1.741 & 1.13 & \z49.93 & \z3.37 & 6.74 & 3434 & 0.420\\
    AD10f3 & 5.554 & 7.721 & 0.4 & 1.665 & 0.93 & \z47.28 & \z2.64 & 5.58 & 3149 & 0.468\\
    AD10f4 & 5.554 & 6.134 & 0.4 & 1.611 & 0.78 & \z44.40 & \z2.10 & 4.62 & 2942 & 0.510\\ [0.3 em]

    AD11f2 & 6.000 & 7.756 & 0.4 & 1.815 & 1.35 & \z52.60 & \z4.16 & 7.90 & 3696 & 0.380\\ [0.3 em]

    AD12f3 & 7.000 & 8.175 & 0.4 & 1.914 & 1.65 & \z56.27 & \z4.23 & 9.30 & 4010 & 0.340\\ 
    AD12f4 & 7.000 & 7.586 & 0.4 & 1.798 & 1.29 & \z52.29 & \z3.99 & 7.64 & 3574 & 0.393\\ [0.3 em]

    AD13f4 & 8.000 & 8.585 & 0.4 & 2.049 & 2.09 & \z61.30 & \z6.75 &11.01 & 4436 & 0.295\\
    \hline
  \end{tabular}
\end{table*}


\subsection{Gravitational Wave Extraction}
\label{sec:wave_extraction}

We employ the Newtonian quadrupole formula in the first-moment of
momentum density formulation as discussed, \eg
in~\cite{dimmelmeier_02_a, shibata_03_a, dimmelmeier_05}. In essence,
we compute the quadrupole wave amplitude $ A_{20}^{\mathrm{E2}} $ of
the $\l=2, m=0$ mode in a multipole expansion of the radiation field
into pure-spin tensor harmonics \cite{thorne_80_a}.  In axisymmetric
AIC, this quadrupole mode provides by far the largest contribution to
the GW emission and other modes are at least one or more orders of
magnitude smaller.  Of course, should nonaxisymmetric instabilities
develop (which we cannot track in our current 2D models), these would
then provide a considerable nonaxisymmetric contribution to the GW
signal.

The GW amplitude is related to the dimensionless GW strain $ h $ in
the equatorial plane by
\begin{equation}
  h = \frac{1}{8} \sqrt{\frac{15}{\pi}} \left(\frac{A^{\mathrm
      E2}_{20}}{r}\right) = 8.8524 \times 10^{-21}
  \left(\frac{A^{\mathrm{E2}}_{20}}{10^3 {\mathrm{\ cm}}}\right)
  \left(\frac{10 {\mathrm{\ kpc}}}{r}\right),
\end{equation}
where $ r $ is the distance to the emitting source.

We point out that although the quadrupole formula is not gauge
invariant and is only valid in the weak-field slow-motion limit, it
yields results that agree very well in phase and to $ \sim 10
\mbox{\,--\,} 20\% $ in amplitude with more sophisticated methods
\cite{shibata_03_a,Nagar2007,baiotti_08_a}.

In order to assess the prospects for detection by current and planned
interferometric detectors, we calculate characteristic quantities for
the GW signal following~\cite{thorne:87}. Performing a Fourier
transform of the dimensionless GW strain $ h $,
\begin{equation}
  \hat{h} =
  \int_{-\infty}^\infty \!\! e^{2 \pi i f t} h \, dt\,,
  \label{eq:waveform_fourier_tranform}
\end{equation}
we can compute the (detector dependent) integrated characteristic
frequency
\begin{equation}
  f_{\mathrm c} =
  \left( \int_0^\infty \!
  \frac{\langle \hat{h}^2 \rangle}{S_h}
  f \, df \right)
  \left( \int_0^\infty \!
  \frac{\langle \hat{h}^2 \rangle}{S_h}
  df \right)^{-1}\!\!\!\!\!\!\!,
  \label{eq:characteristic_frequency}
\end{equation}
and the dimensionless integrated characteristic strain
\begin{equation}
  h_{\mathrm c} =
  \left( 3 \int_0^\infty \!
  \frac{S_{h,{\,\mathrm c}}}{S_h} \langle \hat{h}^2 \rangle
  f \, df \right)^{1/2}\!\!\!\!\!\!\!\!,
  \label{eq:characteristic_amplitude}
\end{equation}
where $ S_h $ is the power spectral density of the detector and $
S_{h,\,{\mathrm c}} = S_h (f_{\mathrm c}) $. We approximate the average
$ \langle \hat{h}^2 \rangle $ over randomly distributed angles by $ (
3 / 2) \, \hat{h}^2 $. From Eqs.~(\ref{eq:characteristic_frequency}),
(\ref{eq:characteristic_amplitude}) the optimal single-detector
signal-to-noise ratio (SNR) can be calculated as  
\begin{equation}
\mathrm{SNR} \equiv \frac{h_{\mathrm c}}{h_{\mathrm{rms}} (f_{\mathrm c})}, 
\end{equation}
where $ h_{\mathrm{rms}} = \sqrt{f S_h} $ is the value of the
root-mean-square strain noise for the detector.


\begin{table*}
  \footnotesize
  \centering
  \caption{Summary of key quantitative results from our AIC simulations. $
    \rho_{\mathrm{max, b}} $ is the maximum density in the core at 
    the time of bounce $ t_{\mathrm{b}} $, $ | h |_{\mathrm{max}} $ is
    the peak value of the GW signal amplitude, while $
    \beta_{\mathrm{ic, b}} $ is the inner core parameter $ \beta $ at
    bounce. Models marked by unfilled/filled circles ($\blt/\crc$)
    undergo a pressure-dominated bounce with/without significant early
    postbounce convection. Models marked with the cross sign ($\tms$)
    undergo centrifugal bounce at subnuclear densities. The values
    left/right of the vertical separator ($ | $) are for the models
    with low/high temperature profiles.} 
  \label{tab:collapse_models}
  \begin{tabular}{@{}l@{~}@{~}c@{~}c@{~}c@{~}c@{~}c@{~~~~}c@{~}@{}l@{~}@{~}c@{~}c@{~}c@{~}c@{~}c@{~}c@{~}}
    \hline \\ [-1 em]
    Collapse & &
    $ \rho_{\mathrm{max,b}} $ &
    $ t_{\mathrm b}$ &
    $ |h|_{\mathrm{max}} $ &
    $ \beta_{\mathrm{ic,b}} $ & 
    Collapse & &
    $ \rho_{\mathrm{max,b}} $ &
    $ t_{\mathrm b}$ &
    $ |h|_{\mathrm{max}} $ &
    $ \beta_{\mathrm{ic,b}} $ & \\
    \raiseentry{model} & &
    [$10^{14} {\mathrm{\ g\ cm }}^{-3} $] &
    [ms] &
    $ \displaystyle \left[ \!\!\!
      \begin{array}{c}
        10^{-21} \\ [-0.2 em]
        {\mathrm{\ at\ 10\ kpc}}
      \end{array}
    \! \right] $ &
    [$ \% $ ] & 
    \raiseentry{model} & &
    [$10^{14} {\mathrm{\ g\ cm}}^{-3} $] &
    [ms] &
    $ \displaystyle \left[ \!\!\!
      \begin{array}{c}
        10^{-21} \\ [-0.2 em]
        {\mathrm{\ at\ 10\ kpc}}
      \end{array}
    \! \right] $ &
    [$ \% $ ] & \\
    \hline \\ [- 0.0 em]
    AU0 & $\crc|\crc$ & $2.782|2.807$ & $214.1|204.9$ & $0.27|0.20$ &  $0.00|0.00$  & AD1f1 & $\blt|\blt$ & $2.718|2.733$ & $217.1|207.2$ & $1.57|1.24$ & $2.06|1.80$ \\
    AU1 & $\blt|\blt$ & $2.697|2.719$ & $214.9|205.5$ & $0.78|0.62$ &  $1.27|1.20$  & AD1f2 & $\blt|\blt$ & $2.759|2.767$ & $215.8|206.2$ & $0.96|0.75$ & $1.26|1.10$ \\
    AU2 & $\blt|\blt$ & $2.628|2.629$ & $216.7|206.9$ & $2.33|1.90$ &  $3.75|3.56$  & AD1f3 & $\crc|\crc$ & $2.787|2.795$ & $214.9|205.5$ & $0.47|0.36$ & $0.59|0.49$ \\
    AU3 & $\blt|\blt$ & $2.610|2.613$ & $217.3|207.4$ & $2.79|2.29$ &  $4.49|4.28$  & AD1f4 & $\crc|\crc$ & $2.803|2.803$ & $214.4|205.0$ & $0.26|0.18$ & $0.14|0.12$ \\
    AU4 & $\blt|\blt$ & $2.525|2.508$ & $221.4|210.8$ & $5.01|4.23$ &  $8.48|8.19$  &       &             &               &               &             &             \\
    AU5 & $\blt|\blt$ & $2.461|2.444$ & $224.2|212.9$ & $5.90|5.05$ & $10.46|10.17$ & AD3f1 & $\blt|\blt$ & $2.560|2.564$ & $222.5|211.5$ & $4.09|3.41$ & $6.26|5.81$ \\ 
        &             &               &               &             &               & AD3f2 & $\blt|\blt$ & $2.583|2.597$ & $220.1|209.6$ & $3.46|2.85$ & $5.23|4.85$ \\
    BU0 & $\crc|\crc$ & $2.805|2.800$ & $100.9|95.3$  & $0.19|0.16$ &  $0.00|0.00$  & AD3f3 & $\blt|\blt$ & $2.622|2.631$ & $218.2|208.1$ & $2.76|2.25$ & $4.17|3.87$ \\
        &             &               &               &             &               & AD3f4 & $\blt|\blt$ & $2.656|2.662$ & $216.7|207.0$ & $2.03|1.63$ & $3.10|2.90$ \\
    CU0 & $\crc|\crc$ & $2.782|2.788$ &  $56.3|51.4$  & $0.20|0.24$ &  $0.00|0.00$  &       &             &               &               &             &             \\
        &             &               &               &             &               & AD6f1 & $\blt|\blt$ & $2.356|2.377$ & $237.6|222.9$ & $6.47|5.77$ & $13.20|12.52$ \\
    DU0 & $\crc|\crc$ & $2.797|2.781$ &  $27.5|25.2$  & $0.34|0.42$ &  $0.00|0.00$  & AD6f2 & $\blt|\blt$ & $2.395|2.399$ & $233.1|219.7$ & $6.41|5.63$ & $12.31|11.79$ \\
    DU1 & $\crc|\crc$ & $2.786|2.777$ &  $27.6|25.2$  & $0.23|0.27$ &  $0.18|0.17$  & AD6f3 & $\blt|\blt$ & $2.421|2.424$ & $229.1|216.7$ & $6.28|5.43$ & $11.47|11.06$ \\
    DU2 & $\crc|\crc$ & $2.774|2.763$ &  $27.6|25.2$  & $0.24|0.31$ &  $0.39|0.38$  & AD6f4 & $\blt|\blt$ & $2.463|2.444$ & $225.7|214.1$ & $5.95|5.12$ & $10.56|10.20$ \\
    DU3 & $\crc|\crc$ & $2.756|2.758$ &  $27.6|25.2$  & $0.31|0.27$ &  $0.53|0.51$  &       &             &               &               &             &               \\
    DU4 & $\blt|\blt$ & $2.728|2.719$ &  $27.6|25.3$  & $0.64|0.55$ &  $1.07|1.03$  & AD9f1 & $\tms|\tms$ & $1.913|1.994$ & $282.0|250.9$ & $3.88|5.77$ & $20.80|18.84$ \\
    DU5 & $\blt|\blt$ & $2.669|2.658$ &  $27.7|25.3$  & $1.21|1.07$ &  $2.04|1.97$  & AD9f2 & $\blt|\blt$ & $2.000|2.073$ & $265.1|241.7$ & $5.25|6.17$ & $20.10|18.09$ \\ 
    DU6 & $\blt|\blt$ & $2.642|2.640$ &  $27.8|25.4$  & $1.97|1.73$ &  $3.27|3.17$  & AD9f3 & $\blt|\blt$ & $2.092|2.115$ & $253.9|234.6$ & $6.96|6.41$ & $18.44|17.45$ \\
    DU7 & $\blt|\blt$ & $2.627|2.619$ &  $27.9|25.5$  & $2.17|1.92$ &  $3.61|3.50$  & AD9f4 & $\blt|\blt$ & $2.159|2.166$ & $244.7|228.3$ & $7.30|6.74$ & $17.67|16.74$ \\ [0.5 em]
 
    AD1 & $\blt|\blt$ & $2.661|2.679$ & $218.7|208.5$ & $2.24|1.75$ &  $2.96|2.62$  & AD10f1 & $\tms|\tms$ & $1.797|1.899$ & $303.4|261.3$ & $3.99|4.14$ & $23.71|21.04$ \\
    AD2 & $\blt|\blt$ & $2.620|2.616$ & $220.1|209.5$ & $2.85|2.30$ &  $3.93|3.55$  & AD10f2 & $\tms|\tms$ & $1.901|1.980$ & $272.4|245.2$ & $4.15|5.91$ & $21.07|19.30$ \\
    AD3 & $\blt|\blt$ & $2.547|2.547$ & $225.5|213.7$ & $4.62|3.93$ &  $7.20|6.71$  & AD10f3 & $\tms|\blt$ & $1.984|2.042$ & $261.2|239.3$ & $5.52|6.39$ & $20.40|18.53$ \\
    AD4 & $\blt|\blt$ & $2.447|2.442$ & $232.9|219.4$ & $5.89|5.19$ & $10.58|10.01$ & AD10f4 & $\blt|\blt$ & $2.096|2.098$ & $249.8|231.7$ & $7.32|6.67$ & $18.70|17.83$ \\
    AD5 & $\blt|\blt$ & $2.361|2.389$ & $241.1|225.5$ & $6.37|5.68$ & $13.09|12.33$ &        &             &               &               &             &               \\
    AD6 & $\blt|\blt$ & $2.324|2.355$ & $246.7|229.7$ & $6.39|5.78$ & $14.03|13.18$ & AD11f2 & $\tms|\tms$ & $1.734|1.845$ & $296.9|257.6$ & $3.78|4.13$ & $24.23|21.65$ \\
    AD7 & $\blt|\blt$ & $2.226|2.228$ & $260.0|238.8$ & $6.00|5.73$ & $16.34|15.23$ &        &             &               &               &             &               \\
    AD8 & $\blt|\blt$ & $2.145|2.167$ & $264.3|240.9$ & $6.01|5.70$ & $17.48|16.30$ & AD12f3 & $\tms|\tms$ & $0.319|1.555$ & $372.3|249.4$ & $1.61|4.08$ & $22.88|23.08$ \\
    AD9 & $\tms|\tms$ & $1.817|1.911$ & $319.7|267.3$ & $3.40|4.00$ & $23.38|20.69$ & AD12f4 & $\tms|\tms$ & $1.432|1.677$ & $298.1|257.9$ & $3.14|4.44$ & $24.58|22.97$ \\
    AD10& $\tms|\tms$ & $1.629|1.790$ & $393.6|284.5$ & $2.36|3.54$ & $24.41|21.25$ &        &             &               &               &             &               \\
        &             &               &               &             &               & AD13f4 & $\tms|\tms$ &$7\!\times\!10^{-4}\!|0.312$\z&$331.8|322.9$&$0.40|2.01$&$15.39|24.02$\\
    DD1 & $\crc|\crc$ & $2.779|2.772$ &  $27.5|25.2$  & $0.46|0.37$ &  $0.70|0.58$  &&&&&&\\
    DD2 & $\blt|\blt$ & $2.684|2.686$ &  $27.7|25.3$  & $1.24|1.06$ &  $1.95|1.81$  &&&&&&\\
    DD3 & $\blt|\blt$ & $2.642|2.638$ &  $27.8|25.5$  & $2.41|2.08$ &  $3.76|3.56$  &&&&&&\\
    DD4 & $\blt|\blt$ & $2.586|2.571$ &  $28.1|25.7$  & $3.78|3.29$ &  $5.93|5.71$  &&&&&&\\
    DD5 & $\blt|\blt$ & $2.526|2.498$ &  $28.4|25.9$  & $5.22|4.61$ &  $8.27|8.09$  &&&&&&\\
    DD6 & $\blt|\blt$ & $2.457|2.425$ &  $28.8|26.3$  & $6.52|5.82$ & $10.57|10.51$ &&&&&&\\
    DD7 & $\blt|\blt$ & $2.389|2.352$ &  $29.2|26.5$  & $7.58|6.81$ & $12.75|12.79$ &&&&&&\\
   \hline
  \end{tabular}
\end{table*}


\section{Collapse Dynamics}
\label{sec:colldyn}

\begin{figure}
  \centerline{\includegraphics[width = 86 mm]{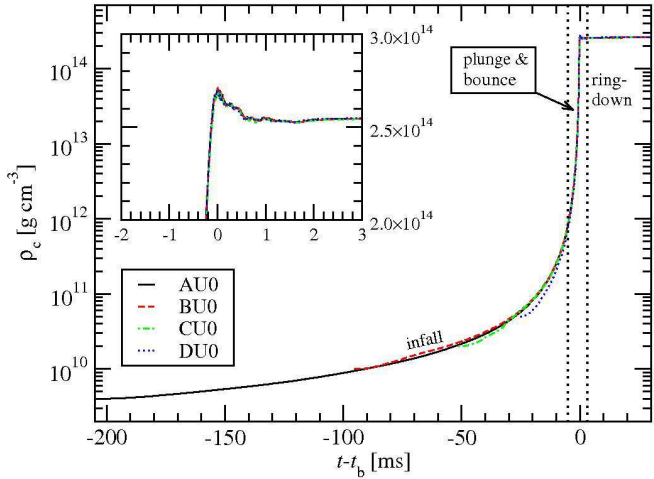}}
  \caption{Evolution of the maximum (central) density for the
    nonrotating low-$T$ models AU0, BU0, CU0 and DU0. The inset plot
    displays a zoomed-in view of the maximum density around the time
    of core bounce on a linear scale.  As clearly discernible from
    this figure, the collapse dynamics in the plunge and bounce phase
    are essentially independent of the initial WD central
    density. Time is normalized to the time of bounce $
    t_\mathrm{b}$.}  
\label{fig:rho_max_t_non_rotating}
\end{figure}

The AIC starts when the progenitor WD reaches its effective
Chandrasekhar mass and pressure support is reduced due to electron
capture in the core. Similar to the case of massive star iron core
collapse (\eg~\cite{ott_04_a, moenchmeyer_91_a, dimmelmeier_02_a,
  zwerger_97_a} and references therein), the collapse evolution can be
divided into three phases:

{\it Infall:} This is the longest phase of collapse and, depending on
model parameters, lasts between $ \sim 25 ~ \mathrm{ms} $ and $ \sim
300 ~ \mathrm{ms} $. The inner part of the WD core (the ``inner
core''), which is in sonic contact, contracts homologously ($ v_r
\propto r $), while the ``outer core'' collapses
supersonically. Fig.~\ref{fig:rho_max_t_non_rotating} shows the time
evolution of the central density for the nonrotating high-$T$ AIC
models. In the infall phase, the core contracts slowly, which is
reflected in the slow increase of $ \rho_{\mathrm c} $.

{\it Plunge and bounce}: The short dynamical ``plunge'' phase sets in
when $\rho_\mathrm{c}$ reaches $ \sim 10^{12} \, {\mathrm{g \,
    cm^{-3}} } $, and the peak radial infall velocity is $ \sim 0.1 c
$.  At this point, neutrinos begin to be trapped in the inner
core. The latter rapidly contracts to reach nuclear densities ($
\rho_{\rm nuc} \simeq 2.7 \times 10^{14} {\rm\ g\ cm}^{-3} $) at which
the nuclear EOS stiffens, decelerating and eventually reversing the
infall of the inner core on a millisecond timescale.  Because of its
large inertia and kinetic energy, the inner core does not come to rest
immediately. It overshoots its equilibrium configuration, then bounces
back, launching a shock wave at its outer edge into the still
infalling outer core. The bounce and the re-expansion of the inner
core is also evident in the time evolution of $\rho_\mathrm{c}$ shown
in Fig.~\ref{fig:rho_max_t_non_rotating} which, at core bounce,
reaches a value of $\sim 2.8 \times 10^{14} \, {\mathrm{g \,
    cm^{-3}}} $ in the nonrotating AIC models, after which the core
slightly re-expands and settles down at $\sim 2.5 \times 10^{14} \,
{\mathrm{g \, cm^{-3}}} $. As pointed out by extensive previous work
(see, \eg \cite{bethe:90, goldreich_80_a, yahil_83_a, van_riper_82_a,
  dimmelmeier_08_a} and references therein), the extent of the inner
core at bounce determines the initial kinetic energy imparted to the
bounce shock, the mass cut for the material that remains to be
dissociated, and the amount of angular momentum that may become
dynamically relevant at bounce.

{\it Ringdown}: Following bounce, the inner core oscillates with a
superposition of various damped oscillation modes with frequencies in
the range of $ 500 - 800 \, \mathrm{Hz} $, exhibiting weak
low-amplitude variations in $ \rho_{\mathrm c} $
(Fig.~\ref{fig:rho_max_t_non_rotating}).  These oscillations
experience rapid damping on a timescale of $10 ~ {\mathrm{ms}} $ due to
the emission of strong sound waves into the postshock region which
steepen into shocks. The newly born PNS thus \emph{rings} down to
its new equilibrium state.

The ringdown phase is coincident with the burst of neutrinos that is
emitted when the bounce shock breaks out of the energy-dependent
neutrinospheres (see, \eg \cite{thompson_03_a,dessart_06_a}). The
neutrino burst removes energy from the postshock regions and enhances
the damping of the PNS ringdown oscillations (\eg \cite{ott:06spin}),
but, due to the limitations of our present scheme (see
Sec.~\ref{sec:deleptonization}), is not accounted for in our models.

\subsection{Nonrotating AIC}
\label{sec:nonrotating_collapse_dynamics}

The set of nonrotating AIC models that we consider here consists of
models AU0, BU0, CU0, DU0. As noted in
Section~\ref{sec:parameter_space:rhoc}, these models have different
central densities with values in the range from $ 4 \times 10^{9} $ to
$ 5 \times 10^{10} \ {\mathrm{g \, cm^{-3}}}$ which, because of the
strong dependence of the WD compactness on the central density,
corresponds to a range of WD radii from $ 1692 $ to $ 813 $ km (see
Fig.~\ref{fig:rho_e_non_rotating}). Once mapped onto our computational
grid and after the initial $\overline{Y}_e(\rho)$ parametrization is
applied (see Sec.~\ref{IEFP}), all WD models start to collapse by
themselves and no additional artificial pressure reduction is
necessary. This is in contrast to previous work that employed a simple
analytic EOS and required an explicit and global change of the
adiabatic exponent to initiate collapse (\eg
\cite{zwerger_97_a,dimmelmeier_02_b}).

The free-fall collapse time $\tau_{\mathrm{ff}}$ of a Newtonian
self-gravitating object of mean density $\rho_\mathrm{mean}$ is
proportional to $\rho_\mathrm{mean}^{-1/2}$. For our set of
spherically-symmetric AIC models we find a scaling $\tau_{\mathrm{ff}}
\propto \rho_{\mathrm c}^{-0.87}$, where $\rho_{\mathrm c}$ is the
precollapse central density of the WD. This stronger scaling is due to
the fact that WD cores are not constant density objects and that the
collapse is not pressureless. Furthermore, the pressure reduction
initiating and accelerating collapse is due primarily to electron
capture which scales roughly with $\rho^{5/3}$
(\eg \cite{bethe:90}). Hence, lower-density WDs collapse only slowly,
spending much of their collapse time near their initial equilibrium
states.

In Fig.~\ref{fig:rho_max_t_non_rotating}, we plot the evolution of the
central densities of the nonrotating high-$T$ models. 
Despite the strong dependence of the collapse times on
the initial central densities, the evolution of $ \rho_{\mathrm c} $
around bounce does not exhibit a dependence on the initial central
density. Moreover, the mass and the size of the inner core is rather
insensitive to the initial value of $ \rho_{\mathrm c}$.

These features, somewhat surprising in the light of the strong
dependence of the collapse times on the initial value of $
\rho_{\mathrm c} $, are a consequence of the fact that the inner core
mass is determined primarily not by hydrodynamics, but by the
thermodynamic and compositional structure of the inner core set by
nuclear and neutrino physics~\cite{bethe:90}. However, an important
role is played also by the fact that an increase (decrease) of the
central density of an equilibrium WD leads to a practically exact
homologous\footnote{For a discussion of homology in the stellar
  structure context, see~\cite{kippenhahn_90}.} contraction
(expansion) of the WD structure in the inner regions ($ m(\varpi)
\lesssim 1\, M_\odot $) in the nonrotating case (this can be seen in
Fig.~\ref{fig:m_vs_r}), at least in the range of central densities
considered in this paper (as we shall see in
Section~\ref{sec:rotating_collapse_dynamics}, this feature also holds
to good accuracy in the case of rotating WDs). These aspects, in
combination with the homologous nature of WD inner-core collapse, make
the size and dynamics of the inner core in the bounce phase
practically independent of the central density of the initial
equilibrium WDs.  Early analytical work \cite{goldreich_80_a,
  yahil_83_a,bethe:90} demonstrated (neglecting thermal corrections
\cite{burrows:83} and rotation) that the mass $M_{\mathrm{ic}} $ of
the inner core is proportional to $Y_e^2$ in the infall phase during
which the fluid pressure is dominated by the contribution of
degenerate electrons. Around bounce, at densities near nuclear matter
density, the nuclear component dominates and the simple $Y_e^2$
dependence does not hold exactly any longer.
As discussed in Sec.~\ref{IEFP}, we adopt the
parametrization $\overline{Y_e}(\rho)$ as extracted from the
simulations of Dessart~et~al.~\cite{dessart_06_a} which predict very
efficient electron capture, resulting in an average inner-core $Y_e$
at bounce of $\sim 0.18$ in the high-$T$ models. This is significantly
lower than in standard iron core collapse where the inner-core $Y_e$
at bounce is expected to be around $\sim 0.25-0.30$
\cite{thompson_03_a,liebendoerfer_05_b,buras_06_a}. In our nonrotating
AIC models, we find inner core masses at bounce $ M_{\mathrm{ic,b}}
\sim 0.27 M_\odot $
\footnote{We define the inner core as the region which is in sonic
  contact at the time of bounce, \ie
  \begin{equation}
    M_{\mathrm{ic,b}} \equiv \int_{|v_r| < c_{\mathrm s}} \rho W d V \ ,  
  \end{equation}
where $ W $ is the Lorentz factor and $dV$ is the invariant 3-volume
element. The bounce time is defined as the time when the radial
velocity of the outer edge of the inner core becomes positive. Note
that such a measure of the inner core is strictly valid only at the
time of bounce.}
(see Fig.~\ref{fig:mic_vs_beta_ic}) which are, as expected,
significantly smaller than in iron core collapse (where $
M_{\mathrm{ic,b}} \sim 0.5 M_\odot $~\cite{buras_06_a, buras_06_b,
  liebendoerfer_05_b}).  Due to their small mass, our AIC inner cores
have less kinetic energy at bounce and reach lower densities than
their iron core counterparts. For example, the nonrotating AIC models
exhibit central densities at bounce of $ \sim 2.8 \times 10^{14} ~
{\mathrm{g ~ cm^{-3}}} $, while in nonrotating iron core collapse,
maximum densities of $ \gtrsim 3 \times 10^{14} ~ {\mathrm{g ~
    cm^{-3}}} $ are generally reached at bounce in simulations (\eg
\cite{dimmelmeier_08_a}). In addition to $M_\mathrm{ic,b}$, the
bounce density depends also on the stiffness of the nuclear EOS
whose variation we do not explore here (see, \eg \cite{dimmelmeier_08_a,
sumiyoshi:05}).

Since the free proton fraction at precollapse and early collapse
densities grows rather rapidly with temperature in the range from $
\sim 10^9 $ to $ \sim 10^{10} $ K (\eg~\cite{bruenn_85_a}), the
efficiency of electron capture is sensitive to the temperature of WD
matter. For example, in the low-$T$ models, the value of $ Y_e $ drops
to $ \simeq 0.32 $ when the density reaches $ 10^{12} \, {\mathrm{g \,
    cm^{-3}}} $, while in the high-$T$ models, we obtain $ Y_e \simeq
0.3 $ at that time. Due to this dependence of $ Y_e $ on $ T $, the
inner core masses of low-$T$ models are larger by $ \sim 10 \% $
compared to those of high-$ T $ models. Moreover, since the electron
degeneracy pressure is proportional to $ \left (Y_e \rho \right)^{4/3}
$ \cite{bethe:90}, the collapse times of the low-$T$ models are longer
by $ \sim 5 \% $.  We find similar systematics in test calculations in
which we modify the $\overline{Y_e}(\rho)$ trajectories of low-$T$
models to yield larger $Y_e$ at bounce (see Sec.~\ref{sec:tempye}). An
increase of the inner-core $Y_e$ by $10\%$ ($20\%$) leads to an
increase of $M_{\mathrm{ic,b}}$ by $\sim 11\%$ ($\sim 25\%$).

It is important to note that the nonrotating AIC models discussed
above as well as all of the other models considered in this study,
experience prompt hydrodynamic explosions. 
The bounce shock, once formed, does slow down, but never
stalls and steadily propagates outwards. While the shock propagation
is insensitive to the initial WD temperature profile, it shows
significant dependence on the initial WD central density: Owing to the
greater initial compactness and the steeper density gradient of the 
higher-density models, the shock propagation in those models is faster
and the shock remains stronger when it reaches the WD surface. For
example, in the lowest-density progenitor model AU0, the shock reaches
the surface within $ \sim 120$ ms after its formation, while in the
highest density model DU0, it needs only $ \sim 80$ ms. We point out
that Dessart~et~al.~\cite{dessart_06_a, dessart_07_a} and previous AIC
studies~\cite{woosley_92_a, fryer_99_a} reported significant shock
stagnation in the postbounce phase of AIC due to the dissociation of
infalling material and neutrino losses from the postshock region. Our
present computational approach includes dissociation (through the EOS,
see Sec. \ref{sec:eos}), but does not account for neutrino losses in
the postbounce phase. Hence, the "prompt" explosions in our models are
most likely an artifact of our incomplete treatment of the postbounce
physics. 

\begin{figure}
  \centerline{\includegraphics[width = 86 mm, angle =
      0]{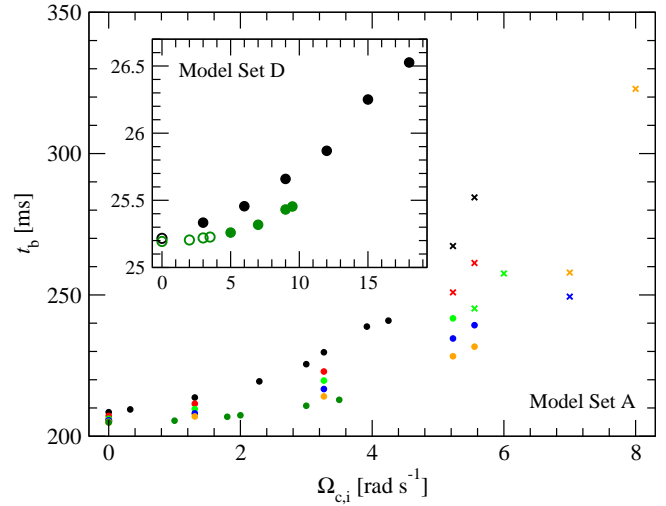}}
  \caption{Times to core bounce from the onset of collapse 
    as a function of initial central angular velocity
    $\Omega_{\mathrm{c,i}}$. Shown are the results of the high-$T$
    model sequence (the low-$T$ models exhibit identical
    systematics). Models denoted by an unfilled (filled) circle
    undergo a pressure-dominated bounce with (without) significant
    prompt postbounce convection.  Models marked by a cross undergo
    centrifugal bounce at subnuclear densities and models marked with
    a small (large) symbol are of set A (D). The colors correspond to
    various precollapse rotational configurations (see the legend in
    Fig.~\ref{fig:rhomaxb_vs_betaicb}).  Note that due to their much
    higher initial compactness, the high-density D models (shown in
    the inset plot) have much shorter collapse times than their
    lower-density A counterparts.}
  \label{fig:collapse_times}
\end{figure}
\begin{figure}
  \centerline{\includegraphics[width = 86 mm]{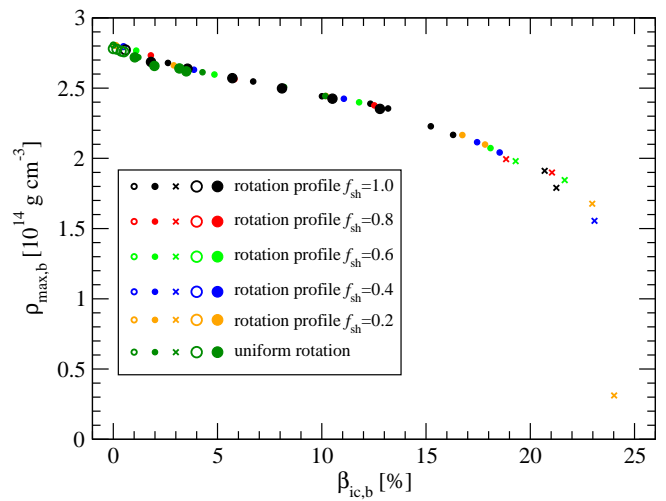}}
  \caption{The maximum density $ \rho_{\mathrm{max,b}} $ at bounce as a
    function of the inner core parameter $ \beta_{\mathrm{ic,b}} $ at
    bounce for the entire set of high-$T$ AIC models. Due to the
    increasing role of centrifugal support, $ \rho_{\mathrm{max,b}} $
    decreases monotonically with increasing rotation (see the main
    text for details). The symbol convention for the various sets is
    explained in the caption of Fig.~\ref{fig:collapse_times}. 
  }
  \label{fig:rhomaxb_vs_betaicb} 
\end{figure}


\subsection{Rotating AIC}
\label{sec:rotating_collapse_dynamics}

The AIC of rotating models proceeds through the same stages as AIC
without rotation and exhibits similar general features, including the
well defined split of the WD into an inner core that is in sonic
contact and collapses quasi-homologously\footnote{The collapse is
  \emph{quasi}-homologous because in this case the relation between
  the infall velocity $v_r$ depends on both the radial coordinate and
  on the polar coordinate~\cite{zwerger_97_a}.}, and a supersonically
infalling outer core. Conservation of angular momentum leads to an
increase of the angular velocity $\Omega \propto \varpi^{-2}$ and of
the centrifugal acceleration $a_{\mathrm{cent}} = \Omega^2 \varpi
\propto \varpi^{-3}$.  The latter has opposite sign to gravitational
acceleration, hence provides increasing \emph{centrifugal support}
during collapse, slowing down the contraction and, if sufficiently
strong, leading to centrifugally-induced core
bounce only slightly above nuclear density or even at subnuclear
density~\cite{tohline_84_a,moenchmeyer_91_a}.

\begin{figure}
  \centerline{\includegraphics[width = 86 mm]{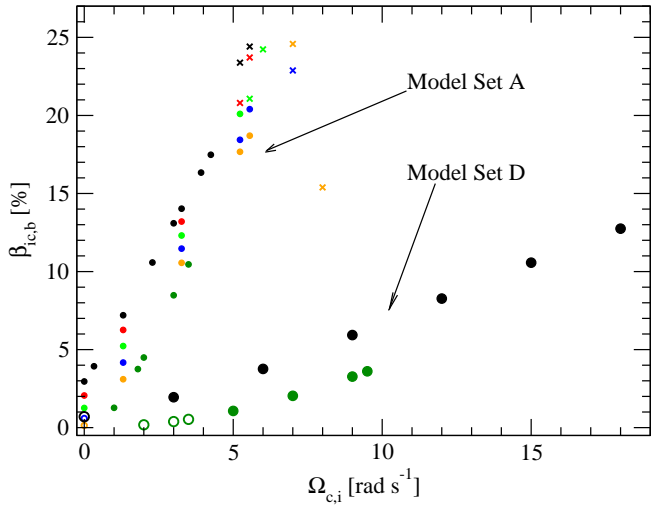}}
  \caption{The inner core parameter $ \beta_{\mathrm{ic,b}} $ at the
    time of bounce for all low-$T$ AIC models plotted as a function of
    the precollapse central angular velocity $ \Omega_{\mathrm{c,i}}
    $. For models with slow to moderately-rapid rotation, $
    \beta_{\mathrm{ic, b}} $ increases roughly linearly with $
    \Omega_{\mathrm{c,i}} $ (at fixed $ \rho_{\mathrm{c,i}} $ and
    rotation law). In more rapidly rotating models, the growth of $
    \beta_{\mathrm{ic, b}} $ saturates at $ \sim 24.5 \, \% $, and
    further increase of the progenitor rotation results in a decrease 
    of $ \beta_{\mathrm{ic, b}} $. Since D models experience less
    spin-up during collapse than A models, an increase of $
    \rho_{\mathrm{c,i}} $ at fixed $ \Omega_{\mathrm{c,i}} $ and
    rotation law results in a decrease of $ \beta_{\mathrm{ic,b}} $. A
    uniformly rotating model with a given $ \Omega_{\mathrm{c,i}} $ and
    $ \rho_{\mathrm{c,i}} $ reaches smaller $ \beta_{\mathrm{ic,b}} $
    than the differentially rotating model with the same $
    \Omega_{\mathrm{c,i}} $ and $ \rho_{\mathrm{c,i}} $. The symbol
    convention for the various sets is explained in the caption of
    Fig.~\ref{fig:collapse_times}.}
  \label{fig:betaic_vs_omega}
\end{figure}

Just as in the case of nonrotating AIC, models of set A collapse more
slowly than D models because of the dependence of the collapse times
on the initial central densities. However, due to centrifugal support,
the collapse times grow with increasing precollapse rotation. This is
visualized in Fig.~\ref{fig:collapse_times} in which we plot the time
to core bounce as a function of the initial central angular velocity
$\Omega_{\mathrm{c,i}}$. The maximum angular velocity of uniformly
rotating models is limited by the WD surface mass-shedding limit and
is $\sim 3.5\,\mathrm{rad\,s}^{-1}$ ($\sim 9.5\,\mathrm{rad\,s}^{-1}$)
in model AU5 (DU7). The effect of rotation on the collapse time of
uniformly rotating models (dark-green symbols in
Fig.~\ref{fig:collapse_times}) is small and the time to core bounce
increases by $\sim5\%$ from zero to maximum precollapse rotation in
model set A. The more compact D models collapse much faster than their
lower initial density A counterparts and, in addition, experience a
smaller spin-up of their more compact inner cores. Hence, uniformly
rotating D models are less affected by rotation and their collapse
times vary by only $\sim 0.8\%$ from zero to maximum rotation.

As mentioned in Sec.~\ref{sec:rotlaw}, WD models that rotate
differentially according to the rotation law of Yoon \&
Langer~\cite{yoon_05, yoon_04} have an 
angular velocity that increases from its central value
$\Omega_{\mathrm{c,i}}$ with $\varpi$ up to a
maximum $\Omega_{\mathrm{max,i}}$ at the cylindrical radius
$\varpi_{\mathrm p}$, 
beyond which $\Omega$ decreases to sub-Keplerian values at the surface
(see Fig.~\ref{fig:precollapse_omega}).  The rate at which $ \Omega $
increases in the WD core is controlled by the shear parameter $
f_{\mathrm{sh}} $, which we choose in the range from $ 0.2 $ to $ 1
$. The case $ f_{\mathrm{sh}} = 0.2 $ corresponds to a \emph{nearly
  uniformly} rotating inner region, while $ f_{\mathrm{sh}} = 1 $
corresponds to strong differential rotation with $ \Omega
(\varpi_{\mathrm p}) / \Omega_{\mathrm{c,i}} \sim 2-3 $. In mass
coordinate, this corresponds to a ratio $ \Omega (M_{\mathrm{ic,b}}) /
\Omega_{\mathrm{c,i}}$ of $ \sim 1.4 - 2.4 $, where $M_{\mathrm{ic,b}}
\simeq 0.3\,M_\odot$ is the approximate mass that constitutes the
inner core at bounce in a nonrotating WD model.

In contrast to uniformly rotating models, differentially rotating WDs
are \emph{not limited} by the mass shedding limit at the surface. As a
result, $\Omega_{\mathrm{c,i}}$ can in principle be increased up to
the point beyond which the precollapse WD inner core becomes fully
centrifugally supported and does not collapse at all. For model set
AD, this maximum of $\Omega_{\mathrm{c,i}}$ is $\sim
8\,{\mathrm{rad\,s}}^{-1}$ (the low-$T$ model AD13f4, which 
becomes centrifugally supported already at a central density
of $\sim 7\times10^{10}\,\mathrm{g\,cm}^{-3}$) 
while the more compact DD models still collapse rapidly at
$\Omega_{\mathrm{c,i}} \sim 18\,{\mathrm{rad\,s}}^{-1}$ (model
DD7). As shown in  
Fig.~\ref{fig:collapse_times}, the most rapidly rotating AD model
(AD13f4) reaches core bounce after a time which is $\sim 55\%$ larger
than a nonrotating A model. For the most rapidly rotating DD model
this difference is only $\sim 5\%$.

In Fig.~\ref{fig:rhomaxb_vs_betaicb} we plot the maximum density
$\rho_{\mathrm{max,b}}$ at bounce as a function of the inner core
parameter $\beta_{\mathrm{ic,b}}$ at bounce. Slowly to
moderately-rapidly rotating WDs that reach $\beta_{\mathrm{ic,b}}
\lesssim 15\%$ are only mildly affected by rotation and their
$\rho_{\mathrm{max,b}}$ decrease roughly linearly with increasing 
$\beta_{\mathrm{ic,b}}$, but stay close to $\rho_{\mathrm{nuc}}$. The
effect of rotation becomes nonlinear in more rapidly rotating WDs. 
Models of our set that reach $\beta_{\mathrm{ic,b}} \gtrsim 18\%$
(\ie AD models with $\Omega_{\mathrm{c,i}} \gtrsim 5\,{\mathrm{rad\,
    s}}^{-1}$) undergo core bounce induced partly or completely
centrifugally at subnuclear densities. 

\begin{figure}
  \centerline{\includegraphics[width = 86 mm, angle = 0]{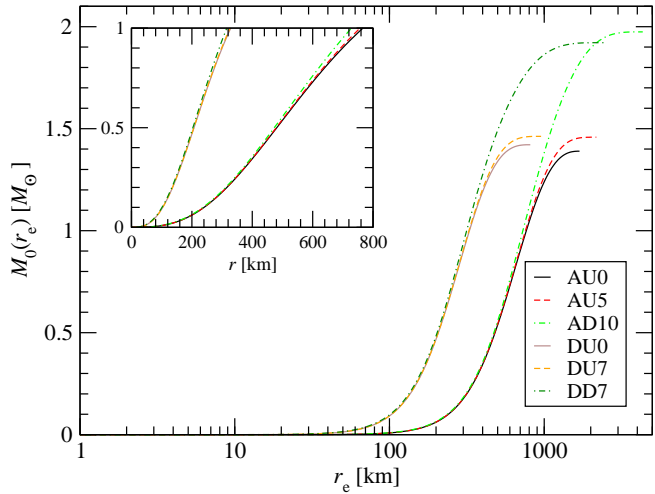}}
  \caption{Mass $M_0 (r_\mathrm{e})$ in units of solar masses of the
    WD inner region plotted as a function of the equatorial radial
    coordinate $ r_\mathrm{e} $ for a number of AIC models with slow
    and rapid rotation as well as high and small central
    densities. The inset plot shows the same on a linear radial
    scale. The mass distribution of the inner $ M_0 (r_\mathrm{e})
    \lesssim 0.5 M_\odot $ region is largely independent of the
    rotational configuration, while an increase (decrease) of the
    central density leads to homologous contraction (expansion) of the
    inner regions. See Sec.~\ref{sec:parameter_space} for details of
    the model setups.}  
  \label{fig:m_vs_r}
\end{figure}

\begin{figure}[t]
  \centerline{\includegraphics[width = 86 mm]{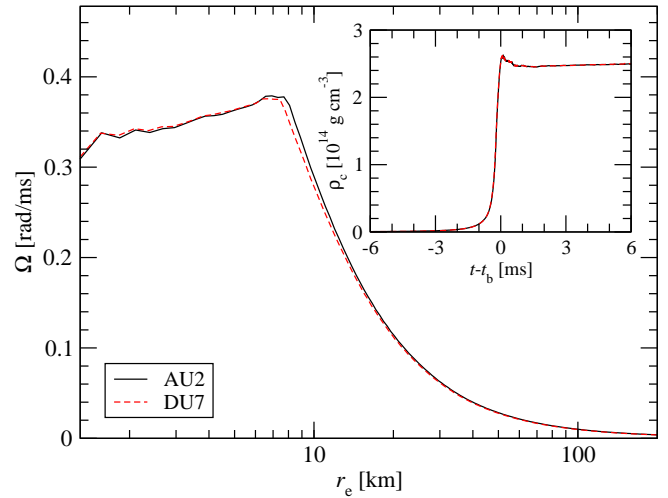}}
  \caption{Angular velocity profiles in the equatorial plane at the
    time of core bounce and time-evolution of the central density $
    \rho_{\mathrm c} $ (in the inset) for models AU2 and DU7. The
    initial angular velocity of model DU7 is larger by a factor of $
    \sim 5.3 $ than that of model AU2, but the latter experiences
    a $\sim 5.3$ greater spin-up during collapse. As a result,
    these models produce inner cores with almost identical rotational
    configurations and similar masses in the bounce phase. This is
    reflected in an identical evolution of the central
    densities at bounce.}
  \label{fig:omega_AU2_DU7}
\end{figure}

As shown in Fig.~\ref{fig:betaic_vs_omega}, $\beta_{\mathrm{ic,b}}$ is
a monotonic function of $\Omega_{\mathrm{c,i}}$, but is very sensitive
to both the rotation law and the initial WD compactness. Our most
rapidly uniformly rotating models AU5 and DU7 (both near the
mass-shedding limit) reach $\beta_{\mathrm{ic,b}}$ of $\sim 10.5\%$
and $\sim 3.6\%$, respectively.  Hence, uniformly rotating WDs always
undergo core bounce due to the stiffening of the nuclear EOS and with
little influence of rotation on the dynamics.

In models where centrifugal effects remain subdominant during
collapse, $\beta_{\mathrm{ic,b}}$ grows practically linearly with
$\Omega_{\mathrm{c,i}}$. This relationship flattens off for models
that become partially or completely centrifugally supported near
bounce. $\beta_{\mathrm{ic,b}}$ grows with increasing rotation up to
$\sim 24.5\%$ (model AD13f4), beyond which any further increase in
precollapse rotation leads to a \emph{decrease} of
$\beta_{\mathrm{ic,b}}$, since the inner core becomes fully
centrifugally supported before reaching high compactness and
spin-up. In other words, there exists a \textit{``centrifugal
  limit''} beyond which centrifugal forces dominate and, as a result,
increasing precollapse rotation leads to a decreasing
$\beta_\mathrm{ic,b}$ at core bounce. This result is analogous to what
previous studies \cite{dimmelmeier_08_a,ott_04_a} found in the
rotating core collapse of massive stars and has consequences for the
appearance of nonaxisymmetric rotational instabilities in PNSs. This
will be discussed in more detail in Sec.~\ref{sec:rotinst}.  

The influence of the precollapse compactness on the dynamics of
rotating AIC can also easily be appreciated from
Fig.~\ref{fig:betaic_vs_omega}.  The higher-density, more-compact WDs
of set D spin up much less than their A counterparts since their inner
cores are already very compact at the onset of collapse. Hence, a
higher-density WD that reaches a given value of
$\beta_{\mathrm{ic,b}}$ must have started out with a larger
$\Omega_{\mathrm{c,i}}$ than a lower-density WD reaching the same
$\beta_{\mathrm{ic,b}}$. For the particular choice of initial central
densities represented by D and A models and in the case of uniform or
near-uniform rotation, the ratio between the $\Omega_{\mathrm{c,i}}$
of a D and A model required to reach the same $\beta_{\mathrm{ic,b}}$
is $\sim 5.3$. This factor can be understood by considering
Fig.~\ref{fig:m_vs_r} in which we plot the enclosed mass as a function
of equatorial radial coordinate $ r_\mathrm{e} $ of selected A and D
initial WD configurations with slow and rapid rotation. The important
thing to notice is that the WD core structure ($M \lesssim 0.5
M_\odot$) is insensitive to the rotational configuration and obeys a
homology relation.  Stated differently, for a model of set D, a
homologous expansion in the radial direction by a factor of $\sim 2.3$
yields an object whose inner part is very similar to a lower-density A
model. In turn, the collapse of A models corresponds to a $\sim 2.3 $
times greater contraction of the WD core compared to their D model
counterparts and a spin up that is greater by a factor of $\sim
(2.3)^2 \simeq 5.3$. This explains the strong dependence of the
inner-core angular velocity and $\beta_{\mathrm{ic,b}}$ on the initial
central density observed in
Fig.~\ref{fig:betaic_vs_omega}. Furthermore, it suggests that one can
find A$-$D model pairs that differ greatly in their precollapse
angular velocities, but yield the same rotational configuration at
bounce.  An example for this is shown in Fig.~\ref{fig:omega_AU2_DU7}
in which we plot for the uniformly rotating model pair AU2-DU7 the
equatorial angular velocity profile at the time of bounce as well as
the evolution of the central density around the time of bounce.  AU2
and DU7 have practically identical angular velocity profiles and their
core structure, core mass and $\beta_{\mathrm{ic,b}}$ agree very
closely. As can be seen in the inset plot of
Fig.~\ref{fig:omega_AU2_DU7}, this results in nearly identical
$\rho_\mathrm{c}$ time evolutions around bounce and demonstrates that
WDs with quite different precollapse structure and rotational setup
can produce identical bounce and postbounce dynamics. This can also
occur for pairs of differentially rotating models and is an important
aspect to keep in mind when interpreting the GW signal from AIC
discussed in Sec.~\ref{sec:GW}.

\begin{figure}
  \centerline{\includegraphics[width = 86 mm]{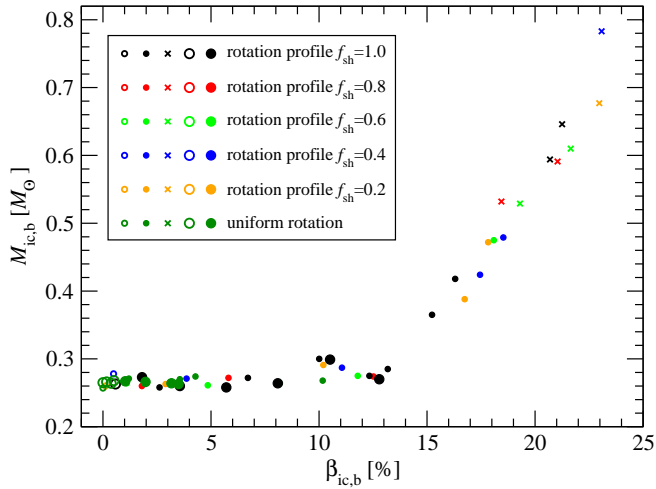}}
  \caption{Mass $ M_{\mathrm{ic,b}} $ of the inner core at bounce for
    all high-$T$ models versus the parameter $ \beta_{\mathrm{ic,b}} $
    of the inner core at bounce. The systematics of $
    M_{\mathrm{ic,b}} $ with $\beta_{\mathrm{ic,b}}$ are identical for
    the set of low-$T$ models, but their $ M_{\mathrm{ic,b}} $ are
    generally $\sim 10\%$ larger. The symbol convention for the
    various sets is explained in the caption of
    Fig.~\ref{fig:collapse_times}.}
  \label{fig:mic_vs_beta_ic}
\end{figure}

\begin{figure*}[t]
  \vspace{0.1cm}  
  \centerline{\epsfxsize = 7.8 cm
              \epsfbox{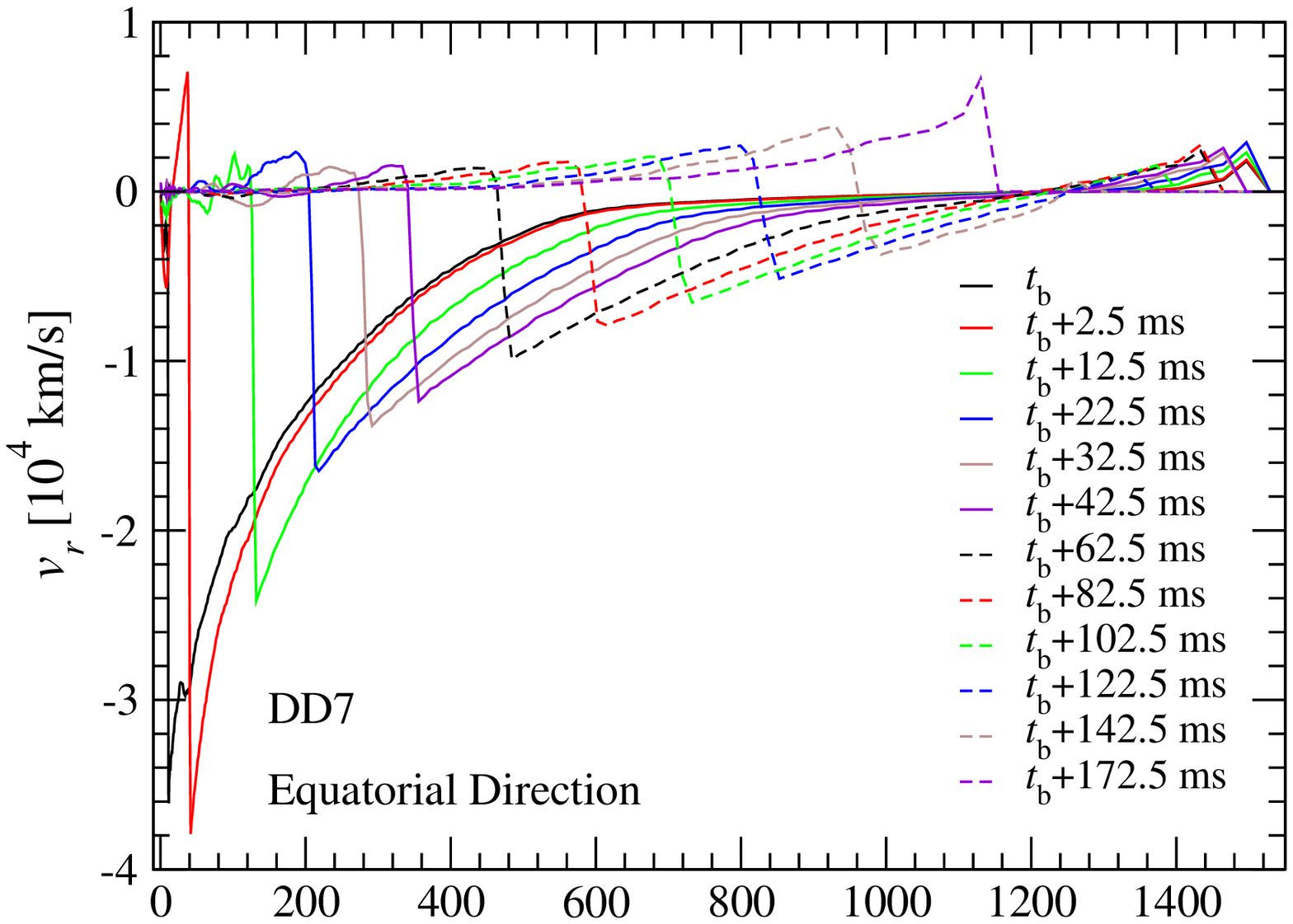}
              \hspace{0.5cm}
              \epsfxsize = 7.8 cm
              \epsfbox{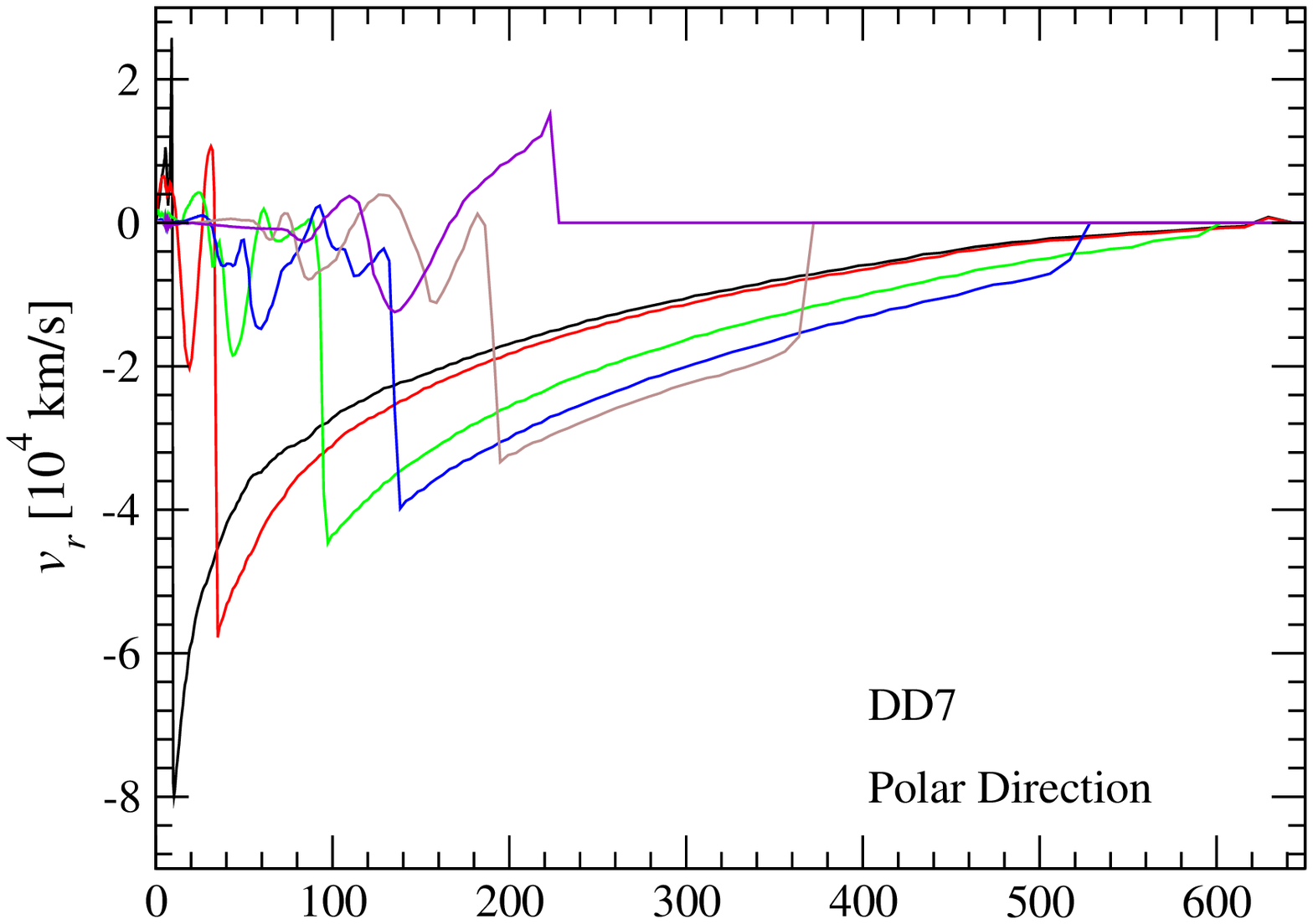}}
  \vspace{0.5cm}  
  \centerline{\epsfxsize = 7.8 cm
              \epsfbox{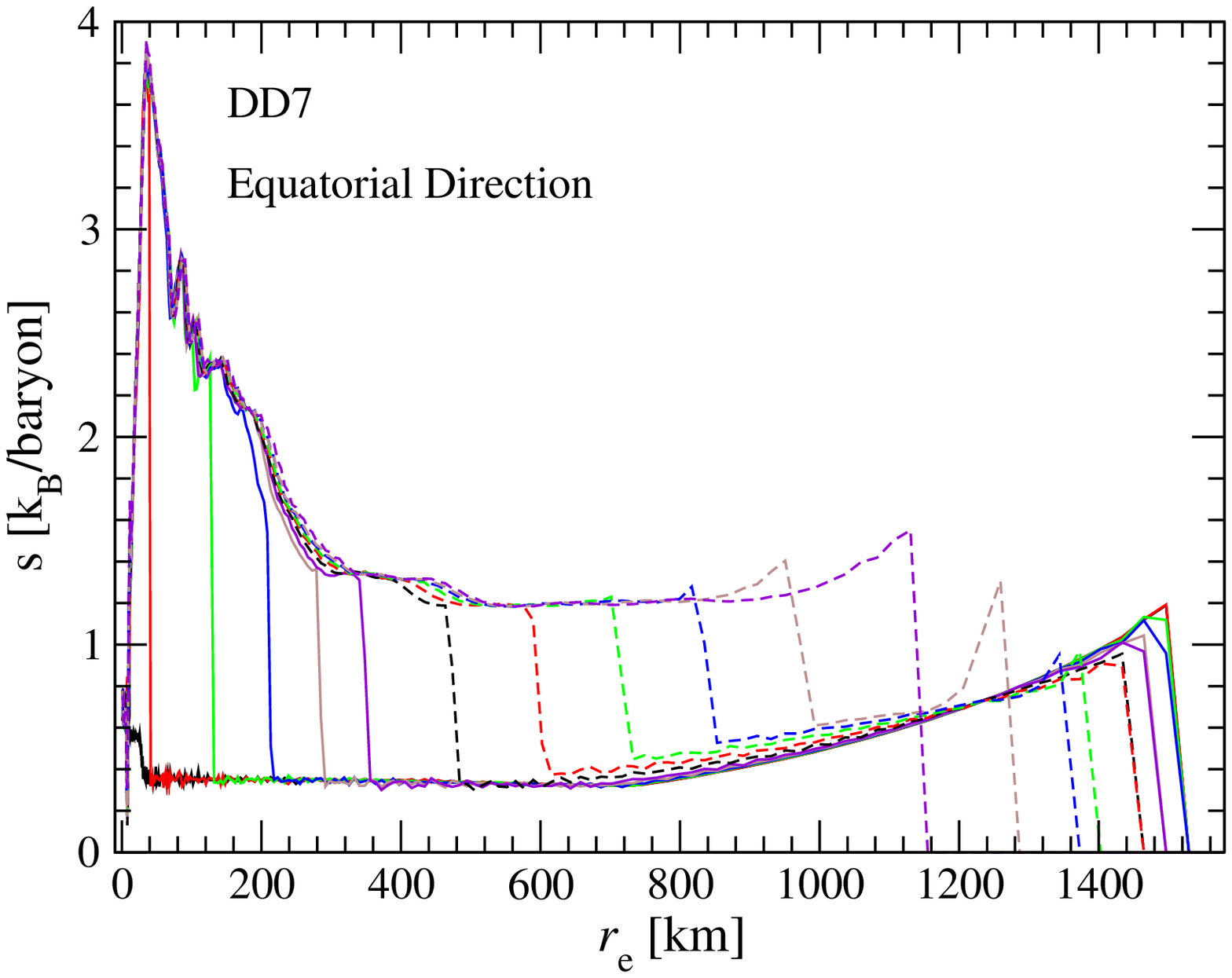}
              \hspace{0.5cm}
              \epsfxsize = 7.8 cm
              \epsfbox{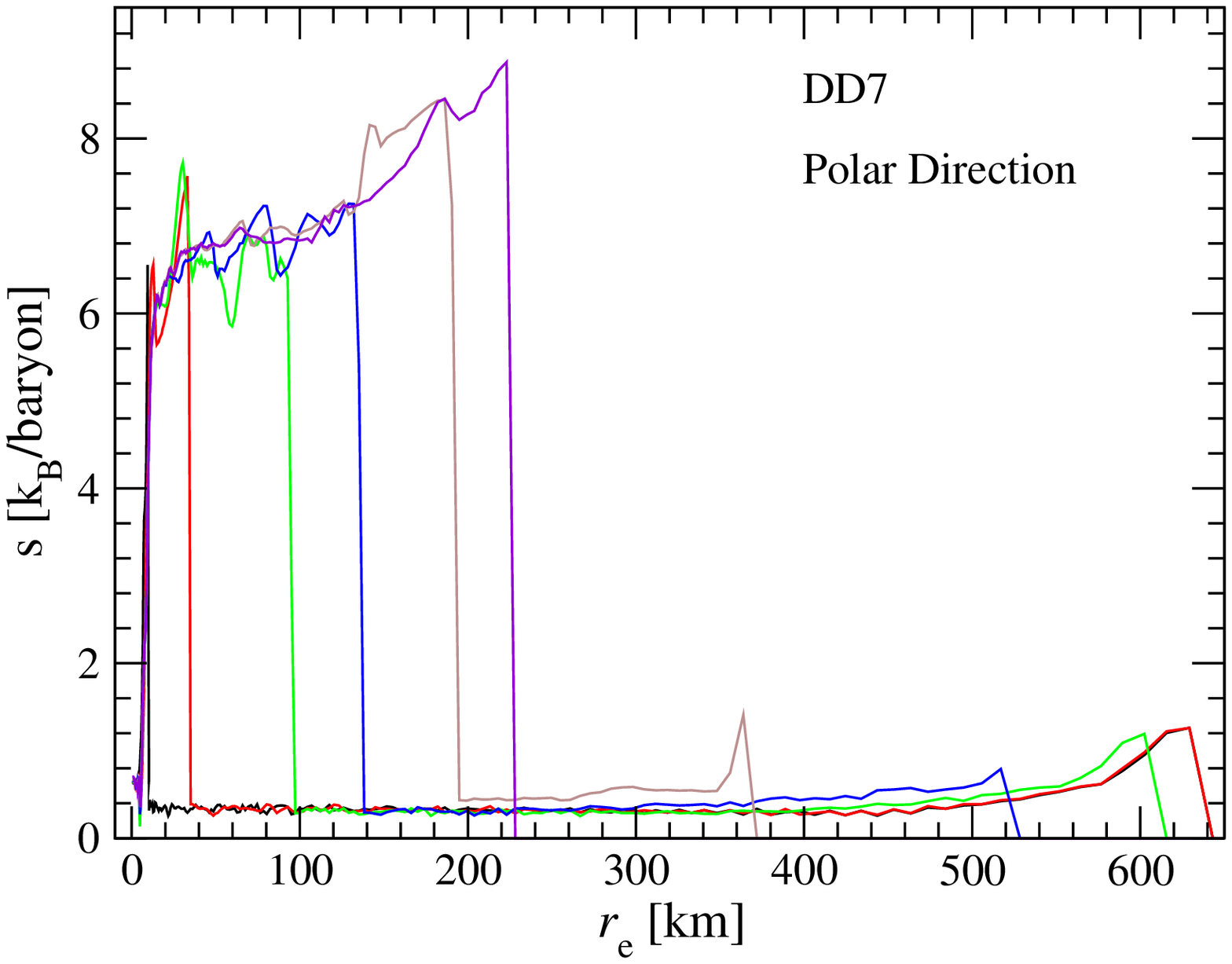}}
  \caption{Profiles of radial velocity (top panels) and specific
    entropy per baryon (bottom panels) at different postbounce times
    for AIC model DD7.}
  \label{fig:shock_propagation_DD}
\end{figure*}

Figure~\ref{fig:mic_vs_beta_ic} shows the mapping between
$\beta_{\mathrm{ic,b}}$ and the inner core mass $M_{\mathrm{ic,b}}$ at
bounce for all high-$T$ models. Rapid (differential) rotation not only
increases the equilibrium mass of WDs (see
Table~\ref{tab:initial_models}), but rotational support also increases
the extent of the region in sonic contact during collapse. Hence, it
may be expected that $M_{\mathrm{ic,b}}$ grows with increasing
rotation. However, for WDs below $\beta_{\mathrm{ic,b}} \lesssim
13\%$, $M_{\mathrm{ic,b}}$ is essentially unaffected by rotation and
stays within $0.02\,M_\odot$ of the nonrotating value of
$0.28\,M_\odot$. Only when the effects of rotation become strong at
$\beta_{\mathrm{ic,b}} \gtrsim 13-18\%$ does $M_{\mathrm{ic,b}}$
increase roughly linearly with $\beta_{\mathrm{ic,b}}$.  WDs that
undergo centrifugal bounce have $\beta_{\mathrm{ic,b}} \gtrsim 18\%$
and correspondingly large inner cores that are more massive than $\sim
0.5\,M_\odot$. Such values of $M_{\mathrm{ic,b}}$ are accessible only
to differentially rotating WDs.

Also for rotating models, the dependence of the AIC dynamics on the
initial temperature of AIC progenitor WDs is simple and
straightforwardly understood from the nonrotating results discussed in
Sec.~\ref{sec:nonrotating_collapse_dynamics}. These show that the
low-$T$ models yield inner cores that are $\sim 10\%$ larger in mass
than in their twice-as-hot high-$T$ counterparts. Due to their larger
mass, the inner cores of collapsing low-$T$ AIC progenitors also
contain a larger amount of angular momentum.  At fixed rotation law
and $\Omega_{\mathrm{c,i}}$, they reach values of
$\beta_{\mathrm{ic,b}}$ that are larger by up to $ \sim 5\%$ (in
absolute value!). Hence, lower-$T$ WDs become affected by centrifugal
support, bounce centrifugally, and reach the centrifugal limit at
lower $\Omega_{\mathrm{c,i}}$ than their higher-$T$ counterparts.
Along the same lines behave test calculations in which we impose
increased inner-core values of $Y_e$ (see
Secs.~\ref{sec:deleptonization} and \ref{sec:tempye}). The increased
$Y_e$ leads to more massive and more extend inner cores which, in
turn, are more likely to experience a centrifugal support.

To conclude our discussion of rotating AIC, we summarize for the
reader that the PNSs born from the set of differentially (uniformly)
rotating AIC models considered here have average angular
velocities\footnote{The average angular velocity $
  \bar{\Omega} $ of the differentially rotating models considered here
  is computed using the approximation $ \bar{\Omega} = J_{\mathrm{ic}} /
  I_{\mathrm{ic}} $, where $ J_{\mathrm{ic}} $ is the inner core angular
  momentum and $ I_{\mathrm{ic}} $ is the (Newtonian) inner core moment
  of inertia.} in the range from $ 0 $ to $ \sim 5\ {\mathrm{rad \,
    ms}}^{-1}$ ($ \sim 3.3\, {\mathrm{rad \, ms}}^{-1}$), while their
pole to equator axis ratios vary from $ 1 $ to $ \sim 0.4 $ ($ \sim
0.6 $)\footnote{The PNS formed in the AIC of WDs is
  surrounded by hot low density material in the early postbounce
  phase, making it hard to define the boundary of the PNS
  unambiguously. For the present rough estimate of the axis ratio, we
  assume a density threshold of $ 10^{12} \, {\mathrm{g \, cm^{-3}}} $
  to mark the boundary of the PNS.}. Some of the rapidly rotating WDs
produce PNSs with a slightly off-center maximum in density, though the
density distribution of the inner regions does not exhibit a
pronounced toroidal geometry. The clearest deviation from a
centrally-peaked density distribution is produced in the case of model
AD10, which reaches $\beta_{\mathrm{ic,b}} \simeq 21.3 ~\% $ ($
\beta_{\mathrm{ic,b}} \simeq 24.4 ~ \% $) in its high-$T$ (low-$T$)
variant. In this model, the point of highest density after bounce is
located at $ r \simeq 0.94 $ km, but the maximum value is larger than
the central density value by only $ \sim 0.3 \, \% $. For models with
less rapid rotation, the off-center maximum is much less pronounced,
and completely disappears for $ \beta_{\mathrm{ic,b}} $
below $ \sim 20 \, \% $.


\subsection{Shock Propagation and the Formation of Quasi-Keplerian
Disks}
\label{sec:shock_disc}


As pointed out earlier (see
Sec.~\ref{sec:nonrotating_collapse_dynamics}), all AIC models
considered in this study undergo weak hydrodynamic explosions. 
This is an artifact of our approach that neglects
postbounce neutrino emission, but is unlikely to 
strongly affect the results presented in this section, since
in the MGFLD simulations of \cite{dessart_06_a}, the shock
stalls only for a very short period and a weak explosion
is quickly initiated by neutrino heating.

In moderately-rapidly and rapidly rotating AIC (with $
\beta_{\mathrm{ic, b}} \gtrsim 5 \, \% $), the shock propagation is
significantly affected by centrifugal effects. The material near the
equatorial plane of rotating WDs experiences considerable centrifugal
support, and its collapse dynamics is slowed down.  As a consequence,
the bounce is less violent and the bounce shock starts out weaker near
the equatorial plane than along the poles. Centrifugal support of
low-latitude material also leads to reduced postbounce mass accretion
rates near the equatorial plane, facilitating steady propagation of
the shock at low latitudes. In the polar direction, where centrifugal
support is absent, the shock propagates even faster due to the steeper
density gradient and smaller polar radius of the WD. This quickly
leads to a prolate deformation of the shock front in all rotating
models and the shock hits the polar WD surface much before it breaks
out of the equatorial envelope. This is shown in
Fig.~\ref{fig:shock_propagation_DD}, where we plot the equatorial and
polar profiles of the radial velocity and specific entropy per baryon
for model DD7 at various postbounce times. Due to the prolateness of
the shock front, it breaks out of the polar surface $ \sim 130 $ ms
before reaching the WD's equatorial surface. Moreover, due to the
anisotropy of the density gradient and the initial shock strength, the
specific entropy of the shock-heated material is larger by a factor of
$ \sim 2 - 3 $ along the polar direction.

The asphericity of the shock front and the anisotropy of the shock
strength become more pronounced in AIC with increasing rotation
\cite{dimmelmeier_02_b, ott:06spin}.
As pointed out in Section~\ref{sec:nonrotating_collapse_dynamics}, due
to the their greater initial compactness and thus steeper density
gradients, the shock propagates faster in D models: In model DD1, for
example, the shock reaches the surface in the equatorial plane within
$ \sim 88 $ ms, while for model AD1, the corresponding time is $ \sim
143 $ ms.

Rapid rotation and, in particular, rapid differential rotation,
increases the maximum allowable WD mass. The most rapidly uniformly
rotating WDs in our model set (\ie models DU7 and AU5) have an
equilibrium mass of $\sim 1.46 M_\odot$, which is only slightly above
$M_\mathrm{Ch}$ in the nonrotating limit. Our most rapidly
differentially rotating WDs (models AD13f2 and DD7), on the other
hand, reach equilibrium masses of up to $\sim 2\,M_\odot$. Much of the
rotationally supported material is situated at low latitudes in the
outer WD core, falls in only slightly during collapse, and forms
a quasi-Keplerian disk-like structure. The equatorial bounce
shock is not sufficiently strong to eject much of the disk material
and ``wraps'' around the disk structure, producing only a small
outflow of outer disk material at $ v_r \lesssim 0.025 c $. This is in
agreement with Dessart et al.~\cite{dessart_06_a}, who first pointed
out that rapidly rotating AIC produces PNSs surrounded by massive
quasi-Keplerian disk-like structures in the early postbounce phase. As
recently investigated by Metzger~et~al.~\cite{metzger:09} (but not
simulated here), the hot disk will experience neutrino-cooling on a
timescale of $ \sim 0.1$ s, driving the disk composition neutron-rich
to reach $ Y_e \sim 0.1 $ \cite{metzger:09, dessart_06_a}, depleting
the pressure support and leading to limited contraction of the inner
parts of the disk. The outer and higher-latitude regions expand with a
neutrino-driven wind \cite{dessart_06_a}. As discussed
by~\cite{metzger:09}, subsequent irradiation of the disk by neutrinos
from the PNS increases its proton-to-neutron ratio, and $ Y_e $ may
reach values as high as $ \sim 0.5 $ by the time the weak interactions
in the disk freeze out. The disk becomes radiatively inefficient, $
\alpha $-particles begin to recombine and a powerful disk wind
develops, blowing off most of the disk's remaining
material. Metzger~et~al.~\cite{metzger:09} argue that, depending on
disk mass, the outflows synthesize of the order of $10^{-3} -
10^{-2}\,M_\odot$ of ${}^{56} {\mathrm{Ni}} $, but very small amounts
of intermediate-mass isotopes, making such AIC explosions
spectroscopically distinct from ${}^{56} {\mathrm{Ni}} $ outflows in
standard core-collapse and thermonuclear SNe.

\begin{table}
  \small
  \centering
  \caption{Summary of properties of the quasi-Keplerian disks formed in
    the set of
    AIC models AD, AU, DD and DU. $ H_{\mathrm{disk}} $ is the thickness
    and $ R_{\mathrm e} $ is the equatorial radius of the disk, while $
    M_{\mathrm{disk}} $ is its mass. These quantities are computed at
    the time when the shock reaches the WD surface in the equatorial
    plane. The disk parameters do not vary significantly between the two
    choices of WD temperature considered in this study.} 
  \label{tab:remnant_disc}
  \begin{tabular}{@{~~}l@{~~~~~}c@{~~~~~}c@{~~~~~}c@{~~~~~}}
    \hline \\ [-1 em]
    Collapse &
    $ R_{\mathrm e} $ &
    $ H_{\mathrm{disk}} / R_{\mathrm e} $ &
    $ M_{\mathrm{disk}} $ \\
    model &
    [km] & &
    [$ M_\odot $] \\
    \hline \\ [-0.5 em]

    AU1  & $347$  & $0.928$ & $\lesssim\!10^{-3}$\\
    AU2  & $401$  & $0.903$ & $\lesssim\!10^{-3}$\\
    AU3  & $447$  & $0.848$ & $\lesssim\!10^{-3}$\\
    AU4  & $732$  & $0.577$ & $0.002$\\
    AU5  & $907$  & $0.484$ & $0.030$\\ [0.5 em]

    DU1  & $248$  & $0.980$ & $\lesssim\!10^{-3}$\\
    DU2  & $249$  & $0.971$ & $\lesssim\!10^{-3}$\\
    DU3  & $249$  & $0.952$ & $\lesssim\!10^{-3}$\\
    DU4  & $261$  & $0.916$ & $\lesssim\!10^{-3}$\\
    DU5  & $291$  & $0.801$ & $\lesssim\!10^{-3}$\\
    DU6  & $332$  & $0.701$ & $\lesssim\!10^{-3}$\\
    DU7  & $350$  & $0.671$ & $0.007$\\ [0.5 em]

    AD1  & $866$  & $0.479$ & $0.030$\\
    AD2  & $935$  & $0.452$ & $0.038$\\
    AD3  & $1118$ & $0.437$ & $0.093$\\
    AD4  & $1321$ & $0.410$ & $0.222$\\
    AD5  & $1558$ & $0.374$ & $0.323$\\
    AD6  & $1638$ & $0.370$ & $0.356$\\
    AD7  & $1784$ & $0.377$ & $0.470$\\
    AD8  & $1912$ & $0.382$ & $0.507$\\
    AD9  & $2278$ & $0.342$ & $0.607$\\
    AD10 & $2700$ & $0.296$ & $0.805$\\ [0.5 em]

    DD1  & $360$  & $0.669$ & $0.005$\\
    DD2  & $402$  & $0.597$ & $0.008$\\
    DD3  & $461$  & $0.525$ & $0.019$\\
    DD4  & $554$  & $0.466$ & $0.054$\\
    DD5  & $670$  & $0.436$ & $0.161$\\
    DD6  & $853$  & $0.374$ & $0.279$\\
    DD7  & $1313$ & $0.255$ & $0.507$\\
    \hline
  \end{tabular}
\end{table}

Our results, summarized in Tab.~\ref{tab:remnant_disc}, show that the
masses and the geometry of the disks produced in AIC are sensitive to
the angular momentum distribution in the precollapse WDs. In models
with uniform rotation below the mass-shedding limit, only a very small
amount of low-latitude material rotates at near-Keplerian angular
velocities. Therefore, most of the outer core material of such models
undergoes significant infall, so that uniformly
rotating WDs will generally produce small disks.  The largest disk
mass for uniform rotation is $ M_{\mathrm{disk}} \sim 0.03 M_\odot
$\footnote{We point out that because the disks do not settle down to
  exact equilibrium right after bounce or not even after shock
  passage, it is hard to introduce an unambiguous definition of the
  disk mass. In the present study, we define the disk as the structure
  that surrounds the PNS at $ \varpi > 20 $ km with densities below $ 
  10^{11} ~ {\mathrm{g ~ cm}^{-3} } $ and angular velocity $ \Omega >
  0.58\, \Omega_{\mathrm K} $. The latter condition ensures that the
  disk cannot contract by more than a factor of $ \sim 3 $ as a result
  of cooling.} and is produced in model AU5 which rotates near the
mass-shedding limit. Since the angular velocity of the outer ($ \varpi
> \varpi_{\mathrm p} $) core of differentially rotating models is set
to reach nearly-Keplerian values
(cf.\ Eq.~(\ref{eqq:omega_outer_core})), most of the outer WD envelope
has substantial centrifugal support and thus the differentially
rotating models yield significantly larger $ M_{\mathrm{disk}} $. For
example, model AD4 which has $ \beta_{\mathrm{ic,b}} $ and total
angular momentum comparable to model AU5 yields a disk mass of $\sim
0.2 M_\odot $.

The total mass of the disk and its equatorial radius (the disk
thickness $ H_{\mathrm{disk}} $) grow with increasing rotation
(see Tab.~\ref{tab:remnant_disc}). Slowly rotating models such as AD1
have little centrifugally supported material and acquire spheroidal
shape soon after bounce, resulting in a disk mass as small as $ \sim
0.03 M_\odot $. More rapidly rotating models such as AD10 produce
significantly more strongly flattened disks, with $ R_\mathrm{e} \sim
2700 $ km, $ H_{\mathrm{disk}} \sim 800 $ km and a disk mass $ 
M_{\mathrm{disk}} $ of $\sim 0.8 M_\odot $. Due to the greater initial
compactness of the higher-density D models, their disks are less
massive and have smaller equatorial radii when compared to A
models. Hence, when considering two WDs of set A and D with the
same total angular momentum, the mass and equatorial extent 
of the disk in the D model will be smaller by a factor of 
$ \sim 1.5 - 2 $.

These results indicate that massive disks of $ M_{\mathrm{disk}} \gtrsim
0.1 M_\odot $ are unlikely to be compatible with the assumption of
uniformly rotating accreting WDs argued for by~\cite{saio_04,
  piro_08_a}. In order to produce disks of appreciable mass
and significant ${}^{56}$Ni-outflows in AIC, the progenitor
must either be an accreting WD obeying a differential rotation law
similar to that proposed by~\cite{yoon_04,yoon_05}, or may
be the remnant of a binary-WD merger event. However, for the 
latter, the differential rotation law is unknown and may be very
different from what we consider here (for
a discussion of binary-WD merger simulations, see, \eg
\cite{rosswog:09} and references therein).


\section{Gravitational wave emission}
\label{sec:GW}

\begin{figure}
  \centerline{\includegraphics[width = 86 mm]{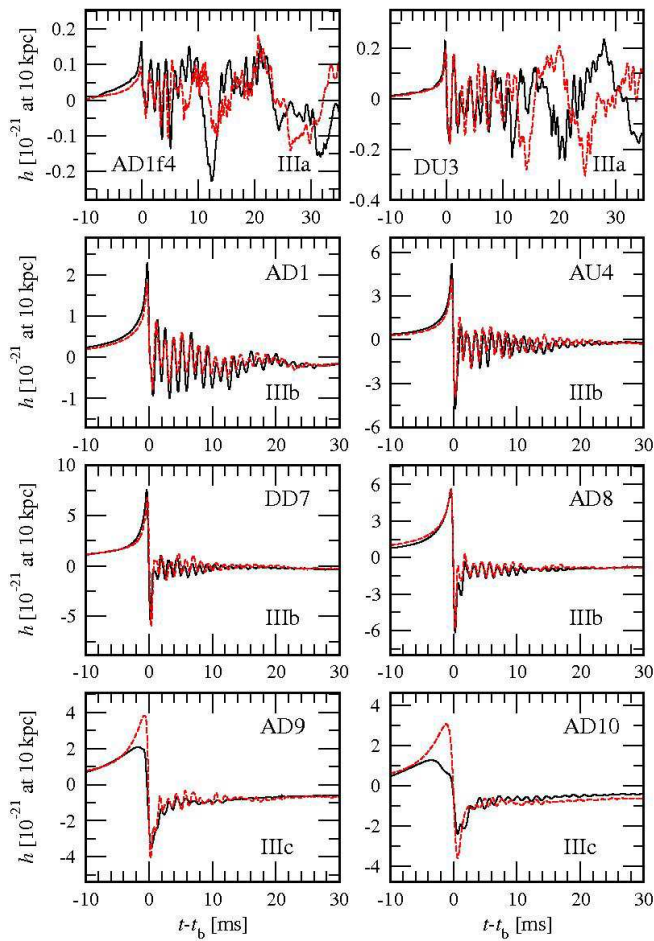}}
  \caption{Evolution of the dimensionless GW strain $ h $ (in units of
    $ 10^{-21} $ at a source distance of $ 10 \ \mathrm{kpc} $) as a
    function of postbounce time for representative models with
    different precollapse rotation profiles, central densities and
    temperatures (low-$T$ models with solid black lines and high-$T$
    with dashed red lines). Models with slow and (almost) uniform
    precollapse rotation (\eg AD1f4 or DU3) develop considerable
    prompt postbounce convection visible as a dominating
    lower-frequency contribution in the waveform. Centrifugal effects
    damp this prompt convection and the waveforms of models with
    moderately-rapid rotation (\eg AD1, AU4, DD7 and AD8) and of
    rapidly rotating models (\eg AD9 or AD10) exhibit no such
    contribution to the signal.}  
  \label{fig:gw_strain_representative_models}
\end{figure}

In Fig.~\ref{fig:gw_strain_representative_models}, we present the time
evolution of the GW strain $ h $ at an assumed source distance of $
10\,\mathrm{kpc} $ for a representative set of AIC models evolved with
the $\overline{Y_e}(\rho)$ parametrization obtained from
\cite{dessart_06_a}.  The GW signals of all models have the same
overall morphology. This general AIC GW signal shape bears strong
resemblance to GW signals that have been classified as ``type~III'' in
the past~\cite{zwerger_97_a, dimmelmeier_02_b, ott:09rev}, but also
has features in common with the GW signals predicted for rotating iron
core collapse (``type~I'', \cite{dimmelmeier_08_a}).

The GW strain $ h $ in our AIC models is positive in the infall phase
and increases monotonically with time, reaching its peak value in the
plunge phase, just $ \sim 0.1\,{\mathrm{ms}}$ before bounce. Then, $ h
$ rapidly decreases, reaching a negative peak value within $ \sim
1-2\,{\mathrm{ms}} $. While the first positive peak is produced by
rapid infall of the inner core, the first negative peak is caused by
the reversal of the infall velocities at bounce. Following the large
negative peak, $ h $ oscillates with smaller amplitude with a damping
time of $ \sim 10 $ ms, reflecting the hydrodynamical ringdown
oscillations of the PNS. Although all AIC models of our baseline set
produce Type~III signals, we can introduce three subtypes whose
individual occurrence depends on the parameter $\beta_{\mathrm{ic,b}}$
of the inner core at bounce:

\textbf{Type IIIa}. In slowly rotating WDs (that reach $
\beta_{\mathrm{ic,b}} \lesssim 0.7 \, \% $), strong prompt convective
overturn develops in the early postbounce phase, adding a
lower-frequency contribution to the regular ringdown signal (\eg
models AD1f4, DU3). The largest-amplitude part of this GW signal type
comes from the prompt convection. Nevertheless, the GW signal produced
by the bouncing centrifugally-deformed inner core is still
discernible, with the first positive peak being generally larger by a
factor of $ \sim 2 $ than the first negative peak. Subsequent
ringdown peaks are smaller by a factor of $ \gtrsim 3 $. We point out
that the observed prompt postbounce convection is most likely
overestimated in our approach, since we do not take into account
neutrino losses and energy deposition by neutrinos in the immediate
postshock region, whose effect will quickly smooth out the negative
entropy gradient left behind by the shock and thus significantly damp
this early convective instability in full postbounce
radiation-hydrodynamics calculations (see, \eg~\cite{mueller_04_a,
  buras_06_a, ott:09rev}).

\textbf{Type IIIb}. In moderately rapidly rotating WDs that reach $
0.7 \lesssim \beta_{\mathrm{ic,b}} \lesssim 18 \, \% $ and  still
experience a pressure-dominated bounce, convection is effectively
suppressed due to a sufficiently large positive specific angular
momentum gradient~(\eg \cite{ott:08sn}). Hence, there is no noticeable
convective contribution to the postbounce GW signal (see, \eg models
AD1, AU4, DD7, AD8). For this signal subtype, the peak GW strain $
|h|_{\mathrm{max}} $ is associated with the first positive peak while
relative values of the amplitudes of the first several peaks are
similar to type~IIIa.

\textbf{Type IIIc}. If rotation is sufficiently rapid and leading to $
\beta_{\mathrm{ic,b}} \gtrsim 18 \, \% $, the core bounces at
subnuclear densities due to strong centrifugal support. This is
reflected in the GW signal by an overall lower-frequency emission and
a significant widening of the bounce peak of the waveform (see, \eg
models AD9, AD10). In some models of this subtype, the negative peak
can be comparable to or slightly exceed that of the first positive
peak in the waveform. This reflects the fact that the plunge
acceleration is apparently reduced more significantly by rotation than
is the re-expansion acceleration at core bounce.
The postbounce ringdown peaks in all type~IIIc models are smaller by
a factor of $ \gtrsim 2 $ compared to the bounce signal. As pointed
out in Sec.~\ref{sec:rotating_collapse_dynamics}, uniformly rotating
models do not rotate sufficiently rapidly to experience centrifugal
bounce. Hence, they do not produce a type~IIIc signal.

\vskip.3cm

The AIC GW signal morphology is affected only slightly by variations
in WD temperature and their resulting changes in the inner core $Y_e$
that are on the few-percent level for the range of precollapse
temperatures considered here. In test calculations with more
substantially increased inner-core values of $Y_e$ (see
Secs.~\ref{sec:deleptonization} and \ref{sec:tempye}) and, in turn,
significantly larger values of $M_\mathrm{ic,b}$, we find signals that
are intermediate between type~III and type I.

Key quantitative results from our model simulations are
summarized in Tables.~\ref{tab:collapse_models} and
\ref{tab:detectability}. The waveform data for all models are
available for download from \cite{stellarcollapseGW}.


\subsection{Peak Gravitational Wave Amplitude}
\label{sec:peak_amplitude}

\begin{figure}
  \centerline{\includegraphics[width = 86 mm]{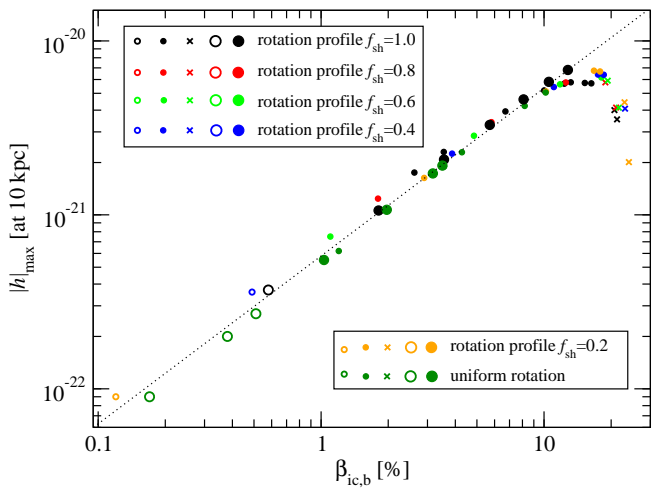}}
  \caption{Peak value $ |h|_{\mathrm{max}} $ of the GW amplitude at a
    source distance of $ 10 $ kpc distance for the \textit{burst}
    signal of all models versus the parameter $ \beta_{\mathrm{ic,b}}
    $ of the inner core at the time of bounce. At slow to 
    moderately rapid rotation, $|h|_{\mathrm{max}} $ scales almost
    linearly with $ \beta_{\mathrm{ic,b}} $ (marked by the
    dotted straight line), while for $ \beta_{\mathrm b} \gtrsim 16\%
    $ centrifugal effects reduce $ |h|_{\mathrm{max}} $. The symbol
    convention for the various sets is explained in the caption of
    Fig.~\ref{fig:collapse_times}.} 
  \label{fig:h_max_vs_beta}
\end{figure}

Across our entire model set, the peak GW amplitude $
|h|_{\mathrm{max}} $ covers a range of almost two orders of magnitude,
from $\sim 10^{-22}$ to $\sim 10^{-20}$ (at distance to the source of
$10\,\mathrm{kpc}$; see
Tab.~\ref{tab:collapse_models}). $|h|_\mathrm{max}$ depends on various
parameters and it is difficult to provide a simple description of its
systematics that encompasses all cases.  In order to gain insight into
how $|h|_\mathrm{max}$ depends on $\Omega_{\mathrm{c,i}} $, on
differential rotation, on the initial $\rho_{\mathrm c}$, on the
precollapse WD temperature, and on the degree of deleptonization in
collapse, we describe below the effects of variations in one of these
parameters while holding all others fixed.

\textit{(i)} In a sequence of precollapse WDs with fixed differential
rotation, $\rho_\mathrm{c}$, and $T_\mathrm{0}$, the peak GW amplitude
$ |h|_{\mathrm{max}} $ increases steeply with $\Omega_{\mathrm{c,i}} $
in slowly rotating models that do not come close to being
centrifugally supported. When centrifugal effects become dynamically
important, $ |h|_{\mathrm{max}} $ saturates at $ \sim 7 \times
10^{-21} $ (at $ 10\,\mathrm{kpc}$) and then decreases with increasing
$ \Omega_{\mathrm{c,i}} $. This reflects the fact that such rapidly
spinning inner cores produced by AIC cannot reach high densities and
high compactness and that the slowed-down collapse decreases the
deceleration at bounce, thus reducing $ |h|_{\mathrm{max}} $ and
pushing the GW emission to lower frequencies.

\begin{figure}
  \centerline{\includegraphics[width = 86 mm]{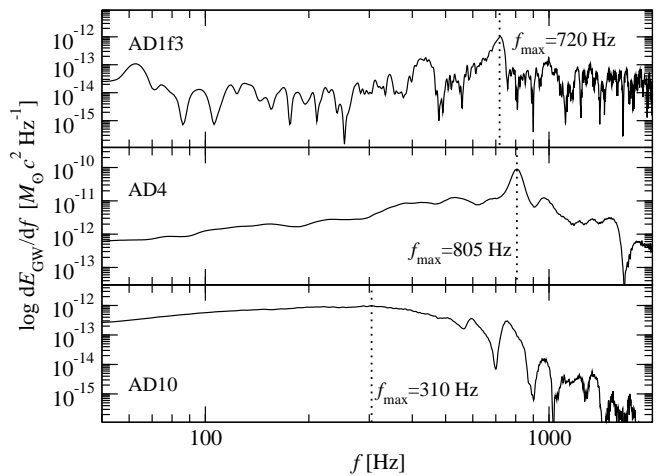}}
  \caption{Spectral energy density of the GW signal for
    representative AIC models AD0f3 (top panel), AD4 (center panel),
    and AD10 (bottom panel). $f_\mathrm{max}$ is the peak frequency
   of the GWs emitted at core bounce.}
  \label{fig:gw_spectrum_representative_models}
\end{figure}

\textit{(ii)} In a sequence of precollapse WDs with fixed $
\Omega_{\mathrm{c, i}} $, $T_\mathrm{0}$, and $\rho_\mathrm{c}$, an
increase in the degree of WD differential rotation leads to an
increase in the amount of angular momentum present in the WD inner
core at bounce. This translates into an increase of $
|h|_{\mathrm{max}} $ in models that do not become centrifugally
supported and experience a pressure-dominated bounce. The transition
to centrifugal bounce is now reached at lower values of
$\Omega_{\mathrm{c, i}} $ (see
Sec.~\ref{sec:rotating_collapse_dynamics}), so that the centrifugal
saturation of $ |h|_{\mathrm{max}} $ described above in \emph{(i)} is
reached at much smaller values of $\Omega_{\mathrm{c, i}} $.

\textit{(iii)} In a sequence of precollapse WDs with fixed $
\Omega_{\mathrm{c, i}} $, $T_\mathrm{0}$, and differential rotation
and varying $\rho_{\mathrm c}$, models with lower (higher) $\rho_\mathrm{c}$
yield larger (smaller) values of $ |h|_{\mathrm{max}} $. This is
because models that are initially less compact spin up more during
collapse (\cf the discussion in
Sec.~\ref{sec:rotating_collapse_dynamics}). However, this systematics
holds only as long as the model does not become centrifugally supported,
which happens for lower (higher) $\rho_c$ WDs at smaller (greater) $
\Omega_{\mathrm{c, i}} $.

\begin{figure}
  \centerline{\includegraphics[width = 86
      mm]{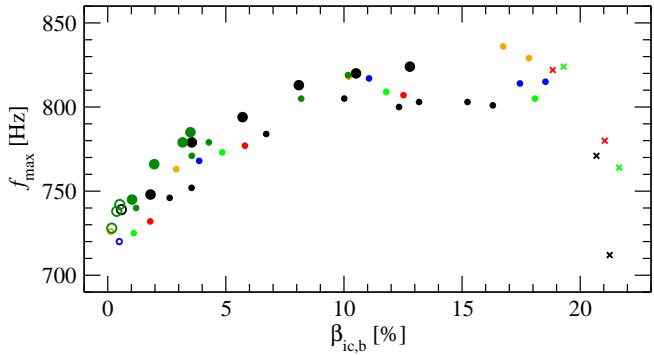}}
  \caption{Frequency $ f_{\mathrm{max}} $ at the maximum of the GW
    spectral energy density in pressure-dominated bounce and a subset
    of centrifugal bounce models versus the parameter $
    \beta_{\mathrm{ic,b}} $ of the inner core at bounce. The symbol 
    convention for the various sets is explained in the caption of
    Fig.~\ref{fig:collapse_times}. }
  \label{fig:fmax_vs_beta}
\end{figure}

\textit{(iv)} When only the WD temperature is varied,
we find that for slowly to moderately-rapidly rotating WDs, high-$T$
models generally reach smaller $ |h|_{\mathrm{max}} $ than their
low-$T$ counterparts. This is because high-$T$ WDs yield
smaller inner cores at bounce, which hold less angular momentum and,
as a consequence, are less centrifugally deformed (see
Tab.~\ref{tab:collapse_models} and in
Fig.~\ref{fig:gw_strain_representative_models}).  However, this
behavior reverses in rapidly rotating WDs for which low-$T$
models are more centrifugally affected and, hence, yield a smaller $
|h|_{\mathrm{max}} $ than their high-$T$ counterparts.

\textit{(v)} If the degree of deleptonization is decreased by an
ad-hoc increase of inner-core $Y_e$ (see
Secs.~\ref{sec:deleptonization} and \ref{sec:tempye}) and all else is
kept fixed, $M_\mathrm{ic,b}$ increases and for slowly to moderately
rapidly rotating WDs, $ |h|_{\mathrm{max}} $ increases. As for the
low-$T$ case discussed in the above, this behavior reverses in rapidly
rotating WDs for which high-$Y_e$ models are more centrifugally
affected and yield smaller $ |h|_{\mathrm{max}} $ than their
lower-$Y_e$ counterparts.

To demonstrate the dependence of $ |h|_{\mathrm{max}} $ on the overall
rotation of the inner core at bounce, we plot in
Fig.~\ref{fig:h_max_vs_beta} $ |h|_{\mathrm{max}} $ as a function of
the inner core parameter $ \beta_{\mathrm{ic,b}} $ at bounce
for our high-$T$ models.  $|h|_{\mathrm{max}}$ depends primarily on
$\beta_{\mathrm{ic,b}}$ and is rather independent of the particular
precollapse configuration that leads to a given
$\beta_{\mathrm{ic,b}}$. For small $\beta_{\mathrm{ic,b}}$ far away
from the centrifugal limit, we find $|h|_{\mathrm{max}} \propto
\beta_\mathrm{ic,b}^{0.74}$, where we have obtained the exponent by a
power-law fit of high-$T$ models with $1\% \lesssim
\beta_\mathrm{ic,b} \lesssim 13\%$. This finding is in qualitative
agreement with what~\cite{dimmelmeier_08_a} saw for iron core
collapse.  The overall maximum of $|h|_{\mathrm{max}}$ is reached in
WDs that yield $\beta_{\mathrm{ic,b}} \sim 16\%$, beyond which $
|h|_{\mathrm{max}} $ decreases with increasing
$\beta_{\mathrm{ic,b}}$.


\subsection{Gravitational-Wave Energy Spectrum}
\label{sec:gw_spectrum}

\begin{figure*}[t]
  \begin{center}
  \hspace*{-.4cm}\includegraphics[width = 63 mm]{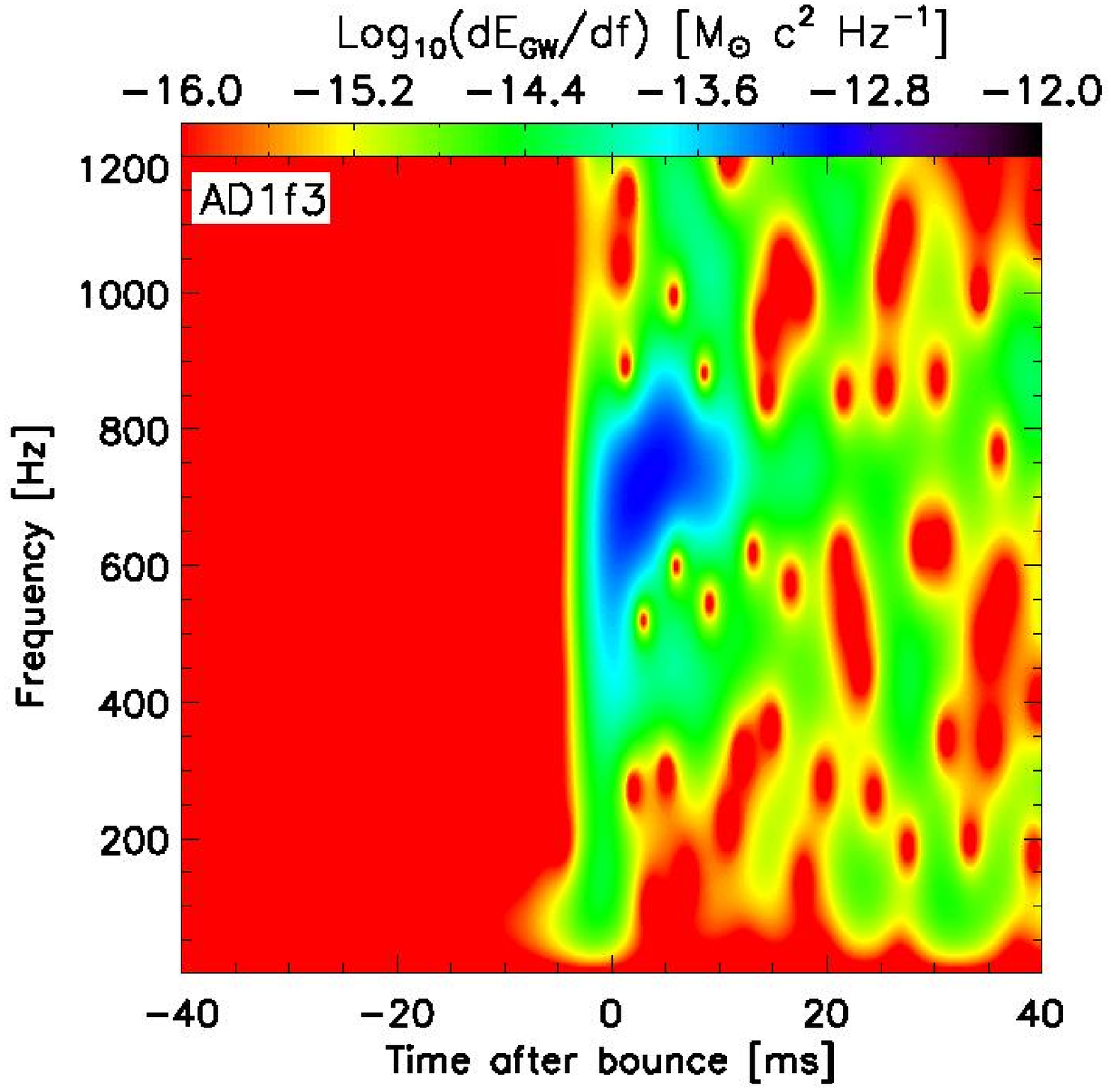}
  \hspace*{-.25cm}\includegraphics[width = 63 mm]{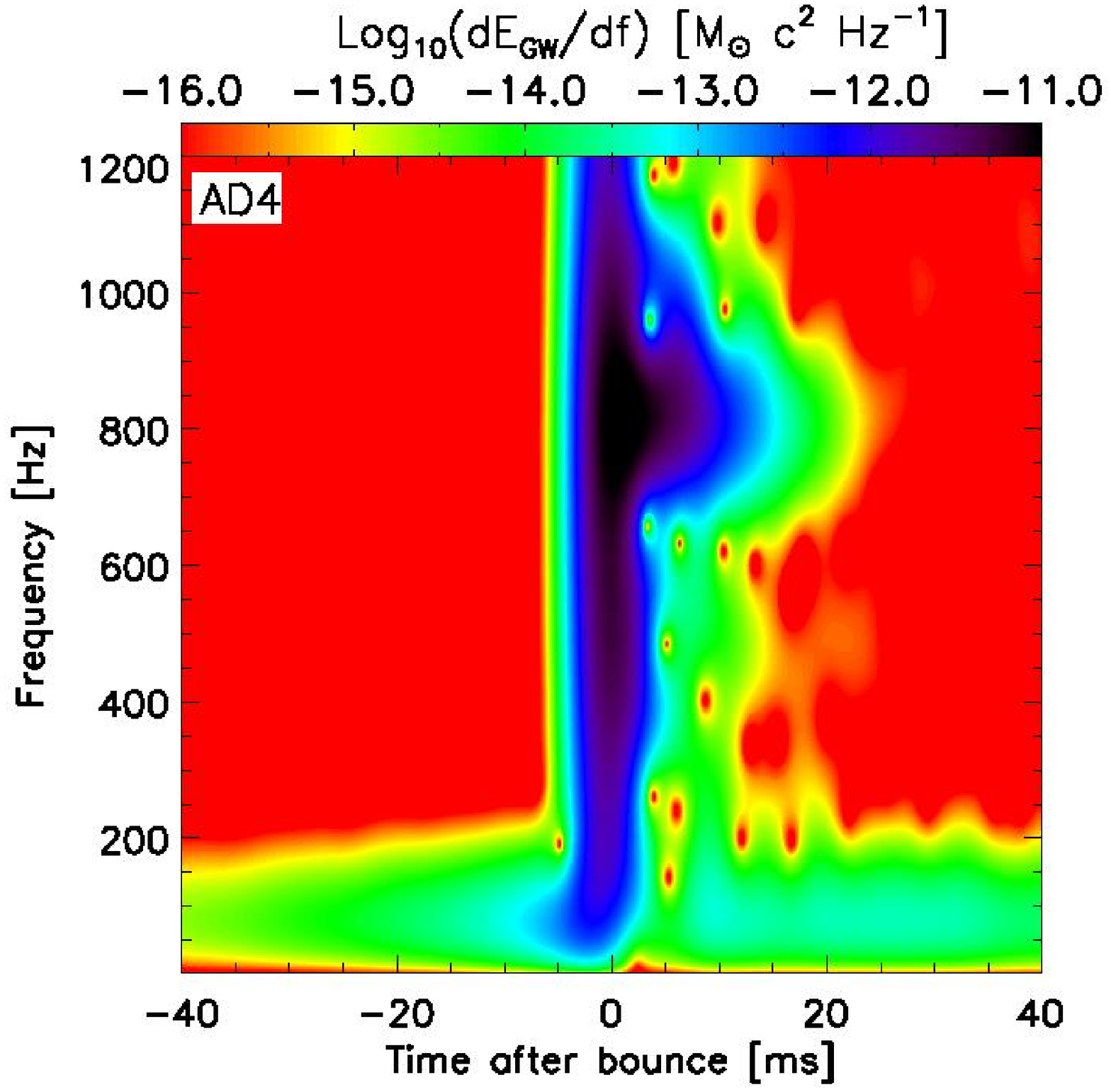}
  \hspace*{-.25cm}\includegraphics[width = 63 mm]{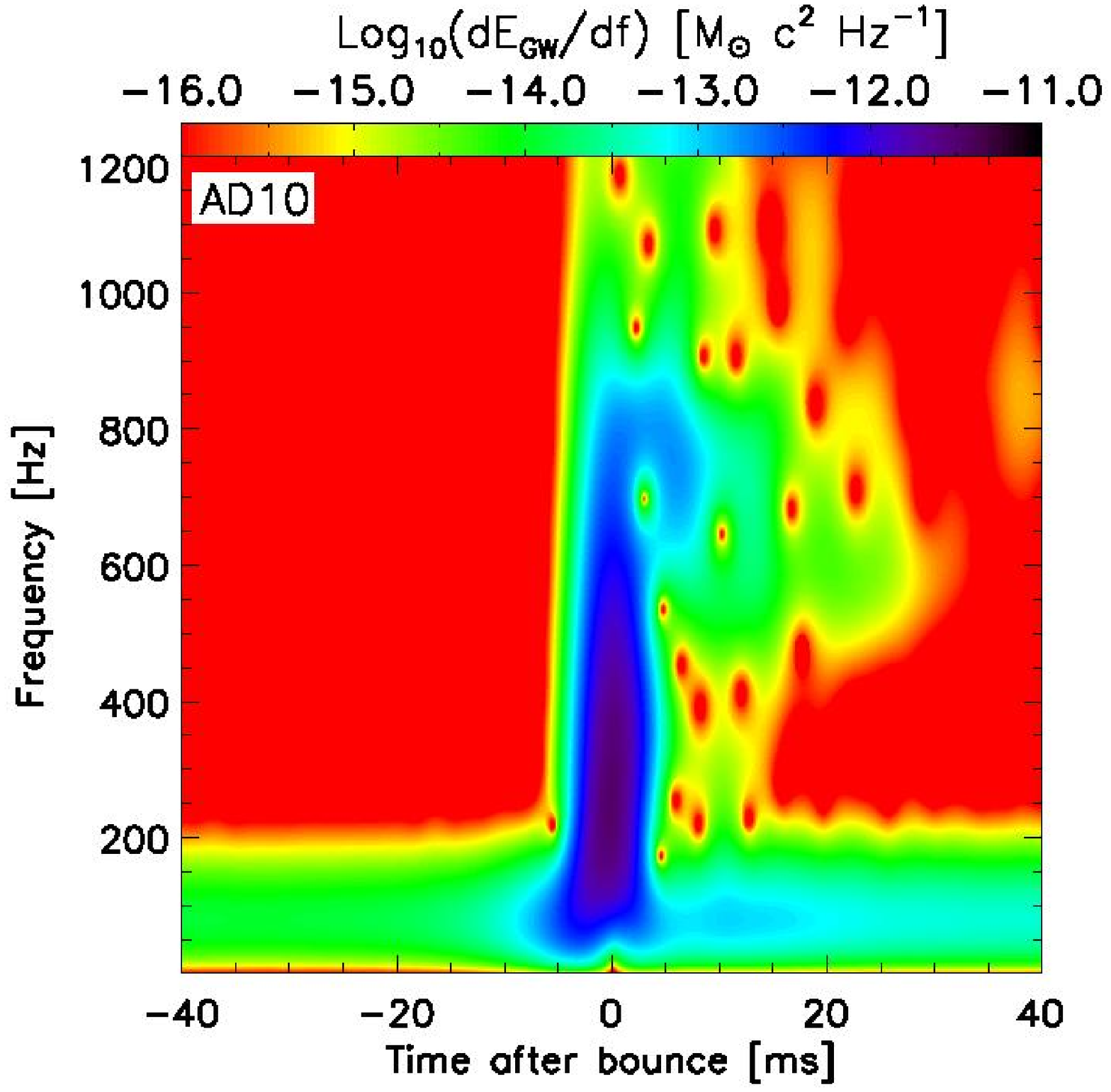}\hspace*{-.3cm}
  \end{center}
  \caption{Time-frequency colormaps of the GW signals of models AD1f3
    (type~IIIa), AD4 (type~IIIb), and AD10 (type~IIIc, see also
    Fig.~\ref{fig:gw_spectrum_representative_models}). Plotted is the
    ``instantaneous'' spectral GW energy density $dE_{\mathrm{GW}} /
    df$ in a $2$-ms Gaussian window as a function of postbounce
    time. Note the prebounce low-frequency contribution in the
    moderately-rapidly rotating models (model AD4, center panel) and
    rapidly rotating models (model AD10, right panel). The range of
    the colormap of the left panel (model AD1f3) is smaller by one dex
    than those of the other panels.}
  \label{fig:gw_tf}
\end{figure*}

The total energy emitted in GWs is in the range of $ \sim 10^{-10}
M_\odot c^2 \lesssim E_{\mathrm{GW}} \lesssim 2 \times 10^{-8} M_\odot
c^2 $ in the entire set of models considered in this article.  In
Fig.~\ref{fig:gw_spectrum_representative_models}, we plot the GW
spectral energy density $ d E_{\mathrm{GW}} / d f $ of the three
models AD1f3, AD4 and AD10 as representative examples of the three
signal subtypes IIIa - IIIc. The top panel shows model AD1f3 as a
representative pressure-dominated bounce model with prompt
convection. In such models, there is a strong structured, but broad
contribution to the spectrum at low frequencies. The integral of such
a contribution (which is present in all models with slow rotation) can
exceed that from core bounce in these models. This is not the case in
model AD1f3, whose GW burst from bounce is the one leading to the peak
at $f_{\mathrm{max}} = 720\,\mathrm{Hz}$.

The central panel of Fig.~\ref{fig:gw_spectrum_representative_models}
depicts $ d E_{\mathrm{GW}} / d f $ of model AD4 as a representative
pressure-dominated bounce model in which no significant postbounce
convection occurs. The spectrum of this model exhibits a distinct and
narrow high-frequency peak at $ f_{\mathrm{max}} \sim 805 $
Hz. Finally, the bottom panel of
Fig.~\ref{fig:gw_spectrum_representative_models} refers to model A10
that experiences centrifugal bounce.  In this model, the dynamics is
dominated by centrifugal effects, leading to low frequency emission
and $f_{\mathrm{max}} = 310\,\mathrm{Hz}$, but higher-frequency
components are still discernible and are most likely related to
prolonged higher-frequency GW emission from the PNS ringdown.

In Fig.~\ref{fig:fmax_vs_beta}, we plot the peak frequencies $
f_{\mathrm{max}} $ of the GW energy spectrum as a function of the
inner core parameter $ \beta_{\mathrm{ic,b}} $ for high-$T$ AIC models
(the low-$T$ and higher-$Y_e$ models show the same overall
systematics). In models that undergo pressure-dominated bounce, $
f_{\mathrm{max}} $ increases nearly linearly with $
\beta_{\mathrm{ic,b}} $ in the region $ \beta_{\mathrm{ic,b}} \lesssim
10 \ \% $, while at $ \beta_{\mathrm{ic,b}} $ in the range of $ 10
\ \% \lesssim \beta_{\mathrm{ic,b}} \lesssim 20 \ \% $, the growth of
$ f_{\mathrm{max}} $ saturates at $ \sim 800 $ Hz and
$f_{\mathrm{max}}$ does not change significantly with further increase
of rotation. For very rapid rotation ($ \beta_{\mathrm{ic,b}} \gtrsim
20 \ \% $), $ f_{\mathrm{max}} $ decreases steeply with $
\beta_{\mathrm{ic,b}} $, reaching a value of $ \sim 400 $ Hz at
$\beta_{\mathrm{ic,b}} \simeq 23 \, \% $ (not shown in the figure, see
Table~\ref{tab:detectability}).

While it is straightforward to understand the systematics of
$f_{\mathrm{max}}$ at high $ \beta_{\mathrm{ic,b}} $ where centrifugal
effects slow down collapse and thus naturally push the GW emission to
low frequencies, the increase of $f_{\mathrm{max}}$ with rotation at
low to intermediate $ \beta_{\mathrm{ic,b}} $ is less intuitive. If one
assumes that the dominant GW emission at core bounce in all models is
due to the quadrupole component of the fundamental quasi-radial mode
of the inner core, one would expect a monotonic decrease of
$f_\mathrm{max}$ with increasing rotation and, hence, decreasing mean
core density (see, \eg \cite{andersson:03}). A possible explanation
for the increase of $f_\mathrm{max}$ at slow to moderately-rapid
rotation is that the primary GW emission in these models is due to the
fundamental quadrupole $^2\! f$-mode, whose frequency may increase 
with rotation. This has been demonstrated by
Dimmelmeier~et~al.~\cite{dimmelmeier:06} who studied oscillation modes 
of sequences of $\gamma = 2$ polytropes. To confirm this
interpretation, and following the technique of mode-recycling outlined
in~\cite{dimmelmeier:06}, we perturb a subset of our postbounce
cores with the eigenfunction of the $^2 \!  f$-mode of a Newtonian
nonrotating neutron star. As expected, we find that the resulting
dynamics of the postbounce core is dominated by a single oscillation
mode with a frequency that matches within $\lesssim
10\%$ the peak frequency $ f_{\mathrm{max}} $ of $dE_{\mathrm{GW}}/df$
of the corresponding slowly or moderately rapidly rotating AIC
model. The interesting details of the mode structure of the inner
cores of AIC and iron core collapse will receive further scrutiny in a
subsequent publication.

Finally, in Fig.~\ref{fig:gw_tf}, we provide time-frequency analyses
of the GW signals of the same representative models shown in
Fig.~\ref{fig:gw_spectrum_representative_models}. The analysis is
carried out with a short-time Fourier transform employing a Gaussian
window with a width of $2\,{\mathrm{ms}}$ and a sampling interval of
$0.2\,{\mathrm{ms}}$. In all three cases, the core bounce is clearly
visible and marked by a broadband increase of the emitted energy. The
slowly-rotating model AD1f3 emits its strongest burst at
$600-800\,\mathrm{Hz}$ ($f_{\mathrm{max}} = 720\,\mathrm{Hz}$) and
subsequently exhibits broadband emission with significant power at
lower frequencies due to prompt convection.  Model AD4 is more
rapidly rotating and shows significant pre-bounce low-frequency
emission due to its increased rotational deformation. At bounce, a
strong burst, again with power at all frequencies, but primarily at
frequencies about its $f_{\mathrm{max}} = 805\,\mathrm{Hz}$, is
emitted. Much of the postbounce $E_{\mathrm{GW}}$ is emitted through
ringdown oscillations at $f_{\mathrm{max}}$ that may be related to the
$^2\! f$ mode of this model's PNS. Finally, in the rapidly rotating
and centrifugally bouncing model AD10, we observe again low-frequency
emission before bounce, but only a small increase of the primary
emission band at bounce to $\sim 200 -
400\,\mathrm{Hz}$. Nevertheless, there is still an appreciable energy
emitted from higher-frequency components of the dynamics at bounce and
postbounce times.


\subsection{Comparing GW signals from AIC \\ and Iron Core Collapse}
\label{sec:gw_types_comparison}

Recent studies~\cite{dimmelmeier_07_a, ott_07_a, ott_07_b,
  dimmelmeier_08_a} have shown that the collapse of rotating iron
cores (ICC) produces GW signals of uniform morphology (so-called
``type I'' signals, see, \eg \cite{zwerger_97_a}) that generically
show one pronounced spike associated with core bounce with a
subsequent ringdown and are similar to the type~III signals found here
for AIC. As in the AIC case, the GW signal of ICC has subtypes for
slow, moderately rapid, and very rapid rotation. For comparing AIC and
ICC GW signals, we chose three representative AIC models and then pick
three ICC models with similar $\beta_{\mathrm{ic,b}}$ from the study
of Dimmelmeier~et~al.~\cite{dimmelmeier_08_a} whose waveforms are
freely available from~\cite{wave_catalog}. This should ensure that we
compare collapse models that are similarly affected by centrifugal
effects for a one-to-one comparison. However, one should keep in mind
that the inner core masses $M_{\mathrm{ic,b}}$ at bounce of ICC models
are generally larger by $\sim 0.2 - 0.3\,M_\odot$ than in our AIC
models (see Sec.~\ref{sec:nonrotating_collapse_dynamics}).

\begin{figure*}
  \centerline{\includegraphics[width = 150 mm]{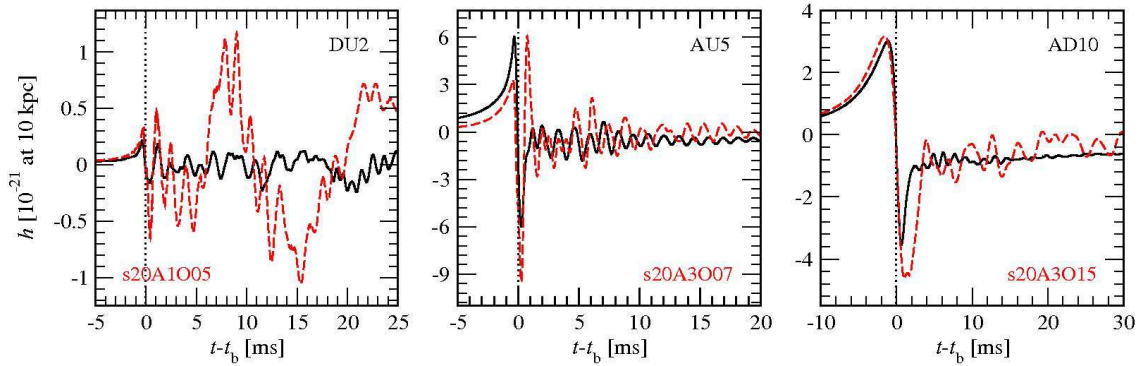}}
  \caption{Evolution of GW signals in the high-$ T $ AIC models DU2,
    AU5, AD10 (black solid lines) and three massive star iron
    core-collapse models s20A1O05, s20A3O07 and s20A3O15 (red
      dashed lines) from the model set of Dimmelmeier et
    al.~\cite{dimmelmeier_08_a}. The AIC model DU2 and the iron core
    collapse model s20A1O05 undergo pressure-dominated bounce with
    significant prompt convection. Models AU5 and s20A3O7 experience
    pressure-dominated bounce without significant convection, and
    models AD10 and s20A3O15 undergo centrifugal bounce. The inner
    cores of models DU2, AU5 and AD10 (s20A1O05, s20A3O07 and
    s20A3O15) reach values of $\beta_{\mathrm{ic,b}}$ of about $ 0.4\%,
    \, 10.2\% $ and $ 21.3\%  $ ($ 0.7\%, \, 10.1\%, $ and $ 21.6\%
    $). Times are given relative to the time of core bounce which we
    mark with a vertical line.}
  \label{fig:gw_strain_ICC_AIC}
\end{figure*}

In Fig.~\ref{fig:gw_strain_ICC_AIC} we present this comparison and
plot the GW signals of the high-$T$ AIC models DU2 (slow rotation,
type~IIIa), AU5 (moderately-rapid rotation, type~IIIb), and AD10 (very
rapid rotation, type~IIIc). In the same order, we superpose the GW
signals of the Dimmelmeier~et~al.~\cite{dimmelmeier_08_a} models
s20A1O05, s20A3O07 and s20A3O15. These models started with the
precollapse iron core of a $20\,M_\odot$ star and were run with the
same code, EOS, and deleptonization algorithm as our AIC models,
though with different, ICC specific $Y_e (\rho)$
trajectories.

The left panel of Fig.~\ref{fig:gw_strain_ICC_AIC} compares the slowly
rotating models DU2 and s20A1O05 that undergo pressure dominated
bounce and exhibit strong postbounce convection. As pointed out
before, the latter is most likely overestimated in our current
approach as well as in Dimmelmeier~et~al.'s.  Note that the width of
the waveform peaks associated with core bounce is very similar,
indicating very similar emission frequencies. Model s20A1O05 exhibits
a significantly larger signal amplitude at bounce.  This is due to
s20A1O05's larger $M_{\mathrm{ic,b}}$ but also to the fact that its
$\beta_{\mathrm{ic,b}}$ is $\sim 0.7\%$ compared to the $\sim 0.4\%$
of DU2 (a closer match was not available from
\cite{wave_catalog}). The prompt convection in model s20A1O05 is more
vigorous and generates a larger-amplitude GW signal than in model DU2.
This is due to the much steeper density gradient in the WD core that
allows the AIC shock to remain stronger out to larger radii.  Hence,
it leaves behind a shallower negative entropy gradient, leading to
weaker convection and postbounce GW emission.

In the central panel of Fig.~\ref{fig:gw_strain_ICC_AIC} we compare
two moderately-rapidly rotating models with nearly identical
$\beta_{\mathrm{ic}}$ of $\sim 10\%$. Both models show a pre-bounce
rise due to the inner core's accelerated collapse in the plunge
phase. The AIC inner core, owing to its lower $Y_e$ and weaker
pressure support, experiences greater acceleration and emits a
higher-amplitude signal than its ICC counterpart in this phase. At
bounce, the stiff nuclear EOS decelerates the inner core, leading to
the large negative peak in the GW signal. Because of the more massive
inner core in ICC and since the EOS governing the dynamics is
identical in both models, the magnitude of this peak is greater in the
ICC model. Following bounce, the ICC model's GW signal exhibits a
large positive peak of comparable or larger amplitude than the
pre-bounce maximum. This peak is due to the re-contraction of the ICC
inner core after the first strong expansion after bounce. With
increasing rotation, this re-contraction and the associated feature in
the waveform become less pronounced.  On the other hand, due to its
smaller inertia, the AIC inner core does not significantly overshoot
its new postbounce equilibrium during the postbounce expansion. Hence,
there is no appreciable postbounce re-contraction and no such large
positive postbounce peak in the waveform.

Example waveforms of AIC and ICC models experiencing core
bounce governed by centrifugal forces are shown in the right panel of
Fig.~\ref{fig:gw_strain_ICC_AIC}. In this case, the pre-bounce plunge
dynamics are significantly slowed down by centrifugal effects and the
GW signal evolution is nearly identical in AIC and ICC. At bounce, the
more massive inner core of the ICC model leads to a larger and broader
negative peak in the waveform and its ringdown signal exhibits larger 
amplitudes than in its AIC counterpart.

Finally, we consider AIC models with variations in the inner-core
$Y_e$ due either to different precollapse WD temperatures or ad-hoc
changes of the $\overline{Y_e}(\rho)$ parametrization (see
Secs.~\ref{sec:deleptonization} and \ref{sec:tempye}).  Lower-$T$ WDs
yield larger inner-core values of $Y_e$ and, in turn, larger
$M_\mathrm{ic,b}$ and GW signals that are closer to their iron-core
counterparts. The same is true for models in which we impose an
increased inner-core $Y_e$: AIC models with inner-core $Y_e$ $10\%$
larger than predicted by~\cite{dessart_06_a} still show clear type~III
signal morphology while models with $20\%$ larger $Y_e$ fall in
between type~III and type I.

To summarize this comparison: rotating AIC and rotating ICC lead to
qualitatively and quantitatively fairly similar GW signals that most
likely could not be distinguished by only considering general signal
characteristics, such as maximum amplitudes, characteristic
frequencies and durations. A detailed knowledge of the actual waveform
would be necessary, but even in this case, a distinction between AIC
and ICC on the basis of the comparison presented here would be
difficult.  It could only be made for moderately-rapidly spinning
cores based on the presence (ICC, type Ib) or absence (AIC, type~IIIb)
of a first large positive peak in the waveform, but, again,
\emph{only} if AIC inner cores indeed have significantly smaller $Y_e$
than their iron-core counterparts. ICC and AIC waveforms of types
Ia/IIIa and Ic/IIIc are very similar.  Additional astrophysical
information concerning the distance to the source and its orientation
as well as knowledge of the neutrino and electromagnetic signatures
will most likely be necessary to distinguish between AIC and ICC.


\subsection{Detection Prospects for the Gravitational Wave 
  Signal from AIC}
\label{sec:detectability}

In order to assess the detection prospects for the GW  signal
from AIC, we evaluate the  characteristic signal
frequency $ f_{\mathrm c} $ and the dimensionless characteristic
GW amplitude $ h_{\mathrm c} $. Both quantities are
detector-dependent and are computed using
Eq.~(\ref{eq:characteristic_frequency}) and
(\ref{eq:characteristic_amplitude}), respectively. 

\begin{figure}
  \centerline{\includegraphics[width = 86 mm, angle = 0]{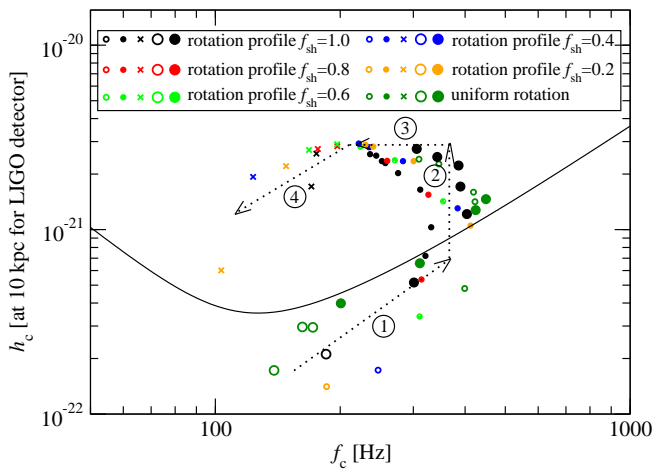}}
  \caption{Detector-dependent characteristic amplitudes of the GW
    signals of all models at an assumed distance of $ 10 $ kpc. The
    symbol convention for the various sets is described in the caption
    of Fig.~\ref{fig:collapse_times}. See the text for a discussion of
    the numbers and arrows.}
  \label{fig:detectability_ligo}
\end{figure}

In Fig.~\ref{fig:detectability_ligo}, we show $ h_{\mathrm c} $ for
all models as a function of $f_{\mathrm c} $ for an initial 4-km LIGO
detector, assuming a source distance of $ 10 $ kpc. For comparison
with detector sensitivity, we include initial LIGO's design
$h_{\mathrm{rms}}$ curve~\cite{ligo}. The 
distribution of our set of models in this figure obeys simple
systematics. A number of very slowly rotating models that undergo
pressure-dominated bounce with prompt convection (type~IIIa) form a
cluster in frequency in one region (near arrow 1). These models have
the overall lowest values of $ h_{\mathrm c} $ and exhibit low values
of $ f_{\mathrm c} $ in the range of $ 130 - 350\,\mathrm{Hz} $. Both
$ f_{\mathrm c} $ and $ h_{\mathrm c} $ grow with increasing rotation
(along arrow 1). For the pressure-dominated bounce models without
significant prompt convection (type~IIIb), $ h_{\mathrm c} $ 
grows with increasing rotation (along arrow 2), now at practically
constant $ f_{\mathrm c}$ of $ \sim 350 $ Hz. Even for these models, $
f_{\mathrm c} $ is always lower than the typical peak frequency $
f_{\mathrm{max}} \sim 700-800\,\mathrm{Hz}$ of their spectral GW
energy densities.  This is due to the specific characteristics
of the LIGO detector, whose highest sensitivity is around
$100\,\mathrm{Hz}$, thus leading to a systematic decrease
of $f_\mathrm{c}$ with respect to $f_\mathrm{max}$.

In more rapidly rotating models, centrifugal effects become more
important, leading to greater rotational deformation of the inner
core, but also slowing down the dynamics around core bounce,
ultimately limiting $h_{\mathrm c}$ and reducing $f_{\mathrm c}$
(along arrow 3). Models that rotate so rapidly that they undergo
centrifugal bounce (type~IIIc) cluster in a separate region in the $
h_{\mathrm c} - f_{\mathrm c}$ plane (along arrow 4), somewhat below
the maximum value of $ h_{\mathrm c} $ and at considerably lower $
f_{\mathrm c} $. The systematics for the lower-$ T $ models and for
other detectors is very similar. Not surprisingly, given the analogies
in the two signals, a similar behavior of $ h_{\mathrm c} $ and $
f_{\mathrm c} $ was observed in the context of rotating iron core
collapse~\cite{dimmelmeier_08_a}.

\begin{figure}
  \centerline{\includegraphics[width = 86 mm, angle = 0]{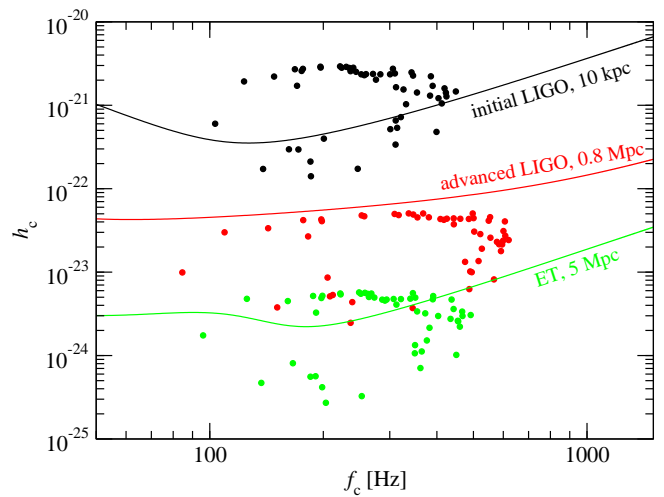}}
  \caption{Location of the GW signals from core bounce in the $
    h_{\mathrm c} $--$ f_{\mathrm c} $ plane relative to the
    sensitivity curves of various interferometer detectors (as
    color-coded) for an extended set of models AD. The sources are at a
    distance of $ 10 $ kpc for LIGO, $ 0.8 $ Mpc for Advanced LIGO,
    and $ 5 $ Mpc for the Einstein Telescope.}
  \label{fig:detectability_ligo_ligoII_et}
\end{figure}

Figure~\ref{fig:detectability_ligo_ligoII_et} provides the same type
of information shown in Fig.~\ref{fig:detectability_ligo} but also for
the advanced LIGO detector when the source is at $0.8$ kpc (\eg within
the Andromeda galaxy), or for the proposed Einstein Telescope
(ET)~\cite{ET} and a source distance of $5$~Mpc. Initial LIGO is
sensitive only to GWs coming from a moderately-rapidly or 
rapidly rotating AIC event in the Milky Way, but its advanced version will
probably be able to reveal sources also outside the Galaxy,
although only within the local group.  Finally,
third-generation detectors such as ET, may be sensitive enough to
detect some AIC events out to $ \sim 5$ Mpc.

\begin{figure}
  \centerline{\includegraphics[width = 86 mm, angle = 0]{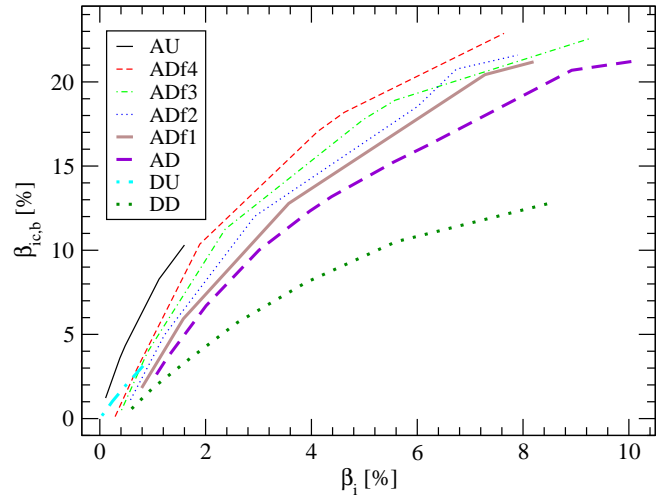}}
  \caption{The inner core parameter $ \beta_{\mathrm{ic,b}} $ at
    bounce is plotted as a function of the precollapse parameter $
    \beta_{\mathrm i} $ for high-$T$ models. Due to different central
    densities and rotation profiles of the precollapse WD models,
    there is no one-to-one correspondence between $
    \beta_{\mathrm{ic,b}} $ and $ \beta_{\mathrm i} $. Hence, although
    one can extract $ \beta_{\mathrm{ic,b}} $ accurately from the
    bounce AIC GW signal, it is impossible to put strong constraints
    on $ \beta_{\mathrm i} $ using the GW signal.}  
  \label{fig:betaicb_vs_betai}
\end{figure}

As pointed out in Secs.~\ref{sec:peak_amplitude}
and~\ref{sec:gw_spectrum}, the GW signal amplitudes and the spectral
GW energy distribution is determined primarily by 
$ \beta_{\mathrm{ic,b}} $. Hence, given the
systematics shown in Fig.~\ref{fig:detectability_ligo}, one may be
optimistic about being able to infer $\beta_{\mathrm{ic,b}}$ to good
precision from the observation of GWs from a rotating AIC event. For
example, as demonstrated in Fig.~\ref{fig:fmax_vs_beta}, even the
knowledge of only $ f_{\mathrm{max}} $ can put some constraints on
$\beta_{\mathrm{ic,b}}$. However, inferring accurately the properties
of the progenitor WD using exclusively information provided by GWs may
be extremely difficult given the highly degenerate dependence of
$\beta_{\mathrm{ic,b}}$ on the various precollapse WD model parameters
discussed in Sec.~\ref{sec:rotating_collapse_dynamics}. To elaborate
on this point, we show in Fig.~\ref{fig:betaicb_vs_betai} the relation
between $\beta_{\mathrm{ic,b}}$ and the precollapse WD parameter
$\beta_{\mathrm i}$. Even if GWs can provide good constraints on
$\beta_{\mathrm{ic,b}}$, a rather large variety of models with
different initial rotational properties would be able to lead to that
same $\beta_{\mathrm{ic,b}}$ and additional astrophysical information
on the progenitor will be needed to determine the precollapse
rotational configuration.  The only exception to this is the
possibility of ruling out uniform WD progenitor rotation if
$\beta_\mathrm{ic,b} \gtrsim 18\% $ (\cf
Sec.~\ref{sec:rotating_collapse_dynamics}).

\begin{table}
  \small
  \centering
  \caption{GW signal characteristics for the high-$T$
     AIC models: $ E_\mathrm{GW}$ is the total GW energy, $
     f_{\mathrm{max}} $ is the peak frequency of the GW energy
     spectrum, $ \Delta f_{50} $ is the frequency interval around $
     f_{\mathrm{max}} $ that emits $ 50 \ \% $ of $ E_\mathrm{GW}
     $. The nonrotating models are omitted here.}
  \label{tab:detectability}
  \begin{tabular}{@{}l@{~~~~}c@{~~~~}c@{~~~~}c@{~~~~}}
    \hline \\ [-1 em]
    AIC &
    $ E_{\mathrm{GW}} $ &
    $ f_{\mathrm{max}} $ &
    $ \Delta f_{50} $ \\
    model &
    [$10^{-9} M_\odot c^2 $] &
    [Hz] &
    [Hz] \\
    \hline \\ [-0.5 em]

    AU1 & 1.1  & 742.7 & 31\\
    AU2 & 5.7  & 782.7 & 28\\
    AU3 & 7.8  & 786.7 & 27\\
    AU4 & 15.7 & 816.0 & 49\\
    AU5 & 17.0 & 831.9 & 120\\ [0.3 em]

    DU1 & 0.1  & 768.0 & 343\\
    DU2 & 0.3  & 770.0 & 556\\
    DU3 & 0.4  & 747.0 & 473\\
    DU4 & 0.9  & 745.0 & 304\\
    DU5 & 2.0  & 765.7 & 26\\
    DU6 & 4.6  & 778.3 & 21\\
    DU7 & 5.8  & 788.3 & 20\\ [0.3 em]

    AD1 & 2.2  & 752.6 & 33\\
    AD2 & 3.7  & 765.4 & 30\\
    AD3 & 8.7  & 790.5 & 37\\
    AD4 & 11.8 & 812.5 & 115\\
    AD5 & 14.2 & 813.6 & 173\\
    AD6 & 15.0 & 815.0 & 201\\
    AD7 & 15.5 & 815.1 & 220\\
    AD8 & 13.8 & 811.0 & 245\\
    AD9 & 1.8  & 806.0 & 413\\
    AD10& 1.0  & 304.0 & 126\\ [0.3 em]

    DD1 & 0.3  & 740.0 & 324\\
    DD2 & 1.4  & 746.7 & 62\\
    DD3 & 4.6  & 780.1 & 21\\
    DD4 & 9.3  & 793.6 & 22\\
    DD5 & 15.2 & 813.7 & 21\\
    DD6 & 18.5 & 820.2 & 48\\
    DD7 & 22.3 & 826.9 & 152\\ [0.3 em]

    AD1f1 & 1.3  & 745.7 & 54\\
    AD1f2 & 0.6  & 731.4 & 69\\
    AD1f3 & 0.2  & 726.5 & 315\\
    AD1f4 & 0.1  & 737.0 & 476\\ [0.3 em]

    AD3f1 & 7.7  & 787.0 & 36\\
    AD3f2 & 7.0  & 781.2 & 31\\
    AD3f3 & 5.5  & 777.2 & 27\\
    AD3f4 & 3.7  & 769.8 & 27\\ [0.3 em]

    AD6f1 & 15.7 & 818.0 & 175\\
    AD6f2 & 16.0 & 819.5 & 165\\
    AD6f3 & 16.0 & 822.6 & 145\\
    AD6f4 & 15.8 & 827.6 & 123\\ [0.3 em]

    AD9f1 & 5.4  & 805.5 & 323\\
    AD9f2 & 11.8 & 813.0 & 273\\
    AD9f3 & 15.9 & 833.0 & 263\\
    AD9f4 & 19.5 & 844.0 & 254\\ [0.3 em]

    AD10f1 & 1.9  & 808.0 & 137\\
    AD10f2 & 5.8  & 809.0 & 333\\
    AD10f3 & 12.1 & 826.6 & 269\\
    AD10f4 & 16.5 & 840.0 & 263\\ [0.3 em]

    AD11f2 & 3.4  & 794.0 & 105\\ [0.3 em]

    AD12f3 & 0.8  & 165.5 & 44\\
    AD12f4 & 1.4  & 231.0 & 73\\ [0.3 em]

    AD13f4 & 0.07 & 62.5 & 60\\ 
    \hline
  \end{tabular}
\end{table}


\section{Prospects for nonaxisymmetric rotational instabilities} 
\label{sec:rotinst}

Nonaxisymmetric rotational instabilities in PNSs formed in AIC or iron
core collapse have long been proposed as strong and possibly
long-lasting sources of GWs (see, \eg \cite{ott:09rev} for a recent
review). The postbounce GW emission by nonaxisymmetric deformations of
rapidly rotating PNSs could be of similar amplitude as the signal from
core bounce and, due to its potentially much longer duration, could
exceed it in emitted energy (\eg~\cite{ott_07_b,
  ott_07_a,scheidegger:08}). Moreover, since the characteristic GW
amplitude $ h_{\mathrm c} $ scales with the square root of the number
of cycles, the persistence of the nonaxisymmetric dynamics for many
rotation periods can drastically increase the chances for detection.

The simulations presented in this paper impose axisymmetry, hence we
are unable to track the formation and evolution of rotationally
induced nonaxisymmetric structures. Nonetheless, since the dynamical
high-$ \beta $ instability can develop only at $ \beta $ above $
\beta_{\mathrm{dyn}} \simeq 0.25$ \cite{baiotti_07_a,Manca07}, we can
still assess the prospects for such instabilities by studying the
values of $ \beta $ reached by our AIC models. Moreover, as we shall
see below, the analysis of the rotational configuration of the newly
formed PNS can give a rough idea about the outlook also for low-$
\beta $ instabilities.

As shown in Sec.~\ref{sec:rotating_collapse_dynamics}, for not very
rapidly rotating models, the parameter $ \beta_{\mathrm{ic, b}} $ of
the inner core at bounce increases with the progenitor rotation and
saturates at $ \sim 24.5 \, \% $ (see
Fig.~\ref{fig:betaic_vs_omega}). Immediately after bounce, the inner
core re-expands and, after undergoing several damped oscillations,
settles into a new quasi-equilibrium state with a $ \beta_{\mathrm{ic,
    pb}} $ typically smaller by $ \sim 3 \, \% $ (in relative value)
than that at bounce.  The highest value of $ \beta_{\mathrm{ic, pb}} $
of our entire model set is $\sim 24\,\%$ (observed in model AD12f4)
and most other rapidly rotating models reach values of $
\beta_{\mathrm{ic, pb}} $ that are well below this value (\cf
Tab.~\ref{tab:collapse_models}). Hence, we do not expect the
high-$\beta$ instability to occur immediately after bounce in most AIC
events.

On the other hand, the matter around the PNS experiences rapid
neutrino-cooling (not modeled by our approach) and the PNS contracts
significantly already in the early postbounce phase. This results in
spin-up and in a substantial increase of
$\beta_{\mathrm{ic,pb}}$. Using the VULCAN/2D code,
Ott~\cite{ott:06phd} studied the postbounce evolution of the PNS
rotation of the Dessart~et~al.\ AIC models \cite{dessart_06_a}. He
found that, in the case of the rapidly rotating $ 1.92 M_\odot $
model, the postbounce contraction leads to a growth of $
\beta_{\mathrm{ic,pb}} $ by $ \sim 50 \, \% $ from $ \sim 14 \ \% $ to
$ \sim 22 \ \% $ in the initial $ \sim 50 \ {\mathrm{ms}} $ after
bounce. We expect that a similar increase of the parameter
$\beta_{\mathrm{ic,pb}}$ should take place also for the rapidly
rotating AIC models considered here. More specifically, if we assume
that $ \beta $ increases by $ \sim 50\ \% $ within $ \sim 50
\ {\mathrm{ms}} $ after bounce, we surmise that AIC models with $
\beta_{\mathrm{ic,b}} \gtrsim 17 \, \% $ at bounce should reach $ 
\beta_{\mathrm{ic, pb}} \gtrsim \beta_{\mathrm{dyn}} $ within this
postbounce interval and thus become subject to the high-$\beta$
dynamical instability.

As mentioned in Sec.~\ref{sec:rotating_collapse_dynamics}, uniformly
rotating WDs cannot reach $ \beta_\mathrm{ic,pb} $ in excess of $ \sim
10.5\,\% $. Hence, they are unlikely to become subject to the
high-$\beta$ dynamical nonaxisymmetric instability, but may contract
and spin up to $ \beta \ge \beta_\mathrm{sec} \simeq 14\,\% $ at which
they, in principle, could experience a secular nonaxisymmetric
instability in the late postbounce phase. However, other processes,
e.g., MHD dynamos and instabilities (see, \eg \cite{balbus_91_a,
  cerda_07_a}) may limit and/or decrease the PNS spin on the long
timescale needed by a secular instability to grow.

In addition to the prospects for the high-$\beta$ instability, the
situation appears favorable for the low-$\beta$ instability as
well. The latter can occur at much lower values of $\beta$ as long as
the PNS has significant \emph{differential} rotation (see, \eg
\cite{shibata:04a,watts:05,saijo:06,cerda_07_b,ott_05_a,scheidegger:08,ott_07_b}
and references therein).  While this instability's true nature is not
yet understood, a necessary condition for its development seems to be
the existence of a corotation point inside the star, \ie a point where
the mode pattern speed coincides with the local angular
velocity~\cite{watts:05, saijo_06_a}. Bearing in mind that the lowest
order unstable modes have pattern speeds of the order of the
characteristic Keplerian angular velocity $
\mathcal{O}(\Omega_{\mathrm{char}})$~\cite{centrella_01_a}, we can
easily verify whether such a criterion is ever satisfied in our
models. Assuming a characteristic mass of the early postbounce PNS of
$ \sim 0.8 M_\odot $ and a radius of $ \sim 20 $ km, we obtain a
characteristic Keplerian angular velocity of $ \Omega_{\mathrm{char}}
\sim 4 \ \mathrm{ rad \ ms}^{-1} $. Because most AIC models that reach
$\beta_{\mathrm{ic,pb}} \gtrsim 15 \, \% $ have a peak value of $
\Omega \gtrsim 5 \ \mathrm{rad \ ms}^{-1}$, it is straightforward to
conclude that these models will have a corotation point and, hence,
that the low-$ \beta $ instability may be a generic feature of
rapidly rotating AIC. We note that even uniformly rotating precollapse
models have strong differential rotation in the postshock region
outside the inner core. However, further investigation is needed to
infer whether such models may be also be subject to low-$ \beta $
dynamical instability.

As a concluding remark we stress that the above discussion is based on
simple order-of-magnitude estimates and is therefore rather
inaccurate. Reliable estimates can be made only by performing
numerical simulations in 3D that adequately treat the postbounce
deleptonization and contraction of the PNS and that investigate the
dependence of the instability on $ \beta_{\mathrm{ic, pb}} $, on the
degree of differential rotation, and on the thermodynamic and MHD
properties of the PNS. Finally, these calculations will also establish
what is the effective long-term dynamics of the bar-mode
deformation. In simulations of isolated
polytropes~\cite{baiotti_07_a,Manca07} and from perturbative
calculations~\cite{saijo2008}, it was found that coupling among
different modes tends to counteract the bar-mode instability on a
dynamical timescale after its development.  It is yet unclear whether
this behavior will be preserved also in the AIC scenario, where
infalling material with high specific angular momentum may lead to
significant changes. This will be the subject of future
investigations.


\section{Summary and Conclusions}
\label{sec:summary}

In this paper we have presented the first general-relativistic
simulations of the axisymmetric AIC of massive white dwarfs to
protoneutron stars. Using the general-relativistic hydrodynamics code
\coconut, we performed \nummodel\ baseline model calculations, each
starting from a 2D equilibrium configuration, using a
finite-temperature microphysical EOS, and a simple, yet effective
parametrization scheme of the electron fraction $Y_e$ that provides an
approximate description of deleptonization valid in the collapse,
bounce, and very early postbounce phases.  The precollapse structure
and rotational configuration of WDs that experience AIC is essentially
unconstrained. This prompted us to carry out this work. With our large
set of model calculations, we have investigated the effects on the AIC
evolution of variations in precollapse central density, temperature,
central angular velocity, differential rotation, and deleptonization
in collapse.  The inclusion of general relativity enabled us to
correctly describe the AIC dynamics and our extended model set allowed
us for the first time to study systematically GW emission in the AIC
context.

We find that the overall dynamics in the collapse phase of AIC events
is similar to what has long been established for rotating iron core
collapse. A universal division in homologously collapsing inner core
and supersonically infalling outer core obtains and the
self-similarity of the collapse nearly completely washes out any
precollapse differences in stellar structure in the limit of slow
rotation. Due to the high degeneracy of the electrons in the cores of
AIC progenitor WDs, electron capture is predicted to be strong already
in early phases of collapse \cite{dessart_06_a}, leading to a low
trapped lepton fraction and consequently small inner core masses
$M_\mathrm{ic,b}$ at bounce of around $0.3\,M_\odot$ which decrease
somewhat with increasing precollapse WD temperature due to the
temperature dependent abundance of free protons.  Test calculations
motivated by potential systematic biases of the AIC
$\overline{Y_e}(\rho)$ trajectories obtained from \cite{dessart_06_a}
(see Secs.~\ref{sec:deleptonization} and \ref{sec:tempye}) with
inner-core values of $Y_e$ increased by $\sim 10\%$ and $\sim 20\%$
yielded values $M_\mathrm{ic,b}$ larger by $\sim 11\%$ and $\sim
25\%$.

Our simulations show that rotation can have a profound influence on
the AIC dynamics, but will \emph{always} stay subdominant in the
collapse of uniformly rotating WDs whose initial angular velocity is
constrained by the Keplerian limit of surface rotation. In rapidly
differentially rotating WDs, on the other hand, centrifugal support
can dominated the plunge phase of AIC and lead to core bounce at
subnuclear densities. We find that the parameter $\beta_\mathrm{ic,b}
= (E_\mathrm{rot}/|W|)_\mathrm{ic,b}$ of the inner core at bounce
provides a unique mapping between inner core rotation and late-time
collapse and bounce dynamics, but the mapping between precollapse
configurations and $\beta_\mathrm{ic,b}$ is highly degenerate, i.e.,
multiple, in many cases very different precollapse configurations of
varying initial compactness and total angular momentum, can yield
practically identical $\beta_\mathrm{ic,b}$ and corresponding
collapse/bounce dynamics. 

Recent phenomenological work presented in
\cite{metzger:09,metzger:09b} on the potential EM display of an AIC
event has argued for both uniform WD rotation
\cite{metzger:09,piro_08_a} and massive quasi-Keplerian accretion
disks left behind at low latitudes after AIC shock 
passage. The analysis of our extensive model set, on the other hand,
shows that uniformly rotating WDs produce no disks at all or, in
extreme cases that are near mass shedding at the precollapse stage,
only very small disks ($M_\mathrm{disk} \lesssim 0.03 M_\odot$). Only
rapidly differentially rotating WDs yield the large disk masses needed
to produce the enhanced EM signature proposed in
\cite{metzger:09,metzger:09b}.

An important focus of this work has been on the GW signature of
AIC. GWs, due to their inherently multi-D nature, are ideal messengers
for the rotational dynamics of AIC.  We find that all AIC models
following our standard $\overline{Y_e}(\rho)$ parametrizations yield
GW signals of a generic morphology which has been classified
previously as \emph{type~III}
\cite{zwerger_97_a,dimmelmeier_02_b,ott:06phd}.  This signal type is
due primarily to the small inner core masses at bounce obtained in
these models.  We distinguish
between three subtypes of AIC GW signals. Type IIIa occurs for
$\beta_\mathrm{ic,b} \lesssim 0.7\%$ (slow rotation), is due in part
to early postbounce prompt convection and results in peak GW
amplitudes $|h_\mathrm{max}| \lesssim 5 \times 10^{-22}$ (at
$10\,\mathrm{kpc}$) and emitted energies $E_\mathrm{GW} \lesssim
\mathrm{few}\,\times\,10^{-9}\,M_\odot \, c^2$. Most of our AIC models
produce type~IIIb GW signals that occur for $0.7\lesssim
\beta_\mathrm{ic,b} \lesssim 18\%$ (moderate/moderately rapid
rotation) and yield $6\times10^{-22} \lesssim
|h_\mathrm{max}|\,(\mathrm{at}\,10\,\mathrm{kpc}) \lesssim
8\times10^{-21}$ and emitted energies of $9\times10^{-10} M_\odot\,c^2
\lesssim E_\mathrm{GW} \lesssim 2\times10^{-8}\,M_\odot\,c^2$.
Rotation remains subdominant in type~IIIa and type~IIIb models and we
find that there is a monotonic and near-linear relationship between
maximum GW amplitude and the rotation of the inner core which is best
described by the power law $|h_\mathrm{max}| \propto 10^{-21}
\beta_\mathrm{ic,b}^{0.74}$. Furthermore, we find that the frequencies
$f_\mathrm{max}$ at which the GW spectral energy densities of type
IIIa and IIIb models peak are in a rather narrow range from $\sim
720\,\mathrm{Hz}$ to $\sim 840\,\mathrm{Hz}$ and exhibit a monotonic
growth from the lower to the upper end of this range with increasing
rotation. This finding suggests that the GW emission in these models
is driven by the fundamental quadrupole (${}^2\!f$) mode of the inner
core.

In the dynamics of AIC models that reach $\beta_\mathrm{ic,b} \gtrsim
18\%$, centrifugal effects become dominant and lead to core bounce at
subnuclear densities. Such models must be differentially rotating at
the onset of collapse and produce type~IIIc GW signals with maximum
amplitudes of $4.0\times10^{-22} \lesssim
|h_\mathrm{max}|\,\mathrm{(at\, 10\,\mathrm{kpc})} \lesssim
5.5\times10^{-21}$, emitted energies of $ 10^{-10} M_\odot\,c^2
\lesssim E_\mathrm{GW} \lesssim 10^{-8} M_\odot\, c^2$, and peak
frequencies of $62\, \mathrm{Hz} \lesssim f_\mathrm{max} \lesssim 800
\, \mathrm{Hz}$. In contrast too type~IIIa and IIIb models, in
type~IIIc models, $|h_\mathrm{max}|$, $E_\mathrm{GW}$, and
$f_\mathrm{max}$ decrease monotonically with increasing
$\beta_\mathrm{ic,b}$.

Combining the information from signal morphology, $|h_\mathrm{max}|$,
$E_\mathrm{GW}$ and $f_\mathrm{max}$, we conclude that already
first-generation interferometer GW detectors should be able to infer
the rotation of the inner core at bounce (as measured by
$\beta_\mathrm{ic,b}$) from a Galactic AIC event. Due to the
degenerate dependence of $\beta_\mathrm{ic,b}$ on initial model
parameters, this can put only loose constraints on the structure and
rotational configuration of the progenitor WD.  However, the
observation of an AIC with $\beta_\mathrm{ic,b} \gtrsim 18\%$ would
rule out uniform progenitor rotation.

Studying the configurations of the protoneutron stars formed in our
AIC models, we find that none of them are likely to experience the
high-$\beta$ nonaxisymmetric bar-mode instability at very early
postbounce times. We estimate, however, that all models that reach
$\beta_\mathrm{ic,b} \gtrsim 17\%$ will contract and reach the
instability threshold within $\sim 50\,\mathrm{ms}$ after bounce.
Less rapidly spinning models will require more time or will go
unstable to the low-$\beta$ instability. The latter requires strong
differential rotation which is ubiquitous in the outer PNS and in the
postshock region of our AIC models.  AIC progenitors, due to their
evolution through accretion or formation through merger, are
predestined to be rapidly rotating and form PNSs that are likely to
become subject to nonaxisymmetric instabilities. This is in contrast
to the precollapse iron cores of ordinary massive stars that are
expected to be mostly slowly-spinning
objects~\cite{heger:05,ott:06spin}. We conclude that the appearance of
nonaxisymmetric dynamics driven by either the low-$\beta$ or
high-$\beta$ instability and the resulting great enhancement of the GW
signature may be a generic aspect of AIC and must be investigated in
3D models.

The comparison of the GW signals of our axisymmetric AIC models with
the gravitational waveforms of the iron core collapse models of
Dimmelmeier~et~al.~\cite{dimmelmeier_08_a} reveals that the overall
characteristics of the signals are rather similar. It appears unlikely
that AIC and iron core collapse could be distinguished on the basis of
the axisymmetric parts of their GW signals alone, unless detailed
knowledge of the signal time series as well as of source orientation
and distance is available to break observational degeneracies.

The results of our AIC simulations presented in this paper and the
conclusions that we have drawn on their basis demonstrate the complex
and in many cases degenerate dependence of AIC outcomes and observational
signatures on initial conditions. The observation of GWs from an AIC
event can provide important information on the rotational dynamics of
AIC.  However, to lift degeneracies in model parameters and gain
full insight, GW observations must be complemented by observations of
neutrinos and electromagnetic waves.  These multi-messenger
observations require underpinning by comprehensive and robust
computational models that have no symmetry constraints and include all
the necessary physics to predict neutrino, electromagnetic, and GW
signatures.

As a point of caution, we note that the generic type~III GW signal
morphology observed in our AIC models is due to the small inner-core
values of $Y_e$ and consequently small inner core masses predicted by
the $\overline{Y_e}(\rho)$ parametrization obtained from the
approximate Newtonian radiation-hydrodynamic simulations of
\cite{dessart_06_a}.  Tests with artificially reduced deleptonization
show that the signal shape becomes a mixture of type~III found in our
study and type I observed in rotating iron core collapse
\cite{dimmelmeier_08_a} if the $Y_e$ in the inner core is larger by
$\sim 20\%$.  In a follow-up study, we will employ
$\overline{Y_e}(\rho)$ data from improved general-relativistic
radiation-hydrodynamics simulations \cite{mueller:09phd} to better
constrain the present uncertainties of the AIC inner-core electron
fraction.

Although performed using general-relativistic hydrodynamics, the
calculations discussed here are limited to conformally-flat spacetimes
and axisymmetry. We ignored postbounce deleptonization, neutrino
cooling, and neutrino heating.  We also neglected nuclear burning,
employed only a single finite-temperature nuclear EOS, and were forced
to impose ad-hoc initial temperature and electron fraction
distributions onto our precollapse WD models in rotational
equilibrium.  Future studies must overcome the remaining limitations
to build accurate models of AIC.  Importantly, extensive future 3D
radiation-hydrodynamic simulations are needed to address the range of
possible, in many cases probably nonaxisymmetric, postbounce
evolutions of AIC and to make detailed predictions of their signatures
in GWs, neutrinos, and in the electromagnetic spectrum.


\section{Acknowledgements}

It is a pleasure to thank Alessandro Bressan, Adam Burrows, Frank
L\"offler, John Miller, Stephan Rosswog, Nikolaos Stergioulas,
Sung-Chul Yoon, Shin Yoshida, and Burkhard Zink for helpful comments
and discussions.  This work was supported by the Deutsche
Forschungsgemeinschaft through the Transregional Collaborative
Research Centers SFB/TR~27 ``Neutrinos and Beyond'', SFB/TR~7
``Gravitational Wave Astronomy'', and the Cluster of Excellence
EXC~153 ``Origin and Structure of the Universe''
(\texttt{http://www.universe-cluster.de}). CDO acknowledges partial
support by the National Science Foundation under grant
no.\ AST-0855535.  The simulations were performed on the compute
clusters of the Albert Einstein Institute, on machines of the
Louisiana Optical Network Initiative under allocation LONI\_numrel04
and on the NSF Teragrid under allocation TG-MCA02N014.

\end{document}